\documentclass[aps,pra,10pt,twocolumn,floatfix,nofootinbib,superscriptaddress,longbibliography,nobibnotes]{revtex4-2}

\usepackage[titles]{tocloft}
\usepackage{longtable}
\usepackage{setspace}

\usepackage{amsmath}
\usepackage{amssymb}
\usepackage{mathtools}
\usepackage{wasysym}
\usepackage{newtxmath}
\usepackage{xcolor}
\usepackage{tikz}
\usepackage{tikz-cd}

\usepackage{array} 
\usepackage[T1]{fontenc}
\usepackage{newtxtext}
\usepackage{dsfont}                 
\usepackage{bbold}                  
\usepackage[normalem]{ulem}         

\usepackage[version=4]{mhchem}

\usepackage{enumerate}
\usepackage[shortlabels]{enumitem}

\usepackage[dvipsnames,table]{xcolor}
\usepackage{graphicx}
\graphicspath{{figs/}}              

\usepackage[flushleft]{threeparttable}
\makeatletter 
\g@addto@macro\TPT@defaults{\footnotesize} 
\makeatother
\usepackage{multirow}

\usepackage{physics}  
\usepackage{csquotes}               
\usepackage{comment}
\usepackage{placeins}

\usepackage[caption=false]{subfig}
\usepackage[
    colorlinks=true,
    linkcolor=magenta,
    citecolor=magenta,
    linktocpage=true,
    urlcolor=magenta
]{hyperref}
\usepackage[capitalize]{cleveref}

\usepackage{orcidlink}

\usepackage{needspace}
\makeatletter
\let\ORI@startsection\@startsection
\newlength{\sec@reserve}
\renewcommand{\@startsection}[6]{%
  \par
  \if@nobreak
    \nobreak                      
  \else
    \ifnum#2<2
      \setlength{\sec@reserve}{6\baselineskip}
    \else
      \setlength{\sec@reserve}{4\baselineskip}
    \fi
    \Needspace*{\sec@reserve}%
  \fi
  \ORI@startsection{#1}{#2}{#3}{#4}{#5}{#6\interlinepenalty\@M}%
}
\makeatother

\DeclareMathAlphabet{\mathbbold}{U}{bbold}{m}{n}
\DeclareMathOperator{\diag}{diag}
\DeclareMathOperator{\sign}{sign}
\DeclareMathOperator{\nullop}{null}
\DeclareMathOperator{\Eu}{Eu}
\DeclareMathOperator{\Pf}{Pf}

\def\imi{\mathrm{i}}
\def\imj{\mathrm{j}}
\def\imk{\mathrm{k}}

	\def\ket#1{|{#1}\rangle}		
	\def\braket#1#2{\langle{#1}|{#2}\rangle}				

\def\intg{\mathbbold{Z}}					
\def\ztwo{\mathbbold{Z}_2}					
\def\unit{\mathbbold{1}}					
\def\reals{\mathbbold{R}}					
\def\cmplx{\mathbbold{C}}					
\def\quats{\mathbbold{H}}					
\def\bs#1{\boldsymbol{#1}}

\cftsetindents{section}{0em}{2.05em}
\cftsetindents{subsection}{0.75em}{1.65em}
\cftsetindents{subsubsection}{1.5em}{1.25em}

\definecolor{TB}{rgb}{0.93,0.47,0.2}

\usepackage{bm}

\definecolor{ZG}{rgb}{0.0, 0.56, 0.15}

\usepackage{bm}

\definecolor{apoorv}{rgb}{0.0, 0.5, 0.0}

\makeatletter
\let\oldfnsymbol\@fnsymbol
\def\@fnsymbol#1{\ensuremath{\ifcase#1\or \dagger\or \ddagger\or \mathsection\or \mathparagraph\or \|\or \dagger\dagger\or \ddagger\ddagger\else\@ctrerr\fi}}
\makeatother

\allowdisplaybreaks

\begin{document}


\title{Topological characterization of multifold band degeneracies in Altland-Zirnbauer symmetry classes}

\author{Askar Iliasov\textsuperscript{*}\,\orcidlink{0000-0003-2409-7292}}
\email{askar.iliasov@physik.uzh.ch}
\affiliation{Department of Physics, University of Zurich, Winterthurerstrasse 190, 8057 Zurich, Switzerland}

\author{Zolt\'{a}n Guba\textsuperscript{*}\,\orcidlink{0000-0002-6130-1064}}
\email{zoltan.guba@physik.uzh.ch}
\affiliation{Department of Physics, University of Zurich, Winterthurerstrasse 190, 8057 Zurich, Switzerland}

\author{Tsuneya Yoshida\,\orcidlink{0000-0002-2276-1009}}
\affiliation{Department of Physics, Kyoto University, Kyoto 606-8502, Japan}
\affiliation{Institute for Theoretical Physics, ETH Zurich, 8093 Zurich, Switzerland}

\author{Apoorv Tiwari\,\orcidlink{0000-0003-4742-775X}
}
\affiliation{Center for Quantum Mathematics, University of Southern Denmark}
\affiliation{Danish Institute of Advanced Study, University of Southern Denmark}

\author{Tom\'{a}\v{s} Bzdu\v{s}ek\,\orcidlink{0000-0001-6904-5264}}
\email{tomas.bzdusek@uzh.ch}
\affiliation{Department of Physics, University of Zurich, Winterthurerstrasse 190, 8057 Zurich, Switzerland}


\begin{abstract}
Topological degeneracies of energy bands in crystalline matter are conventionally characterized by invariants computed on an enclosing sphere over which the spectrum remains gapped, a program completed for minimal degeneracies in all ten Altland-Zirnbauer (AZ) symmetry classes.
Higher-order degeneracies have instead been studied almost exclusively under crystalline-symmetry protection.
Here, we characterize generic $n$-fold degeneracies stabilized solely by AZ symmetries acting locally in momentum space.
Their codimension grows quadratically with $n$, placing multifold nodes in parameter spaces combining momenta with tuning parameters or synthetic dimensions.
Crucially, the enclosing-sphere paradigm faces a fundamental obstruction: two $(n\,{-}\,1)$-fold degeneracy loci emanating from the $n$-fold node necessarily pierce every choice of enclosing sphere, leaving no uniform spectral gap and thus no standard homotopy classification.
We elevate this obstruction into the diagnostic itself.
Namely, on the two nodal manifolds where the loci cross the sphere, complementary spectral gaps are restored, admitting conventional band invariants (Chern numbers, Stiefel-Whitney classes, and winding numbers).
This observation establishes a general two-way correspondence: (1)~the multifold node is topologically protected whenever the associated nodal manifolds are robustly linked, and (2)~invariants on cycles of one manifold encode their linking numbers with cycles of the other.
Carrying out this program for minimal models of all ten AZ classes, we recast multifold band topology as the topology of linked nodal manifolds and lay the foundation for characterizing multifold nodes in models with arbitrarily many bands.
\end{abstract}

\maketitle

\makeatletter
\let\@fnsymbol\oldfnsymbol
\makeatother

\renewcommand{\thefootnote}{\fnsymbol{footnote}}
\footnotetext{These authors contributed equally.}
\renewcommand{\thefootnote}{\arabic{footnote}}
\setcounter{footnote}{0}



\setcounter{tocdepth}{1}
\onecolumngrid

\vspace{-5mm}

\begin{center}
    \begin{minipage}{0.68\textwidth} 
    \makeatletter
    \renewcommand{\baselinestretch}{1.17}
    \makeatletter
    {\small 
    \tableofcontents 
    }
    \makeatother
    \renewcommand{\baselinestretch}{1.0}
    \end{minipage}
\end{center}

\vspace{-3mm}
\twocolumngrid


\section{Introduction}

Which degeneracies of energy levels are robust, and which can be lifted by a perturbation?
First addressed by von Neumann and Wigner nearly a century ago~\cite{vonNeumann:1929}, this question has since evolved into a central organizing principle of topological band theory.
Depending on symmetry and dimensionality, the electron energy bands of crystalline solids (as well as their photonic~\cite{Lu:2013,Lu:2015} and acoustic~\cite{Xiao:2015,He:2018} counterparts) may touch at isolated points, along lines, or over entire surfaces in the three-dimensional momentum space, defining the rich families of nodal-point~\cite{Murakami:2007,Wan:2011,Young:2012,Yang:2015b,Armitage:2018}, nodal-line~\cite{Burkov:2011,Weng:2015b,Fang:2016,Schoop:2016,Bzdusek:2016}, and nodal-surface~\cite{Zhong:2016,Agterberg:2017,Wu:2018} semimetals and superconductors.
Such band nodes are typically not accidental: they carry quantized topological charges that obstruct their removal by symmetry-preserving perturbations~\cite{Horava:2005,Zhao:2013,Fang:2015,Bzdusek:2017,Volovik:2003}, and that translate into distinctive observables, ranging from anomalous electromagnetic responses~\cite{Nielsen:1983,Zyuzin:2012,Son:2013,Hosur:2013,Huang:2015,Xiong:2015} to unconventional boundary modes such as Fermi arcs~\cite{Xu:2015,Lv:2015} and drumhead surface states~\cite{Heikkila:2011,Kim:2015,Bian:2016}.
Determining which band degeneracies are topologically protected---under which symmetries, and by which invariants---has become a foundational program~of~the~field.

The canonical characterization of a band node, temporarily disregarding the role of spatial symmetries, rests on two ingredients. 
The first is codimension counting: via the classic counting argument~\cite{vonNeumann:1929}, 
one determines the number $\delta$ of real parameters that must be tuned to bring energy bands into coincidence, given the antiunitary symmetries that act locally in momentum space and that are organized into the tenfold way of Altland and Zirnbauer (AZ)~\cite{Altland:1997}.
In a momentum (or, more generally, parameter) space of dimension $\delta$, band degeneracies then generically occur at isolated points. 
The second ingredient is the enclosing sphere: since the spectrum remains gapped everywhere on a small sphere $S^{\ell}$, $\ell=\delta-1$, surrounding the node, the restricted Hamiltonian defines a map from $S^{\ell}$ to a classifying space of gapped Hamiltonians, and the topological charge of the node is an element of the corresponding homotopy group~\cite{Zhao:2016,Bzdusek:2017}. 
For \emph{minimal} degeneracies---twofold band touchings, supplemented by Kramers or particle-hole doubling where applicable---this program was brought to completion for all ten AZ classes in Ref.~\citenum{Bzdusek:2017}, in the setting where time-reversal and particle-hole symmetry preserve the sign of $\bs{k}$.
The latter assumption is automatic when the components of $\bs{k}$ are tuning parameters rather than crystal momenta, and it equally describes momentum-space nodes protected by symmetries composed with spatial inversion, which has motivated the name AZ$+\mathcal{I}$ classes~\cite{Bzdusek:2017}. 
The resulting invariants include the Chern number of Weyl points, the quantized Berry phase and the $\mathbb{Z}_2$ monopole charge of nodal-line rings~\cite{Fang:2015}, the winding numbers of chiral-symmetric nodes, and the Pfaffian sign that stabilizes nodal (Bogoliubov-Fermi) surfaces~\cite{Agterberg:2017,Zhong:2016}.
The topological characterization of band nodes is further enriched if spatial symmetries, in particular mirror and rotation symmetries, are taken into consideration~\cite{Michel:1999,Fang:2012,Chiu:2014,Wieder:2016,Bradlyn:2016,Zhu:2016,Lenggenhager:2022a,Yu:2022}.

Remarkably, even for twofold degeneracies, these enclosing-sphere charges carry information beyond a single energy gap.
In spinless $\mathcal{P}\mathcal{T}$-symmetric systems (symmetry class AI), a nodal ring exhibits a nontrivial monopole charge (computable as the second Stiefel-Whitney invariant on the enclosing two-sphere) if and only if it is linked with an odd number of nodal lines formed in the adjacent energy gap~\cite{Ahn:2018}.
Analogous correspondences between higher-dimensional monopole charges and linking were established for doubly charged nodal surfaces of centrosymmetric superconductors in Bogoliubov-de Gennes (BdG) symmetry
classes~\cite{Kim:2021}, and for Weyl nodal surfaces of five-dimensional class-A systems, whose second Chern number counts their linking with nodal surfaces in the adjacent gap~\cite{Lian:2016}.
All of the above correspondences were mathematically established with the help of cohomological tools, suggesting the indispensable role of cohomology groups besides homotopy groups in the full characterization of band nodes.
In parallel, the discovery of non-Abelian frame charges and of the reciprocal braiding of nodes residing in consecutive energy gaps~\cite{Ahn:2019,Wu:2019,Bouhon:2020,Tiwari:2020} made explicit that band nodes of individual gaps cannot, in general, be characterized in isolation, and stimulated a broad exploration of multigap topological
phases~\cite{Bouhon:2020b,Lim:2023,Davoyan:2024,Jankowski:2024,Jankowski:2024b,Jankowski:2025b,Jankowski:2025,Lim:2025}.
These developments, which are closest in spirit to the present work, single out linking as a fundamental, yet so far only sporadically explored~\cite{Moore:2008,Yan:2017,Chen:2017,Lenggenhager:2021} ingredient of nodal band topology. In the present work, linking takes center stage: it becomes the very diagnostic of multifold band degeneracies.

Band degeneracies of order higher than two have, until now, been investigated almost exclusively under the protection of crystalline symmetry.
Triple nodal points arise along high-symmetry lines of a three-dimensional Brillouin zone where a two-dimensional irreducible representation (irrep) of the little group crosses a one-dimensional irrep~\cite{Heikkila:2015,Zhu:2016}; their complete classification, according to the absence (type A) or presence (type B) of attached nodal-line arcs, has been achieved both for spinful~\cite{Zhu:2016} and for
spinless~\cite{Lenggenhager:2022a} band structures.
Threefold degeneracies can alternatively be pinned to high-symmetry points as three-dimensional irreps~\cite{Bradlyn:2016,Feng:2021,Tian:2021}, or occur at zero momentum among Goldstone modes~\cite{Park:2021}. 
Similarly, fourfold Dirac points formed by crossings of Kramers-degenerate bands have been classified for all space groups~\cite{Yang:2014}, and nonsymmorphic symmetries may enforce band degeneracies of even higher
order~\cite{Wieder:2016,Bradlyn:2016}.
Many of these crystalline multifold fermions have also been studied experimentally: angle-resolved photoemission has resolved triple-point fermions in \ce{MoP} and \ce{WC}~\cite{Lv:2017,Ma:2018}, fourfold Dirac points in \ce{Na3Bi} and \ce{Cd3As2}~\cite{Liu:2014,Liu:2014b,Neupane:2014,Borisenko:2014}, and higher-fold chiral fermions with long Fermi-arc states in the \ce{CoSi} structural family~\cite{Rao:2019,Sanchez:2019,Schroter:2019,Takane:2019,Schroter:2020}.
In all of these settings, the crystalline symmetry plays a dual role: it lowers the codimension of the multifold degeneracy, thereby making it accessible in three spatial dimensions, but it simultaneously confines the degeneracy to high-symmetry points or lines inside the Brillouin zone, with its protection expressed through the representation theory of little groups~\cite{Bradley:1972}.

Here, we pose a complementary question: what characterizes and protects an $n$-fold band degeneracy when no crystalline symmetry is invoked, so that its stability derives solely from the AZ symmetries?
As a first result, we derive the codimension $\delta^{\mathrm{CL}}_{(n)}$ of a generic $n$-fold degeneracy in each symmetry class (with Cartan label CL, and with Kramers and particle-hole doublings accounted for where appropriate), thus generalizing the analysis of Refs.~\citenum{vonNeumann:1929} and~\citenum{Bzdusek:2017}.
We show that the codimension in question grows \emph{quadratically} with $n$ for all AZ symmetry classes [Table~\ref{tab:codims-WD+balanced}].
While this places generic multifold nodes out of reach of conventional three-dimensional band structures, the lowest instances require only a handful of control parameters: for example, a threefold degeneracy of spinless $\mathcal{P}\mathcal{T}$-symmetric bands (class AI) forms at codimension five, and a fourfold zero-energy degeneracy in the chiral-symmetric class BDI forms at codimension four. Such parameter counts are attainable by supplementing momenta with tuning parameters such as hopping amplitudes and strain, by considering adiabatic pumping cycles~\cite{Kraus:2012,Lohse:2018,Zilberberg:2018}, or by exploiting synthetic dimensions~\cite{Celi:2014,Yuan:2018,Ozawa:2019}.
Along these routes, four-dimensional topological responses have already been measured in cold-atom, photonic, and circuit platforms~\cite{Lohse:2018,Zilberberg:2018,Wang:2020}.
Viewed within such enlarged parameter spaces, the crystalline realizations reviewed above emerge
as symmetry-constrained slices of the universal setting
studied in the present work, with fine tuning traded for point-group representation theory.

Extending the enclosing-sphere paradigm to multifold nodal points, however, runs into a fundamental obstruction that is absent in the study of the minimal band degeneracies. 
Namely, while an $n$-fold node can still be enclosed by a sphere $S^{\ell}$ with  $\ell=\delta^{\mathrm{CL}}_{(n)}-1$, the spectrum on this sphere is \emph{not gapped}.
The reason is that the $n$-fold degeneracy arises as the intersection of loci supporting degeneracies of lower order (i.e., the loci $\mathcal{L}_1$ and $\mathcal{L}_2$ on which, respectively, the lower and the upper $n-1$ of the $n$ bands coincide), and these loci inevitably pierce every sphere that encloses the node [Fig.~\ref{fig:main_scheme}]. 
With no energy gap maintained uniformly over $S^{\ell}$, the Hamiltonian on the enclosing sphere does not define a map to any of the standard single-gap or multigap classifying spaces, and the established homotopy-theoretic classification machinery becomes inapplicable.
Although the minimal (i.e., $n$-band) model admits a formal assignment of a topological charge, namely the integer winding number $\pi_{\ell}(S^{\ell})=\mathbb{Z}$ of the map $S^{\ell}\to S^{\ell}$ defined by the suitably normalized Hamiltonian, this assignment is an artifact of the minimal setting, in which the space of normalized Hamiltonians is itself a sphere. Once additional bands are included, this identification is lost. Moreover, being defined at the level of Hamiltonian matrix elements, the winding number lacks a formulation in terms of gauge-invariant data extracted from the Bloch states, and hence provides no route to diagnosing multifold nodes in realistic many-band~settings.

The key insight of our work is to turn the obstruction itself into the diagnostic. 
Specifically, narrowing our attention to the intersections of the degeneracy loci with the enclosing sphere, $M_1=\mathcal{L}_1\cap S^{\ell}$ and $M_2=\mathcal{L}_2\cap S^{\ell}$, henceforth dubbed \emph{nodal manifolds}, we find that a complementary spectral gap is restored on each of them, allowing their characterization with well-defined topological band invariants.
We then establish a two-way correspondence.
On one hand, we show that the multifold node is topologically protected if the nodal manifolds $M_1$ and $M_2$ remain linked inside the enclosing sphere $S^\ell$ under all admissible perturbations; the underlying notion of linking is made precise by resorting to a homological cell-decomposition of the nodal manifolds. 
On the other hand, we reveal that the familiar band invariants (including Stiefel-Whitney classes, Chern numbers, and winding numbers) evaluated on nontrivial cycles of one nodal manifold directly encode the linking numbers with nontrivial cycles of the complementary nodal manifold.
Intuitively, the nodal manifold in one gap acts as a monopole source of curvature, the quantized flux of which is registered as a topological invariant on the cycles of the other nodal manifold. 
The topology of a multifold nodal point is thereby recast as the topology of linked nodal manifolds, which, owing to the robustness of band invariants derived from Bloch bundles, remains a faithful diagnostic of $n$-fold band nodes in models with arbitrarily many~bands.

We develop this program across all ten AZ symmetry classes, focusing on the lowest-order multifold nodes beyond the minimal degeneracies of Ref.~\citenum{Bzdusek:2017}. 
For each class, we first adapt the von Neumann-Wigner counting [Table~\ref{tab:codims-WD+balanced}].
Second, adopting the minimal (i.e., $n$-band) Hamiltonian, we explicitly identify the nodal manifolds [Table~\ref{tab:nodal_manifolds}].
Finally, we extract nontrivial cycles of the nodal manifolds in the minimal model, and for the cycles of sufficiently low dimension, we explicitly determine (using either analytical arguments or numerical computations) the topological band invariants carried by the degenerate bands [Table~\ref{tab:nodal_manifolds}].
We find that nodal manifolds at non-zero energy (labeled $M_1$, $M_2$, and $M_\varnothing$) coincide with certain well-known classifying spaces of single-gap Hamiltonians, while nodal manifolds at zero energy (labeled $M_\medcirc$) correspond to classifying spaces of certain chiral or particle-hole-symmetric Hamiltonians with two energy gaps that, to our knowledge, have not been previously investigated.
As a consistency check, we also show that our framework reproduces, in its lowest-dimensional corollaries, the established linking phenomenology of twofold nodes~\cite{Ahn:2018,Kim:2021,Lian:2016}.

Although we focus on the simplest nontrivial settings, i.e., on $n$-fold points in $n$-band models, our construction charts a broader program: the correspondence between band invariants and linking extends naturally to models with more than $n$ bands, where a multifold node is protected by a network of pairwise linked nodal manifolds, as well as to multifold nodal lines and surfaces, and to characterizations rooted in cobordism theory.
Moreover, the quadratic growth of the codimension stands in marked contrast to the \emph{linear} growth established for multifold exceptional points of non-Hermitian Hamiltonians~\cite{Delplace:2021,Yoshida:2025}, implying constraints on Hermitian nodal structures, such as the inability of higher-fold degeneracy loci to cross, that lack non-Hermitian counterparts.

The remainder of the paper is organized as follows. 
In Sec.~\ref{sec:general-section} we present the general strategy to characterize multifold band degeneracies: we first outline the codimension analysis and the definition of the nodal manifolds, and subsequently formulate higher-dimensional linking together with a differential-form argument that identifies band invariants with linking numbers.
The section concludes with a brief overview of the results and methods aimed at readers who choose to skip the mathematical discussion.
Sections~\ref{sec:class-AI}--\ref{sec:class-DIII} then carry out the outlined program case by case: we begin with the Wigner-Dyson classes $\textrm{AI}$, $\textrm{A}$, and $\textrm{AII}$ (Secs.~\ref{sec:class-AI}--\ref{sec:class-AII}), treating the class $\textrm{AI}$ with a particular level of detail owing to its low codimension, transparent geometry, and relevance to spinless band structures.
We continue with the chiral classes $\textrm{BDI}$, $\textrm{AIII}$, and $\textrm{CII}$ (Secs.~\ref{sec:class-BDI}--\ref{sec:class-CII}) and with the BdG classes $\textrm{D}$, $\textrm{C}$, $\textrm{CI}$, $\textrm{DIII}$ (Secs.~\ref{sec:D_class}--\ref{sec:class-DIII}).
Section~\ref{sec:conclude} presents a concise summary of our findings and elaborates on the prospective research directions anticipated above, including the extension to higher-fold degeneracies. Technical material (including explicit perturbative counting of codimensions, derivation of classifying spaces with zero-energy states, formulation of linking numbers via Thom forms, and clarifying remarks on the notion of robust linking) is delegated~to~Appendices~\ref{app:explicit_counting}--\ref{app:cobordism-transform}.

\section{General strategy}\label{sec:general-section}
\subsection{Topological characterization of nodal points \texorpdfstring{\\}{} via enclosing sphere}\label{sec:general-enclosing-sphere}

In this section, we detail the main questions we aim to address, as well as the strategies we utilize. 
We begin with the explanation of how $n$-fold nodal points in $n\times n$ matrices can be characterized by elementary homotopy groups of spheres.
Subsequently, we establish a correspondence between the presence of $n$-fold nodal points and the linking numbers of appropriately defined manifolds supporting $(n-1)$-fold band degeneracies in the neighborhood of the $n$-fold nodal point.
The general correspondence between (cohomological) bundle invariants, linking, and multifold nodal points is expected to generalize to $n$-fold degeneracies in models with more than $n$ bands, although the precise classification result may then be modified by additional unstable or reduced stable invariants.
Such a more complete characterization requires further investigation and lies beyond the scope of the present work.

To characterize $n$-fold nodal points, we first derive the codimension for forming an $n$-fold degeneracy in each of the ten available Altland-Zirnbauer symmetry classes~\cite{Altland:1997}. 
This codimension does not depend on the total number of bands, i.e., this result applies to both the minimal case of $n$ band models as well as the more general case of models with more than $n$ bands.
For the three Wigner-Dyson symmetry classes, a derivation based on the von Neumann-Wigner counting~\cite{vonNeumann:1929} is available.
Meanwhile, for the three chiral symmetry and four Bogoliubov-de Gennes (BdG) classes, we present an adapted version of such counting arguments.
We also confirm that it is possible to correctly reproduce the codimension of an $n$-degeneracy by counting the number of linearly independent perturbations that allow us to split the $n$-fold degeneracy in a minimal (i.e., $n$-band) model (see Appendix.~\ref{app:explicit_counting}).

When discussing symmetric classes, we assume that  Altland-Zirnbauer symmetries leave the parameter $\bs{k}$ invariant (i.e., that time-reversal and particle-hole symmetries do not change the sign of $\boldsymbol{k}$).
The absence of the sign flip can be interpreted either as (1)~$\bs{k}$ not being a crystal momentum, but rather a set of system parameters (such as hopping amplitudes) that remain unchanged under the AZ symmetry, or (2)~as compositions of the $\bs{k}$-flipping symmetries with spatial inversion (i.e., with the unitary that maps $\bs{k}\mapsto - \bs{k}$). 
The latter interpretation has been adopted by Ref.~\citenum{Bzdusek:2017}, which dubbed the resulting scheme as $\textrm{AZ}+\mathcal{I}$ symmetry classes; however, from the mathematical perspective, the cases~(1) and~(2) are indistinguishable and equivalent. 
Our results thus extend the characterization of minimal band degeneracies in $\textrm{AZ}+\mathcal{I}$ classes by Ref.~\citenum{Bzdusek:2017} to the significantly more intricate case of \emph{multifold} band degeneracies in the same symmetry setups.

\begin{figure}[t]
    \centering
    \includegraphics[width = \linewidth]{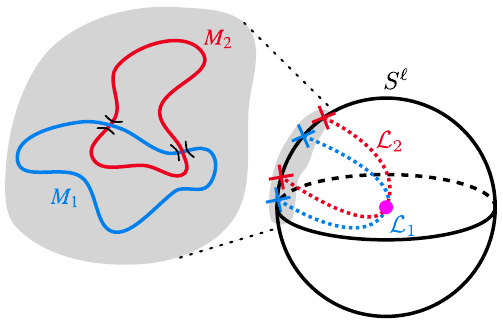}
    \caption{A schematic illustration of our approach to topological characterization of an $n$-fold nodal point in a minimal (i.e., $n$-band) model. The nodal point, which lies at
    the intersection of the
    loci $\mathcal{L}_1$ and $\mathcal{L}_2$ of a lower-degree band degeneracy, is enclosed by a sphere $S^{\ell}$, where $\ell+1$ is the codimension of forming an $n$-fold nodal point. 
    We study 
    topological band invariants on the nodal manifolds $M_1=\mathcal{L}_1\cap S^{\ell}$ and $M_2=\mathcal{L}_2\cap S^{\ell}$. 
    With this approach, we characterize topological protection of the multifold nodal point as well as the linking of the nodal manifolds, which enables us to establish a correspondence between the two characteristics.
    }
\label{fig:main_scheme}
\end{figure}

The following observation is already indicative of the topological invariant that stabilizes the $n$-fold degeneracy in minimal models.
Namely, let the codimension for forming the degeneracy be denoted as $\delta^{\textrm{CL}}_{(n)}$, where $\textrm{CL}$ is a short-hand notation for the Cartan label of the respective symmetry class.
Setting $\ell=\delta^{\textrm{CL}}_{(n)}-1$, the nodal point can be enclosed by an $S^{\ell}$ sphere in parameter space.
For minimal models, the space of symmetry allowed perturbations that split the $n$-fold degeneracy is a real vector space of dimension $\delta^{\textrm{CL}}_{(n)}$, on which a unit sphere of (suitably normalized) perturbations is $S^{\ell}$.
Then the restricted Hamiltonian defines a map $S^{\ell}\to S^{\ell}$, for which the relevant invariant is the homotopy group $\pi_{\ell}(S^\ell) =\intg$.

We emphasize that, in classes with spinful time-reversal symmetry, we do not count the Kramers doubling toward the degree $n$ of a degeneracy; for example, $\delta_{(2)}^\textrm{AII}$ is the codimension of forming a fourfold degenerate Dirac point as the touching of two Kramers-degenerate bands. 
The Kramers degeneracy can be elegantly dropped from the mathematical description by mapping the spin-$\frac{1}{2}$ degree of freedom to quaternion numbers~\cite{Hatsugai:2010}.
The quaternion description of spinful Hamiltonians is detailed when introducing the symplectic class ($\textrm{AII}$) in Sec.~\ref{sec:class-AII}, and is then adopted in the later discussion of certain chiral symmetry classes. 
Similar halving considerations apply to the zero-energy degeneracy in classes with particle-hole-symmetric spectra.
As an example, $\delta^{\textrm{BDI}}_{(2)}$ denotes the codimension of forming fourfold degeneracy at zero energy. 
In the case of $\textrm{DIII}$ and $\textrm{CII}$ classes, both the Kramers and the particle-hole doubling arise.
Therefore, the physical order of the degeneracy is four times larger than the index listed in the label.
For instance, $\delta^{\textrm{DIII}}_{(2)}$ corresponds to eight-fold degeneracy at zero energy.
The table with the results of the codimension analysis is shown in Table~\ref{tab:codims-WD+balanced}.

\begin{table*}
    \centering
    \begin{threeparttable}
    \caption{Codimension of forming band degeneracies of various order in the individual symmetry classes (CL). 
    For each symmetry class, we tabulate the action of time-reversal ($\mathcal{T}$) and particle-hole ($\mathcal{C}$) symmetries that act \emph{locally in $\bs{k}$-space} (i.e., we assume that time-reversal and particle-hole symmetries do not change the sign of parameter $\boldsymbol{k}$, cf.~Sec.~\ref{sec:general-enclosing-sphere}), as well as the presence of chiral symmetry ($\mathcal{S}$). 
    The next columns show the codimensions $\delta_{(n)}$ for the degeneracies of the first four orders, concluding with the general formula for an arbitrary value of $n$.
    We here assume the absence of flat bands pinned at $E=0$ for classes with CS or PHS.
    In symmetry classes with a footnote, the symmetry implies additional doubling or quadrupling of the order $n$ of degeneracy corresponding to $\delta_{(n)}$.
    Note that for chiral symmetry classes, the displayed codimension only applies to degeneracies at $E=0$ -- for degeneracies at $E\neq 0$ one should instead read the codimension in the respective non-chiral class.
    For the Wigner-Dyson classes ($\textrm{A}$, $\textrm{AI}$, $\textrm{AII}$), $n=1$ corresponds to absence of a degeneracy (disregarding here the Kramers degeneracy in class $\textrm{AII}$).
    Entries marked by double dagger `$\ddagger$' correspond to the codimension of the lowest-order degeneracy in the corresponding symmetry class as tabulated in Ref.~\onlinecite{Bzdusek:2017}.
    }
    \label{tab:codims-WD+balanced}
    \begin{ruledtabular}
    \begin{tabular}{cccccccccccc}
    \textrm{CL}                      & & ~$\mathcal{T}$~  & ~$\mathcal{C}$~  & ~$\mathcal{S}$~  & & ~$\delta_{(1)}$~ & ~$\delta_{(2)}$~ & ~$\delta_{(3)}$ & ~$\delta_{(4)}$~ & & ~$\delta_{(n)}$~           \\ \hline
    \hyperref[sec:class-AI]{AI}                    & & ~$1$~  & ~$0$~  & ~$0$~ & & ~$0$~            & ~$2^\ddagger$~   &       ~$5$~     &  ~$9$~           & & ~$\tfrac{1}{2}(n-1)(n+2)$~ \\ 
    \hyperref[sec:class-A]{A}                     & & ~$0$~	& ~$0$~  & ~$0$~ & & ~$0$~            & ~$3^\ddagger$~   &       ~$8$~     &  ~$15$~          & & ~$n^2 - 1$~                \\ 
    \hyperref[sec:class-AII]{AII}$^{\,\textrm{b}}$    & & ~$-1$~ & ~$0$~  & ~$0$~ & & ~$0$~            & ~$5^\ddagger$~   &       ~$14$~    &  ~$27$~          & & ~$(n-1)(2n+1)$~            \\ \hline
    \hyperref[sec:class-BDI]{BDI}$^{\,\textrm{a}}$    & & ~$1$~  & ~$1$~  & ~$1$~ & & ~$1^\ddagger$~   & ~$4$~            &       ~$9$~     &  ~$16$~          & & ~$n^2$~                    \\
    \hyperref[sec:class-AIII]{AIII}$^{\,\textrm{a}}$   & & ~$0$~  & ~$0$~  & ~$1$~ & & ~$2^\ddagger$~   & ~$8$~            &       ~$18$~    &  ~$32$~          & & ~$2n^2$~                   \\ 
    \hyperref[sec:class-CII]{CII}$^{\,\textrm{c}}$    & & ~$-1$~ & ~$-1$~ & ~$1$~ & & ~$4^\ddagger$~   & ~$16$~           &       ~$36$~    &  ~$64$~          & & ~$4n^2$                \\
    \hyperref[sec:D_class]{D}$^{\,\textrm{a}}$      & & ~$0$~  & ~$1$~  & ~$0$~ & & ~$1^\ddagger$~   & ~$6$~            &       ~$15$~    &  ~$28$~          & & ~$n(2n-1)$~                \\
    \hyperref[sec:class-C]{C}$^{\,\textrm{a}}$      & & ~$0$~  & ~$-1$~ & ~$0$~ & & ~$3^\ddagger$~   & ~$10$~           &       ~$21$~    &  ~$36$~          & & ~$n(2n+1)$           \\
    \hyperref[sec:CI_class]{CI}$^{\,\textrm{a}}$     & & ~$1$~  & ~$-1$~ & ~$1$~ & & ~$2^\ddagger$~   & ~$6$~            &       ~$12$~    &  ~$20$~          & & ~$n^2+n$~                  \\
    \hyperref[sec:class-DIII]{DIII}$^{\,\textrm{c}}$   & & ~$-1$~ & ~$1$~  & ~$1$~ & & ~$2^\ddagger$~   & ~$12$~           &       ~$30$~    &  ~$56$~          & & ~$2n(2n-1)$~               \\
    \end{tabular}
    \end{ruledtabular}   
    \begin{tablenotes}
        \item[a] Particle-hole doubling needs to be included to obtain the physical order of the energy degeneracy. 
        \item[b] Kramers doubling needs to be included to obtain the physical order of the energy degeneracy.
        \item[c] Both particle-hole doubling and the Kramer doubling needs to be included to obtain the physical order of the energy degeneracy.
    \end{tablenotes}
    \end{threeparttable}
\end{table*}

Having established that the classifying space of $n$-fold nodal points in $n$-band Hamiltonians is a sphere $S^{\ell}$, we turn to the next major question.  
We aim to understand (\emph{i})~the relationship between the nontrivial topology of the nodal points and the linking of certain appropriately defined nodal manifolds attached to the nodal point, as well as (\emph{ii})~their connection to topological band invariants that can be defined on these nodal manifolds.
Our strategy, summarized in Fig.~\ref{fig:main_scheme}, is to enclose the $n$-fold nodal point with a sphere $S^{\ell}$ in the parameter space. 
For minimal models, the $n$-fold degeneracy can be understood as an intersection of two loci $\mathcal{L}_1$ and $\mathcal{L}_2$ that support $(n-1)$-fold degeneracy inside the parameter space.
In more general models, the $n$-degeneracy still arises as an intersection of the loci of $(n-1)$-degeneracies spanned by different (though clearly overlapping) groups of bands that can be defined in a similar spirit as in the minimal $n$-band model.
We then define nodal manifolds $M_1=\mathcal{L}_1\cap S^{\ell}$ and $M_2=\mathcal{L}_2\cap S^{\ell}$ as the intersection of the degeneracy loci with the enclosing sphere.
On each of these nodal manifolds, there exists a spectral gap that allows us to assign topological band invariants to its Bloch bundle.

Let us point out that this approach constitutes a generalization of the technique formerly used to topologically characterize \emph{minimally degenerate nodal points} (i.e., those corresponding to the first non-zero entry in each row of Table~\ref{tab:codims-WD+balanced})~\cite{Bzdusek:2017}.
Specifically, in that case, the loci $\mathcal{L}_1$ and $\mathcal{L}_2$ capture individual non-degenerate bands, and as such trivially span the entire parameter space, so that both $M_1$ and $M_2$ correspond to the entire enclosing sphere, thus enabling a topological classification in terms of homotopy groups of classifying spaces of single-gap Hamiltonians.
In the remainder of the paper, we will focus on multifold nodal points that are of the lowest order beyond the minimally degenerate case of Ref.~\citenum{Bzdusek:2017}, which correspond to the second non-zero entry in each row of Table~\ref{tab:codims-WD+balanced} (i.e., those with codimension $\delta_{(3)}$ for the Wigner-Dyson symmetry classes and those with $\delta_{(2)}$ for the chiral and BdG classes).

\subsection{Linking of nodal manifolds}
\label{sec:general_linking}

We now argue that the topological stability of the multifold nodal point manifests as the linking of nodal manifolds $M_1$ and $M_2$. 
To reveal this connection, let us assume that the multifold nodal point is topologically trivial.
Then one can split the multifold degeneracy, which results in non-intersecting degeneracy loci $\mathcal{L}_1$ and $\mathcal{L}_2$, allowing us to move them slightly apart. 
The topological characterization should not depend on the radius of the enclosing sphere; therefore, one can take it to be sufficiently small.
However, since the degeneracy loci are non-intersecting, the manifolds $M_1$ and $M_2$ vanish (and are therefore trivially unlinked) for a sufficiently small radius of $S^\ell$. 
Assuming that the linking is robust under continuous deformations of the Hamiltonian, the preceding argument can be reversed. 
Indeed, if $M_1$ and $M_2$ remain linked on the enclosing sphere, their degeneracy loci must intersect inside $S^\ell$. 
Hence, the multifold nodal point is topologically stable.

Second, we claim that the presence of nontrivial band invariants on $M_1$ and $M_2$ implies nontrivial linking of nodal manifolds $M_1$ and $M_2$. 
This can be understood from the following considerations. Let us assume that $M_1$ and $M_2$ are unlinked in the sense that they have disjoint ball neighborhoods $N_1$ and $N_2$.\footnote{By a ball neighborhood of $M_i \subset S^\ell$ we mean the existence of $\mathcal{N}_i \subset S^\ell$ that (1)~is homeomorphic to an $\ell$-dimensional ball and such that (2)~$M_i \subset \mathcal{N}_i$. The disjointness of the two ball neighborhoods $\mathcal{N}_1 \cap \mathcal{N}_2 = \varnothing$ disallows the linking of $M_1$ and $M_2$.} 
Consider the band bundle $E_1$ over $M_1$ whose topological invariant defines the corresponding band invariant. 
We assume that this bundle is defined using an energy gap whose closing locus on the enclosing sphere is $M_2$. 
Since $\mathcal{N}_1$ does not intersect $M_2$, this gap remains open throughout $\mathcal{N}_1$. 
The band bundle $E_1$ over $M_1$ therefore extends to a vector bundle $\widetilde E_1$ over the entire ball neighborhood $\mathcal{N}_1$. 
Since $\mathcal{N}_1$ is homeomorphic to a ball, it is contractible, and every vector bundle over it is topologically trivial. 
Thus, the band bundle $E_1$ over $M_1$ is also topologically trivial, and all corresponding topological invariants vanish (other than the rank of the bundle, which is of no importance in our considerations). 
The same argument applies to the corresponding band bundle over $M_2$, provided that it is defined using the gap whose closing locus is $M_1$. 
This discussion means that if $M_1$ and $M_2$ are unlinked, they should carry trivial band invariants.

By reversing the above implication, we obtain that if nodal manifolds $M_1$ and $M_2$ do not carry trivial band invariants (i.e., when they support nontrivial band invariants), the nodal manifolds should be linked. 
One can understand this implication as follows. 
If there is a topological invariant on a nodal manifold $M$, then a singularity is needed to trivialize this topology. 
This singularity corresponds to the closing of the energy gap, i.e., a crossing of $M$ with another nodal manifold.

By combining the arguments from the previous three paragraphs, we obtain that nontrivial band invariants on $M_1$ and $M_2$ characterize the linking of $M_1$ and $M_2$ and, assuming that the linking of $M_1$ and $M_2$ persists under admissible deformations, imply the topological protection of a multifold nodal~point.

\begin{figure}[t]
    \centering
    \includegraphics[width=\linewidth]{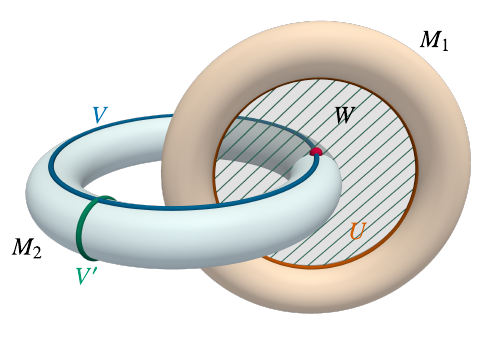}
    \vspace{-10 mm}
    \caption{
    A schematic illustration of the linking between $U\subset M_1$ and $V\subset M_2$. 
    The linking is defined as the intersection number of a submanifold representing a nontrivial homology class in $M_2$, which is a circle $V$ in this case, and the surface $W$ such that $U$ is a boundary of $W$. 
    The manifold $V'$ representing another nontrivial homology class of $M_2$ is unlinked with $U$. 
    }
\label{fig:linking_scheme}
\end{figure}

We can quantify the correspondence between band invariants and linking by means of linking numbers defined for higher-dimensional manifolds. 
For this purpose, we present how to mathematically characterize the linking between non-intersecting manifolds $M_1$ and $M_2$ inside $S^\ell$. 
We rely on the known fact that an integer linking number can be defined for two closed orientable submanifolds $U$ and $V$ in the ambient manifold $L$ if~(1)~$U$ and $V$ are null-homologous (i.e. they can be represented as a boundary of another submanifold in $L$), and~(2)~$\mathrm{dim}(U){+}\mathrm{dim}(V){+}1{=}\mathrm{dim}(L)$~\cite{Horowitz:1990}. 
Conveniently, any submanifold of $S^\ell$ with dimension $0<d<\ell$ is automatically null-homologous in $S^{\ell}$, and we only need to be careful when $d=0$ and $d=\ell$.
However, we find that, generally, $\dim(M_1)+\dim(M_2)+1 > \dim(S^\ell)$, prompting us to quantify the linking of $M_1$ and $M_2$ through the integer linking numbers of \emph{submanifolds} of $M_1$ and $M_2$.

Let us therefore consider closed orientable submanifolds $U\subset M_1$ and $V\subset M_2$ with dimensions $p$ and $q$ such that
\begin{equation}
\label{eq:linkin_constr}
p + q + 1 = \ell.
\end{equation}
Since $M_1$ and $M_2$ do not intersect, a submanifold $U \subset M_1$ that is contractible within\footnote{The contractibility in $M_1$ follows from the homotopy invariance of homology groups~\cite{Hatcher_book}. By this invariance, if the inclusion $i:U\hookrightarrow M_1$ is null-homotopic, then the fundamental class of $U$ has the same image in $H_p(M_1)$ as under a constant map, and therefore vanishes. This means that $U$ is null-homologous in $M_1$.} $M_1$ cannot be linked with a submanifold $V \subset M_2$ that is contractible within $M_2$.
This observation implies that linking can be realized only between the submanifolds generating homology groups\footnote{In general, it is also possible to consider generators of torsion subgroups of homology groups with integer coefficients. However, these are not necessarily representable by manifolds for a general choice of coefficients. A particular exception occurs for the homology groups with $\mathbb{Z}_2$ coefficients, where the generators of homology groups can still always be represented by manifolds, but not necessarily \textit{orientable} ones~\cite{Sullivan:2004}. We discuss the linking of nonorientable manifolds and the corresponding $\mathbb{Z}_2$ linking numbers later in Sec.~\ref{sec:general_remarks}.} 
$H_{p}(M_1,\mathbb{R})$ and $H_{q}(M_2,\mathbb{R})$, where $p$ and $q$ satisfy the constraint in Eq.~\eqref{eq:linkin_constr}. 
Therefore, we can define the set of linking invariants of $M_1$ and $M_2$ as the set of linking numbers between submanifolds generating $H_{p}(M_1,\mathbb{R})$ and $H_{q}(M_2,\mathbb{R})$. 
To calculate the linking numbers, we can choose a representative for each generator and consider all possible pairings between them. 
A schematic illustration of the linking of $M_1$ and $M_2$ is shown in Fig.~\ref{fig:linking_scheme}.

The canonical way to define the linking numbers employs the intersection number, which is defined as follows~\cite{Brennan:2023, Horowitz:1990, Seifert_book}. 
Consider two compact orientable (possibly with boundary) submanifolds $A$ and $\mathcal{B}$ of the sphere $S^\ell$, such that $\mathrm{dim}(A){+}\mathrm{dim}(\mathcal{B}){=}\ell$.
Then, generically, $A$ and $\mathcal{B}$ intersect transversally (i.e., the manifolds don't have collinear tangent subspaces at the intersection point) at a finite number of points $\{p_i \}$. 
If the intersection is not transversal, we can slightly perturb the manifolds $A$ and $\mathcal{B}$. 
Therefore, at each point $p_i$, the tangent space of $S^\ell$ is isomorphic to the direct sum of tangent spaces of $A$ and $\mathcal{B}$: $T_{p_i}L=T_{p_i}A\oplus T_{p_i}\mathcal{B}$.
Thus, the orientations of $A$ and $\mathcal{B}$ define an orientation on $L$ at each $p_i$. 
We define a function $\mathrm{sign}(p_i)=\pm 1$, which is equal to $+1$ if the induced orientation is the same as the original orientation on $S^\ell$, and equal to $-1$ if these orientations are opposite. 
The intersection number of $A$ and $\mathcal{B}$ is the sum of all signs of intersection points:
\begin{subequations}
\begin{equation}
\label{eq:intersection_number}
\mathrm{Int}(A,\mathcal{B})=\sum_{i}\mathrm{sign}(p_i).
\end{equation}
In turn, the linking number of submanifolds $A$ and $B$ in $L$ is defined as
\begin{equation}
\label{eq:linkin_number}
\mathrm{Lk}(A,B)=\mathrm{Int}(A,\mathcal{B}),
\end{equation}
\end{subequations}
where $\mathcal{B}$ is a manifold such that $B=\partial \mathcal{B}$ is a boundary of $\mathcal{B}$. 
Up to a possible sign change, the linking number is symmetric under the exchange of $A$ and $B$.

In the following subsection, which utilizes more advanced concepts from geometry and topology, we aim to formalize the relation between linking and band invariants of submanifolds of $M_1$ and $M_2$. 
Readers who find the discussion too technical are invited to skip this discussion and continue directly with Sec.~\ref{sec:general_overview}, where we present a brief overview of the mathematical fundamentals. 
The subsequent sections showcase the general construction in the context of specific symmetry classes, starting with the symmetry class $\textrm{AI}$ in Sec.~\ref{sec:class-AI}. 
We also present here a case-by-case description of the linking between the submanifolds of $M_1$ and $M_2$ according to Table~\ref{tab:nodal_manifolds}.

\subsection{Topological band invariants as linking numbers}
\label{sec:general-linking-to-invariants}

To formalize the relation between band and linking invariants, we utilize the language of differential forms~\cite{Bott:1982, Nicolaescu:2020}. 
Here, we begin with a brief formulation and basic properties of the approach. 
For the detailed explanation and the derivation of the main identities, see Appendix~\ref{app:linking}.

\begin{table*}
\centering
\caption{
Cell decompositions of nodal manifolds $M_1=\mathcal{L}_1\cap S^{\ell}$ and $M_2=\mathcal{L}_2\cap S^{\ell}$ on enclosing spheres $S^{\ell}$ for minimal higher-degeneracy models. 
For classes without chiral symmetry, both nodal manifolds $M_1$ and $M_2$ are isomorphic and correspond to standard classifying spaces. 
For chiral classes, we replace the nodal manifold $M_1$ with $M_\medcirc$, i.e., the nodal manifold with a band degenerate at zero energy, while $M_2$ is replaced with $M_\varnothing$, i.e., the nodal manifold with degeneracy at non-zero energy; of these, $M_\varnothing$ can be identified with one of the standard classifying spaces.
In the last column, we show some of the possible linkages between cells of different dimensions together with the appropriate bundle invariants.
For each instance of linking, we first indicate how cells of the two nodal manifolds are selected while following the dimensional condition in Eq.~(\ref{eq:linkin_constr}); subsequently, in parentheses, we specify which band invariant on which $n$-cell signals this instance of the linking. 
We use the following abbreviations for the topological invariants: `nSW' for the $n$th Stiefel-Whitney invariant (in $n$ dimensions), `$C_n$' for the $n$th Chern number (in $2n$ dimensions), `$n$D W\#' for the $n$-dimensional winding number of matrices (for odd values of $n$), and `$\sign[\Pf]$' for the Pfaffian sign invariant (for $0$-dimensional cells). 
Invariants characterized as `chiral' are computed for zero-energy eigenstates with specified eigenvalue of the chiral operator in the chiral classes $\textrm{BDI}$, $\textrm{AIII}$, and $\textrm{CII}$.
In all cases, the band invariants are computed from the bands that are degenerate on the corresponding cell of the nodal manifold (in particular: from the zero-energy eigenstates on the cells of $M_\medcirc$, and from the non-zero-energy eigenstates on the cells of $M_\varnothing$). The invariants marked with the superscript `$\ddagger$' (two in class $\textrm{CII}$, two in class $\textrm{C}$, and one in class $\textrm{DIII}$) are conjectured without an explicit analytic argument or numerical verification.}
\label{tab:nodal_manifolds}
\begin{ruledtabular}
\begin{tabular}{ccccl}
CL & $S^{\ell}$ & 
$M_1$ (resp.~$M_\medcirc$) & 
$M_2$ (resp.~$M_\varnothing$) & Linking and the investigated topological invariants \\
\hline 
\hyperref[sec:class-AI]{AI}    & $S^{\mathcolor{olive}{4}}$  & $\mathbb{R}P^2$, \,  $(0,\mathcolor{purple}{1},\mathcolor{blue}{2})$-cells& $\mathbb{R}P^2$, \,  $(0,\mathcolor{purple}{1},\mathcolor{blue}{2})$-cells
 & $\mathcolor{purple}{1}+\mathcolor{blue}{2}+1=\mathcolor{olive}{4}$ \; (1SW on $\mathcolor{purple}{1}$-cells \&  2SW on $\mathcolor{blue}{2}$-cells)\\\hline
 
\hyperref[sec:class-A]{A}      & $S^{\mathcolor{olive}{7}}$  & $\mathbb{C}P^2$, \,  $(0,\mathcolor{purple}{2},\mathcolor{blue}{4})$-cells& $\mathbb{C}P^2$, \,  $(0,\mathcolor{purple}{2},\mathcolor{blue}{4})$-cells
 & $\mathcolor{purple}{2}+\mathcolor{blue}{4}+1=\mathcolor{olive}{7}$ \; ($C_1$ on $\mathcolor{purple}{2}$-cells \&  $C_2$ on $\mathcolor{blue}{4}$-cells) \\\hline

\hyperref[sec:class-AII]{AII}   & $S^{\mathcolor{olive}{13}}$ & $\mathbb{H}P^2$, \,  $(0,\mathcolor{purple}{4},\mathcolor{blue}{8})$-cells& $\mathbb{H}P^2$, \,  $(0,\mathcolor{purple}{4},\mathcolor{blue}{8})$-cells
 & $\mathcolor{purple}{4}+\mathcolor{blue}{8}+1=\mathcolor{olive}{13}$ \; ($C_2$ on $\mathcolor{purple}{4}$-cells \&  $C_4$ on $\mathcolor{blue}{8}$-cells) \\\hline

\hyperref[sec:class-BDI]{BDI}   & $S^{\mathcolor{olive}{3}}$  & $S^1 \times S^1$&
$S^1 \sqcup S^1\simeq \mathsf{O}(2)$ & $\mathcolor{purple}{1}+\mathcolor{violet}{1}+1=\mathcolor{olive}{3}$ \; (chiral 1SW on $\mathcolor{purple}{1}$-cells \& 1SW on $\mathcolor{violet}{1}$-cells) \\&&$(0,\mathcolor{purple}{1},\mathcolor{blue}{2})$-cells& $(\mathcolor{orange}{0},\mathcolor{violet}{1})$-cells &$\mathcolor{blue}{2}+\mathcolor{orange}{0}+1=\mathcolor{olive}{3}$ \; ($\mathrm{sign}[\mathrm{Pf}]$  on $\mathcolor{orange}{0}$-cells)\\\hline

\hyperref[sec:class-AIII]{AIII}  & $S^{\mathcolor{olive}{7}}$  & $\mathrm{UT}S^3\simeq S^3\times S^2$  &
$S^1\times S^{3}/\mathbb{Z}_2\simeq \mathsf{U}(2)$ \, & $\mathcolor{purple}{2}+\mathcolor{violet}{4}+1=\mathcolor{olive}{7}$ \; (chiral $C_1$ on $\mathcolor{purple}{2}$-cell)\\
&& $(0,\mathcolor{purple}{2},\mathcolor{cyan}{3},\mathcolor{blue}{5})$-cells & $(0,\mathcolor{orange}{1},\mathcolor{magenta}{3},\mathcolor{violet}{4})$-cells& $\mathcolor{cyan}{3}+\mathcolor{magenta}{3}+1=\mathcolor{olive}{7}$ \; (3D W\# on $\mathcolor{magenta}{3}$-cell) \\
 & & & & $\mathcolor{blue}{5}+\mathcolor{orange}{1}+1=\mathcolor{olive}{7}$ \; (1D W\# on $\mathcolor{orange}{1}$-cell)\\\hline

\hyperref[sec:class-CII]{CII}    & $S^{\mathcolor{olive}{15}}$ & $(S^7\times S^7)/\mathsf{Sp}(1)$&
$\mathsf{Sp}(2)$& $\mathcolor{purple}{4}+\mathcolor{violet}{10}+1=\mathcolor{olive}{15}$ \; (chiral $C^{\ddagger}_2$ on $\mathcolor{purple}{4}$-cell) \\ &&$(0,\mathcolor{purple}{4},\mathcolor{cyan}{7},\mathcolor{blue}{11})$-cells &$(0,\mathcolor{orange}{3},\mathcolor{magenta}{7},\mathcolor{violet}{10})$-cells& $\mathcolor{blue}{11}+\mathcolor{orange}{3}+1=\mathcolor{olive}{15}$ \; (3D W\#$^{\ddagger}$ on $\mathcolor{orange}{3}$-cells) \\\hline

\hyperref[sec:D_class]{D}    & $S^{\mathcolor{olive}{5}}$ & $S^2 \times S^2$&
$S^2 \sqcup S^2\simeq \mathsf{O(4)}/\mathsf{U(2)}$& $\mathcolor{purple}{2}+\mathcolor{violet}{2}+1=\mathcolor{olive}{5}$ \; ($C_1$ on $\mathcolor{violet}{2}$-cells) \\ &&$(0,\mathcolor{purple}{2},\mathcolor{blue}{4})$-cells&$(\mathcolor{orange}{0},\mathcolor{violet}{2})$-cells& $\mathcolor{blue}{4}+\mathcolor{orange}{0}+1=\mathcolor{olive}{5}$ \; ($C_2$ on $\mathcolor{blue}{4}$-cell \& $\mathrm{sign}[\mathrm{Pf}]$  on $\mathcolor{orange}{0}$-cells)\\\hline

\hyperref[sec:class-C]{C}   & $S^{\mathcolor{olive}{9}}$ & $\mathsf{SO}(5)/\mathsf{U}(2)\simeq \mathbb{C}P^3$ & $\mathsf{Sp}(2)/\mathsf{U}(2)$ & $\mathcolor{purple}{2}+\mathcolor{violet}{6}+1=\mathcolor{olive}{9}$ \; (chiral $C^{\ddagger}_1$ on $\mathcolor{purple}{2}$-cell) \\ &&$(0,\mathcolor{purple}{2},\mathcolor{cyan}{4},\mathcolor{blue}{6})$-cells  &$(0,\mathcolor{orange}{2},\mathcolor{magenta}{4},\mathcolor{violet}{6})$-cell & $\mathcolor{orange}{2}+\mathcolor{blue}{6}+1=\mathcolor{olive}{9}$ \; ($C^{\ddagger}_1$ on $\mathcolor{orange}{2}$-cell )\\\hline

\hyperref[sec:CI_class]{CI}    & $S^{\mathcolor{olive}{5}}$ & $\mathrm{UT}S^2\simeq\mathbb{R}P^3$&
$S^1\times S^{2}/\mathbb{Z}_2\simeq \mathsf{U}(2)/\mathsf{O}(2)$ \, & $\mathcolor{cyan}{2}+\mathcolor{magenta}{2}+1=\mathcolor{olive}{5}$ \; (2SW on $\mathcolor{cyan}{2}$-cell, 2SW on $\mathcolor{magenta}{2}$-cell)\\ &&$(0,\mathcolor{purple}{1},\mathcolor{cyan}{2},\mathcolor{blue}{3})$-cells  &$(0,\mathcolor{orange}{1},\mathcolor{magenta}{2},\mathcolor{violet}{3})$-cells & $\mathcolor{blue}{3}+\mathcolor{orange}{1}+1=\mathcolor{olive}{5}$ \; (1D W\# on $\mathcolor{orange}{1}$-cell) \\\hline

\hyperref[sec:class-DIII]{DIII}  & $S^{\mathcolor{olive}{11}}$  & $\mathrm{UT}S^5\simeq \mathsf{O}(6)/\mathsf{O}(4)$  &
$S^1\times S^{5}/\mathbb{Z}_2\simeq \mathsf{U}(4)/\mathsf{Sp}(2)$ \, & $\mathcolor{purple}{4}+\mathcolor{violet}{6}+1=\mathcolor{olive}{11}$ \; ($C^{\ddagger}_2$ on $\mathcolor{purple}{4}$-cell) \\
&& $(0,\mathcolor{purple}{4},\mathcolor{cyan}{5},\mathcolor{blue}{9})$-cells & $(0,\mathcolor{orange}{1},\mathcolor{magenta}{5},\mathcolor{violet}{6})$-cells& $\mathcolor{blue}{9}+\mathcolor{orange}{1}+1=\mathcolor{olive}{11}$ \; (1D W\# on $\mathcolor{orange}{1}$-cell)
\end{tabular}
\end{ruledtabular}
\end{table*}

The key formula for the linking number of two manifolds embedded into the sphere $S^{\ell}$ takes the form
\begin{equation}
\label{eq:Lk_diffform}
    \mathrm{Lk}(A,B)=\int_{S^{\ell}}\omega_A \wedge \eta_{B},
\end{equation}
where the form $\eta_B$ is defined in two steps. 
First, we define the form $\theta_{B}$, which is the compactly supported Poincar\'{e} dual of $B$ in the small enough open neighborhood $N_B\subset S^{\ell}$, i.e., $\theta_B\in H^{\ell-q}_{c, \mathrm{dR}} (N_B)$\footnote{For a manifold $X$, the compactly supported de Rham cohomology group $H_{c, \mathrm{dR}}^n(X)$ is defined as the cohomology of the complex of smooth differential forms with compact support,
\begin{equation}
0 \longrightarrow \Omega_c^0(X)\xrightarrow{d}\Omega_c^1(X)\xrightarrow{d}\cdots\xrightarrow{d}\Omega_c^n(X)\xrightarrow{d}\cdots,
\end{equation}
where $\Omega_c^n(X)$ denotes the space of smooth $n$-forms on $X$ whose support is compact. Thus, an element of $H_{c, \mathrm{dR}}^n(X)$ is represented by a closed $n$-form with compact support, modulo the addition of exact compactly supported $n$-forms. If the manifold $X$ is non-compact, the group $H_{c, \mathrm{dR}}^n(X)$ may not be isomorphic to $H^n(X,\mathbb{R})$.}, where $q=\dim(B)$. 
Then, we define the form $\eta_B$ by continuing the form $\theta_B$ to the whole sphere $S^\ell$ by assigning zero values everywhere outside its support. 
Remarkably, the form $\eta_B$ is exact. 
To see that, one can notice that $\eta_B$ is closed because the form $\theta_B$ is closed in $N_B$, i.e. $d\theta_B=0$, and continuation by zero does not affect taking the derivative. 
Then, since all homology groups of $S^{\ell}$ (except those of the lowest and highest degree) are trivial, the dual form $\eta_B$ is cohomologically trivial; hence, it is exact.
This allows us to define the form $\omega_B$ by the relation $d\omega_B=\eta_B$.
The same definitions are accordingly applied for $\eta_A$ and $\omega_A$.

Since $\eta_B$ has compact support on $N_B$, the previous integral can be written as an integral over $N_B$ alone
\begin{subequations}
\begin{equation}
    \mathrm{Lk}(A,B)=\int_{N_B}\omega_A \wedge \eta_{B}.
\end{equation}
Using the fact that the restriction of $\eta_B$ on $N_B$ is the Poincaré dual of $B$, we can rewrite the integral further as the integrals over $B$ and $\mathcal{B}$
\begin{equation}
\label{eq:Lk_onU}
    \mathrm{Lk}(A,B)=\int_{B}\omega_A=\int_{\mathcal{B}} \eta_A,
\end{equation}
\end{subequations}
where the last integral explicitly calculates the intersection number $\mathrm{Int}(A,\mathcal{B})$, with non-zero contributions coming from the intersections between $N_A$ and $\mathcal{B}$. 
Since one can consider neighborhoods $N_A$ and $N_B$ as small as one would like, we adopt the limit where the differential forms localize strictly on submanifolds $A$ and $B$. 
In this case, instead of differential forms with compact support, one considers the differential forms with coefficients in distributions strictly localized on submanifolds of lower dimensions, i.e., analogs of $\delta$-functions called de Rham currents~\cite{deRhambook:1984}.

By applying the framework of differential forms, we can argue that band invariants actually distinguish links between submanifolds of $M_1$ and $M_2$ and, therefore, themselves constitute linking invariants.
First, the nontrivial band invariants can be defined only on the submanifolds generating homology groups of $M_1$ and $M_2$. 
Otherwise, we could collapse a submanifold to a point by a continuous deformation. 
Let us choose a submanifold $U\subset M_1$ of dimension $p$ and an integer-valued band invariant $c$, such as the Chern or winding numbers, represented by an integration of a differential form
\begin{equation}
\label{eq:c(U)}
     c(U)=\int_U \chi.
\end{equation}
Notice that one can continue the form $\chi$ smoothly on $S^{\ell}\backslash M_2$ while preserving its topological properties. 
To do that, one can use the concrete representation of the band invariant via, for example, the Berry connection.
This can be done because a topological form is defined for any continuous deformation of the original system, where the relevant gap does not close.
At the same time, for the form $\chi$ defined on $M_1$, the relevant gap closes only on $M_2$. 
Furthermore, it is well known that $d\chi=0$ for both Chern and winding numbers, which means that $d\chi=0$ in $S^{\ell}\backslash M_2$ where the gap is open. 
From here, we conclude that an invariant $c$ is defined for any submanifold $U'\subset S^{\ell}\backslash M_2$ and that it has the same value for the submanifolds that can be continuously deformed into each other without intersecting $M_2$.

Since $U$ is null-homologous in $S^{\ell}$, we can choose a submanifold $W$ such that $U=\partial W$ is a boundary of $W$. 
Then, by virtue of the Stokes theorem, we can rewrite the band invariant~as\footnote{In a general situation, $W$ may be a (singular or simplicial) \emph{chain}, not a smooth manifold. In this case, the construction becomes more complicated, and one should consider the intersection of chains accordingly. The chain intersection is defined as the sum of the intersections of chains' simplices (counting the simplex multiplicities)~\cite{Seifert_book}, where those can be considered as the intersections of manifolds following the definition in Eq.~\eqref{eq:intersection_number}.}
\begin{equation}
\label{eq:top_stokes}
     c(U)=\int_W d\chi,
\end{equation}
where we assume that there exists such a (potentially singular) form $d\chi$ on the whole ambient sphere $S^\ell$, not only $S^{\ell}\backslash M_2$.
Using this presentation of $c(U)$, one can show that $d\chi$ can indeed be represented by a singular form $d\widetilde\chi$ defined on $S^{\ell}$, i.e., that there exists a form $d\widetilde\chi$ satisfying Eq.~\eqref{eq:top_stokes} for an arbitrary choice of $U$ and $W$. 
Moreover, $d\widetilde\chi$ can be chosen to be a current localized on some collection of submanifolds $\{V_j\}\subset M_2\subset S^{\ell}$ of dimension $q=\ell-1-p$, where on each $V_j$ the current is weighted by an integer. 
This can be shown as follows.

We start by noticing that, for a given $W$, there exists a collection of disjoint regions $\mathcal{I}=\{I_l\}$, where $M_2$ intersects $W$. 
If there are no intersections, then from Eq.~\eqref{eq:top_stokes} it follows immediately that $c(U)=0$. 
Each of $I_l$ can be surrounded by a sphere $S^p_l$ in $W$. 
When the form $\chi$ is integrated over $S^p_l$, the result is an integer $k_l$ independent of the deformation of the sphere $S^p_l$. 
Therefore, we can effectively represent the region $I_l$ by a point charge in $I_l\cap M_2$.

\begin{figure}[t]
    \centering
    \includegraphics[width=\linewidth]{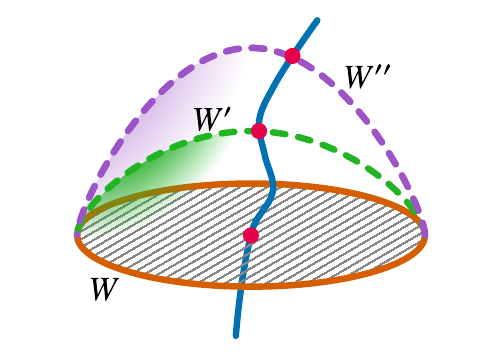}
    \vspace{-10 mm}
    \caption{
    The illustration demonstrates the construction of charged manifolds $V_j$ linked with the manifold $U=\partial W$ by varying the surfaces $W$.
    }
\label{fig:W_variation}
\end{figure}

As the next step, we can continuously vary $W$ for a given $U$ so that the union of all intersection regions covers all of $M_2$.\footnote{Assuming that $U$ is a boundary of a smooth $(p+1)$-dimensional manifold $W$, then for any $s\in S^{\ell}$, we can find a smooth manifold $W'$ bounded by $U$ and containing $s$. We can choose a $(p+1)$-dimensional sphere containing the point $s$ and attach it to $W$ by a thin tube. The resulting manifold is an example of $W'$.} 
For each $W$, we find the corresponding subset $\mathcal{I}$. 
We can choose the representative point charges for each $I_l$ so that the charge locations are continuously dependent on $W$ (an illustration of this approach is shown in Fig.~\ref{fig:W_variation}).
By looking at the evolution of the point charges under such deformations, we construct a collection of charged submanifolds $\{V_j\}\subset M_2$ of dimension $q$, representing the form $d\widetilde{\chi}$, which is defined on the whole ambient sphere $S^{\ell}$.
Moreover, the form $d\widetilde{\chi}$ should respect the orientation on each $V_j$, such that the integral of $d\widetilde{\chi}$ does not change sign during a smooth deformation of $W$; otherwise, $c(U)$ would depend on the choice of $W$. 
It should also be noted that by considering a spherical neighborhood for each point of $V_j$, we have constructed a tubular neighborhood of the submanifold $V_j$, which can be associated with its normal bundle.\footnote{The choice of normal coordinates is not important, since the integrals of $\widetilde{\chi}$ over the surrounding spheres $S_l^p$ depend only on the charge $q$. For convenience, however, one can choose the normal coordinates induced by the metric on the enclosing sphere $S^{\ell}$.}
Finally, we argue that $V_j$ should represent a nontrivial homology class in $H_q(M_2,\mathbb{R})$; otherwise, the submanifolds of $M_1$ and $M_2$ are not linked, and we could retract $U$ to a point in $S^{\ell}$, i.e., we would have $c(U)=0$.

To summarize, we have found that
\begin{equation}
\label{eq:dtildechi}
d\widetilde{\chi}=\sum_j k_j\eta_{V_j}=\sum_j k_j\,d\omega_{V_j},
\end{equation}
where $\eta_{V_j}$ is a current strictly localized on and encoding the orientation of $V_j$ via the normal coordinates, i.e., its Poincaré dual in $S^{\ell}$, and $k_j$ are some integers encoding the charge of $V_j$. 
Equivalently, we have constructed the closed forms $\eta_{V_j}$ on normal bundles associated with $V_j$, such that the integral of $\eta_{V_j}$ over the fiber is equal to $1$. 
The constructed form $\eta_{V_j}$ is called the Thom form of the manifold $V_j$, which is a Poincaré dual of $V_j$ (see Appendix~\ref{app:linking}). 
These observations also confirm the existence of forms $\omega_{V_j}$ in Eq.~\eqref{eq:dtildechi}, as well as our choice of notation for $d\widetilde\chi$, which is indeed an exact form in $S^{\ell}$.\footnote{It is worth noting that the forms $\chi$ and $\widetilde \chi$ may be different in $S^\ell \backslash M_2$. However, they both represent the same element of the cohomology group $H^{p}(S^\ell \backslash M_2,\mathbb{R})$, i.e. $\gamma=\chi -\widetilde\chi$ is an exact form in $S^\ell \backslash M_2$, because they are both closed in $S^{\ell}\backslash M_2$ and give the same integrals over the compact submanifolds of $S^{\ell}\backslash M_2$ due to the Stokes theorem and equality of integrals over surrounding spheres $S^p_l$.}

Since both forms $\chi$ and $\widetilde \chi$ give the same integrals over surrounding spheres $S^p_l$, we can use these forms interchangeably for calculating the band invariant. 
Substituting the expression for $d\widetilde\chi$ into Eq.~\eqref{eq:top_stokes} and using Eq.~\eqref{eq:Lk_onU}, we obtain 
\begin{equation}
     c(U)=\sum_j k_j \int_U \omega_{V_j}=\sum_j k_j\mathrm{Lk}(U,V_j),
\end{equation}
where $V_j$ are submanifolds representing some nontrivial classes in $H(M_2,\mathbb{R})$.\footnote{We anticipate that a in general situation $V_j$ can be taken as homology generators of $M_2$ with $k_j$ being equal to $\pm1$.} 
Therefore, we can conclude that the band invariants represented by smooth differential forms are linear combinations of linking numbers between the submanifolds of the appropriate dimensions.

\subsection{Remarks on formulation of linking}
\label{sec:general_remarks}

The preceding discussion can also be interpreted in a more intuitive way. 
In the case of a Chern number on $M_1$, the differential form $\chi$ is a Berry curvature (or its higher-dimensional analog). 
It is known that the sources of Berry curvature are nodal points (in low dimensions behaving as Dirac monopoles). Therefore, one can think about $M_2$, which is the nodal manifold where the gap closes, as the charged source of the Berry curvature.  
In this picture, the intersection regions $\mathcal{I}\subset W$ behave as analogs of Dirac monopoles.

We should also note that while we have established the correspondence between $\mathbb{Z}$-valued band topological invariants and the linking numbers, the differential forms do not describe $\mathbb{Z}_2$ band invariants, such as Stiefel-Whitney numbers. 
Nevertheless, similar arguments can be applied to this case as well, if one considers the homology groups over $\mathbb{Z}_2$, $H_p(M_1, \mathbb{Z}_2)$ and $H_q(M_2, \mathbb{Z}_2)$ with $p+q+1=\ell$. 
In this case, the intersection number is defined $\mathrm{mod }\,2$, and the generators of homology groups can be represented not only by an orientable, but also by a nonorientable manifold.
The obtained linking number is then a $\mathbb{Z}_2$ invariant. 
Since Stiefel-Whitney classes belong to cohomology groups over $\mathbb{Z}_2$, the preceding discussion allows us to assume that $\mathbb{Z}_2$ band invariants are connected with the $\mathbb{Z}_2$ linking number.

The exact correspondence between the linking numbers and band invariants depends on the symmetry class and on which band invariants are considered. 
Nevertheless, linking invariants give us guidance on what submanifolds can be linked and which band invariants can capture the linking due to the dimension relation $k+p+1=\ell$. 
Since each submanifold generating $\mathbb{Z}$ or $\mathbb{Z}_2$ homology groups corresponds to a cell in the ambient manifold, we can analyze possible pairings of submanifolds using the cell decomposition of $M_1$ and $M_2$.
Table~\ref{tab:nodal_manifolds} shows which cells can produce nontrivial linking for minimal models in most of the symmetric classes with the corresponding band invariants. 
As discussed before, the choice of a $\mathbb{Z}$ or $\mathbb{Z}_2$ invariant can be motivated by the orientability of the corresponding cells. 
However, in some cases, there is no available $\mathbb{Z}$ band invariant even for an orientable manifold, so we use a $\mathbb{Z}_2$ invariant instead.

\subsection{Overview and outline of the results}
\label{sec:general_overview}

Before concluding the section, let us reiterate the key parts of our approach to characterizing multifold band degeneracies protected by Altland-Zirnbauer symmetries.
We generalize the enclosing-sphere approach to the topological characterization of multifold nodal points. 
In the multifold case, there is generally no uniform gap condition on the entire enclosing sphere $S^{\ell}$, and therefore the standard construction has to be modified. 
For minimal $n$-band models, the multifold nodal point arises at the intersection of two degeneracy loci, $\mathcal{L}_1$ and $\mathcal{L}_2$, associated with the $n$-fold degeneracy. 
We define the corresponding nodal manifolds by intersecting these loci with the enclosing sphere as $M_1=\mathcal{L}_1\cap S^{\ell}$ and $M_2=\mathcal{L}_2\cap S^{\ell}$.

On these nodal manifolds, standard band invariants are well defined because the relevant gap condition is restored. 
Moreover, band invariants of different dimensions can also be defined on suitable submanifolds of $M_1$ and $M_2$. 
These invariants characterize the topological protection of the multifold nodal point and can be interpreted as linking numbers between submanifolds of the nodal manifolds. 
In particular, $\mathbb{Z}$-valued band invariants encode ordinary linking numbers for orientable submanifolds, whereas $\mathbb{Z}_2$-valued band invariants give mod-$2$ linking numbers and are naturally applicable in the nonorientable case.

The considered interpretation also requires some care concerning the nature of the linking detected by the band invariants. 
First, the linking may be unstable: owing to fine-tuning of parameters, it can be present on the enclosing sphere and detected by the corresponding band invariants, yet be removed by an arbitrarily small perturbation of the Hamiltonian. 
This non-generic possibility is discussed in Appendix~\ref{app:unstable-linking}. 
Second, even robust linking may be accidental in the sense that, although it persists under small perturbations, its particular configuration can be changed through transformations of the nodal manifolds involving cobordisms, as discussed in Appendix~\ref{app:cobordism-transform}. Nevertheless, in generic, non-fine-tuned situations, a codimension counting ensures that the linking is robust. 
We also note that all linked configurations arising in the considered minimal models are robust by construction since they carry a unit winding number.

In the subsequent discussion, we apply the outlined methodology to all symmetry classes. The choice of the relevant submanifolds and band invariants is guided by the cell decomposition of the nodal manifolds summarized in Table~\ref{tab:nodal_manifolds}.
For each symmetry class, we start with an analysis of codimensions, usually employing a variant of the von Neumann-Wigner argument. 
Note that we assume only the usual `balanced' cases where neither the chiral nor the particle-hole symmetry enforces flat bands at zero energy.
After that, we consider a minimal model (which is topologically nontrivial by construction) and describe the nodal manifolds $M_1$ and $M_2$, if possible with explicit coordinates. 
In many cases, the nodal manifolds can be identified with canonical classifying spaces due to the restrictions on eigenvalues imposed on the enclosing sphere. 
For classes BDI, AIII, and CII, where one needs to analyze the spaces with zero-energy states, the corresponding classifying spaces are not standard. 
For these classes, we present the derivation of properly adjusted classifying spaces in Appendix~\ref{app:class-space-w-zero-states}. 
Finally, we explicitly calculate bundle topological invariants on relevant submanifolds of $M_1$ and $M_2$, if the coordinate description of nodal manifolds admits it. 
For the chiral symmetry classes, when analyzing a manifold $M_\medcirc$ that contains the zero-energy band, we focus on the bundle invariants associated with this degenerate band. 
Nevertheless, the remaining bands also encode information about the linking of nodal manifolds.

To demonstrate our approach, in the next section, we first analyze in detail the class AI, which enjoys one of the lowest codimensions and a transparent description of $M_1$ and $M_2$. 
For the remainder of the symmetric classes, we follow the same approach. 
The calculation of topological invariants is also given for classes A, AII, BDI, AIII, D, CI, and DIII, where the relevant cells admit a suitable parametrization.
For the remaining classes (CII and C), we only calculate the codimensions and describe the nodal manifolds.

\section{Symmetry class \texorpdfstring{$\textrm{AI}$}{AI}}
\label{sec:class-AI}

From this section onward, we present a case-by-case analysis of minimal models within the individual symmetry classes, illustrating how the general strategy discussed in Sec.~\ref{sec:general-section} acquires specific contours. 
We begin with the case of the symmetry class AI, whose exceptionally rich band-topological features have been broadly investigated in one~\cite{Ahn:2018,Wu:2019,Tiwari:2020}, two~\cite{Bzdusek:2017,Tiwari:2020,Bouhon:2020,Bouhon:2020b,Jankowski:2025b}, as well as in three~\cite{Lim:2023,Davoyan:2024,Jankowski:2024,Jankowski:2024b,Jankowski:2025} spatial dimensions.

Our discussion of the symmetry class AI is structured as follows.
We start with reviewing the well-known results for codimension counting by the von Neumann-Wigner argument in full generality. 
Next, we formulate the minimal model, characterize the nodal manifolds $M_1=\mathcal{L}_1\cap S^{\ell}$ and $M_2=\mathcal{L}_2\cap S^{\ell}$.
Finally, we calculate the appropriate bundle invariants, confirming the linking of the nodal manifolds.

The orthogonal class (labeled $\mathrm{AI}$) captures time-reversal-symmetric models with $\mathcal{T}^2 = +\mathbb{1}$, expected for particles with integer spin. 
Without loss of generality, we represent the time-reversal operator as complex conjugation, $\mathcal{T} = \mathcal{K}$.
Then the Hamiltonian $H$ becomes a real symmetric matrix, and the eigenstates can be gauged real. 
By arranging the $N$ eigenstates as columns next to each other, we obtain an orthogonal matrix $\Psi \in \mathsf{O}(N)$.

To understand the dimension of the group $\mathsf{O}(n)$, first note that this is an $n \times n$ matrix with real entries, implying $n^2$ parameters. 
The individual columns of $\mathsf{O}(n)$ are normalized and orthogonal to each other. 
The normalization of each column imposes $n$ constraints on the $n^2$ parameters, and the orthogonality of each pair of distinct eigenstates implies another $n(n-1)/2$ constraints. 
Therefore, 
\begin{equation}
\dim[\mathsf{O}(n)] = n^2 - n -\tfrac{1}{2}n(n-1) = \tfrac{1}{2}n(n-1).
\end{equation}
The von Neumann-Wigner counting~\cite{vonNeumann:1929} states that the dimension of the space of class-$\mathrm{AI}$ Hamiltonians with $N=\sum_{j=1}^k n_j$ bands, where $\{n_j\}_{j=1}^k$ indicates the number of states degenerate at one energy $\varepsilon_j$, is given by
\begin{eqnarray}
d^{\textrm{AI}}_{(n_1,n_2,\ldots,n_k)} &=& k + \dim [\mathsf{O}(N)] - \sum_{j=1}^k \dim [\mathsf{O}(n_j)] \nonumber \\
&=& k +\tfrac{1}{2}N(N-1) - \sum_{j=1}^k \tfrac{1}{2}n_k(n_k-1)
\label{eqn:vNW-AI}
\end{eqnarray}
where $k$ counts the distinct eigenenergies, the second term specifies the eigenstates, and the last term subtracts the redundancy due to the gauge degree of freedom among the degenerate states.

We study the codimension of forming one $n$-fold degeneracy in an otherwise non-degenerate spectrum. 
This is defined as
\begin{equation}
\label{eqn:AI-codim}
\delta^\textrm{AI}_{(n)} = d^\textrm{AI}_{(1,1,1,\ldots)} - d^\textrm{AI}_{(n,1,1,\ldots)} = \tfrac{1}{2}(n-1)(n+2).
\end{equation}
For $n=2$, we have $\delta^\textrm{AI}_{(2)} = 2$, meaning that two parameters need to be tuned to make two bands degenerate. 
The two parameters can be interpreted as coefficients multiplying the two real traceless $2 \times 2$ matrices, $\sigma_x$ and $\sigma_z$, which arise in an effective model capturing the two bands forming the degeneracy.
Therefore, we can assign the two-dimensional degeneracy a winding number, which is an element of 
\begin{equation}
\pi_1(S^1) = \intg , \label{eqn:pi1-S1-AI}
\end{equation}
here, $S^1$ corresponds to the space of normalized perturbations that split the twofold degeneracy. 
It is worth pointing out that this invariant reduces from $\mathbb{Z}$ to $\ztwo$ in models with three or more bands due to braiding with twofold band degeneracies in adjacent energy gaps~\cite{Wu:2019,Bouhon:2020}. 
We return to the stability aspect of the topological invariant in the concluding remarks near the end of the present section.

For $n=3$, we find $\delta^\textrm{AI}_{(3)} = 5$. 
The five parameters that have to be tuned are the five real traceless $3 \times 3$ matrices, which correspond to the five real Gell-Mann matrices. 
We anticipate that the triple degeneracy in a momentum (or other parameter) space with five or more dimensions can be characterized by an element of~$\pi_4(S^4)$.

More generally, for a general $n$, perturbations that split the $n$-fold degeneracy correspond to the real traceless $n \times n$ matrices. 
To count the dimension of this space, notice that such matrices are characterized by $n(n-1)/2$ arbitrary off-diagonal elements, and $n$ diagonal elements subject to a single constraint from tracelessness. 
Therefore, the dimension of this vector space is $n(n-1)/2 + (n-1) = \tfrac{1}{2}(n+2)(n-1)$, exactly matching the codimension determined in Eq.~(\ref{eqn:AI-codim}). 
We anticipate the $n$-fold degeneracy to be characterized by an element of the homotopy group $\pi_{\ell}(S^{\ell})$, where $\ell=\delta^\textrm{AI}_{(n)}-1$.

Next, we turn to the analysis of the minimal model.
We consider the most general three-band Hamiltonian that describes a threefold degeneracy in class $\textrm{AI}$. 
Such Hamiltonians can be expressed as the linear combination of the five real Gell-Mann matrices 
\begin{equation}
    H^\textrm{AI} = \sum_{i \in \mathcal{I}} k_i \Lambda_i,
    \label{eqn:AI-3band-Ham-GM}
\end{equation}
where $\{k_i\}_{i\in\mathcal{I}}$ are five momentum variables\footnote{As clarified in Sec.~\ref{sec:general-enclosing-sphere}, we assume that $\mathcal{T}$ does not flip the sign of the momenta. 
The same statement applies for time-reversal and particle-hole symmetry operators in all symmetry classes discussed in this work.} and $\{\Lambda_i\}_{i\in\mathcal{I}}$ are the five real Gell-Mann matrices. 
We adopt the standard definition,
\setlength{\arraycolsep}{2pt}
\begin{eqnarray}
\Lambda_1 &=&
\left(
\begin{array}{ccc}
0&1&0\\
1&0&0\\
0&0&0
\end{array}
\right),\;\;
\Lambda_2 =
\left(
\begin{array}{ccc}
0&-\imi&0\\
\imi&0&0\\
0&0&0
\end{array}
\right),\;\;
\Lambda_3 =
\left(
\begin{array}{ccc}
1&0&0\\
0&-1&0\\
0&0&0
\end{array}
\right),
\nonumber \\
\Lambda_4 &=&
\left(
\begin{array}{ccc}
0&0&1\\
0&0&0\\
1&0&0
\end{array}
\right),\;\;
\Lambda_5 =
\left(
\begin{array}{ccc}
0&0&-\imi\\
0&0&0\\
\imi&0&0
\end{array}
\right),\;\;
\Lambda_6 =
\left(
\begin{array}{ccc}
0&0&0\\
0&0&1\\
0&1&0
\end{array}
\right),
\label{eqn:GellMann}\\
\Lambda_7 &=&
\left(
\begin{array}{ccc}
0&0&0\\
0&0&-\imi\\
0&\imi&0
\end{array}
\right),\;\;
\Lambda_8 =
\frac{1}{\sqrt{3}}
\left(
\begin{array}{ccc}
1&0&0\\
0&1&0\\
0&0&-2
\end{array}
\right),
\nonumber 
\end{eqnarray}
with the real Gell-Mann matrices corresponding to 
\begin{equation}
\mathcal{I}=\{1,3,4,6,8\}.
\end{equation}
Correspondingly, we write $\bs{k}\,{=}\,(k_1,k_3,k_4,k_6,k_8)$ for the five components of momentum.
Without loss of generality, we dropped the term proportional to the identity matrix as it produces an unimportant overall shift of the spectrum.

The spectrum of the Hamiltonian~(\ref{eqn:AI-3band-Ham-GM}) exhibits a threefold degeneracy at $\boldsymbol{k}=0$, which splits nontrivially away from the threefold degeneracy.
We organize the eigenvalues such that $\varepsilon_1 \leq \varepsilon_2 \leq \varepsilon_3$. 
We define the degeneracy loci $\mathcal{L}_1$ and $\mathcal{L}_2$ such that 
\begin{subequations}
\label{eqn:class-AI-loci}
\begin{eqnarray}
    \mathcal{L}_1 &=& \{ \bs{k} \,|\, \varepsilon_1 = \varepsilon_2 \}, \\
    \mathcal{L}_2 &=& \{ \bs{k} \,|\, \varepsilon_2 = \varepsilon_3 \}.
\end{eqnarray}    
\end{subequations}
Our goal is to determine the intersections $M_1=\mathcal{L}_1\cap S^{4}$ and $M_2=\mathcal{L}_2\cap S^{4}$ of these loci with a four-dimensional sphere $S^4$ of radius $\kappa = 1$.

Since, $\tr[\Lambda_i] = 0$, we have $\varepsilon_1 + \varepsilon_2+ \varepsilon_3=0$ for the Hamiltonian at any $\bs{k}$. 
Furthermore, since the Gell-Mann matrices obey 
\begin{subequations}
\label{eqn:gell-mann-squared}
\begin{equation}
\tr[\Lambda_i\Lambda_j]=2\delta_{ij}, 
\end{equation}
it follows that 
\begin{eqnarray}
\varepsilon_1^2 + \varepsilon_2^2 + \varepsilon_3^2 &=& \tr[(H^\textrm{AI})^2] = \sum_{i,j\in\mathcal{I}} k_i k_j \tr[\Lambda_i \Lambda_j] \nonumber \\
&=& 2 \sum_{i\in\mathcal{I}} k_i^2 = 2,
\end{eqnarray}
\end{subequations}
where in the last step we used the specified radius $\kappa =1$ of the sphere.
The above constraints allow us to specify the nodal manifolds $M_1$ and $M_2$ as
\begin{subequations}
\label{eqn:AI-M1-M2-spectrum}
\begin{eqnarray}
    M_1 &=& \left\{ \bs{k} \,\big|\, \varepsilon_1 = \varepsilon_2 = -{1}/{\sqrt{3}},\;\varepsilon_3 = 2/\sqrt{3}\right\}, \label{eqn:AI-M1-M2-spectrum-M1}\\
    M_2 &=& \left\{ \bs{k} \,\big|\, \varepsilon_1 = -2/\sqrt{3},\;\varepsilon_2 = \varepsilon_3 = 1/\sqrt{3}\right\}.  \label{eqn:AI-M1-M2-spectrum-M2}
\end{eqnarray}
\end{subequations}
For concreteness, we further analyze the manifold $M_1$ in detail.

To specify a concrete Hamiltonian with the spectrum in Eqs.~(\ref{eqn:AI-M1-M2-spectrum-M1}) [and thus to determine the corresponding momenta $k_i$ in Eq.~(\ref{eqn:AI-3band-Ham-GM})], we need to include information about the eigenstates. 
Since we consider Hamiltonians with real coefficients inside a three-dimensional Hilbert space, the eigenvectors constitute an orthonormal $3$-frame in $\mathbb{R}^3$, i.e., with a proper choice of gauge one can achieve $\Psi=\{\ket{\psi_1},\ket{\psi_3},\ket{\psi_3}\} \in \mathsf{SO}(3)$.
In particular, the normalized eigenstate $\ket{\psi_3}$ at energy $\varepsilon_3$ can be any unit vector $\bs{n} \in S^2 \subset \mathbb{R}^3$.
Assuming a standard eigenframe $\Psi_0=\mathbb{1}$ (i.e., one whose first/second/third eigenstate is oriented along the positive $x$/$y$/$z$ axis in the Hilbert space), we obtain a general $\ket{\psi_3}$ [up to orientation, i.e., its $\mathsf{O}(1)$ gauge degree of freedom] by first rotating by an angle $\theta\in[0,\tfrac{\pi}{2}]$ around the $x$-axis, followed by a rotation by $\phi\in[0,2\pi]$ around the $z$-axis,~i.e.,
\begin{subequations}
\label{eqn:AI-eigenstates}
\begin{equation}
\label{eqn:AI-eigenstate-rotation}
\Psi 
= \left(\begin{array}{ccc}
\cos\phi& -\sin\phi & 0 \\
\sin\phi & \cos\phi & 0 \\
0 & 0 & 1
\end{array}\right)
\left(\begin{array}{ccc}
1 & 0 & 0 \\
0 & \cos\theta & -\sin\theta  \\
0 & \sin\theta & \cos\theta 
\end{array}\right)\Psi_0.       
\end{equation}
The resulting eigenframe is expressed as
\begin{equation}
\label{eqn:AI-eigenframe}
\Psi = \left(\begin{array}{ccc}
 \cos \phi  & -\cos \theta  \sin \phi  & \sin \theta  \sin \phi \\
 \sin \phi  & \cos \theta  \cos \phi  & -\sin \theta \cos \phi \\
 0 & \sin \theta  & \cos \theta  \\
\end{array}\right)    
\end{equation}
\end{subequations}
Note that we do not care about the residual $\mathsf{O}(2)$ rotation of $\ket{\psi_1}$ and $\ket{\psi_2}$ around the axis specified by $\ket{\psi_3}$, since their eigenvalues are degenerate on $M_1$, i.e., such rotations drop trivially from
\begin{equation}
\label{eqn:spectral-decomp-AI}
H^\textrm{AI} = \Psi \mathcal{E} \Psi^\top,
\end{equation}
where $\mathcal{E}=(\varepsilon_1,\varepsilon_2,\varepsilon_3)$ is the spectrum specified in Eq.~(\ref{eqn:AI-M1-M2-spectrum}).
The overall $\mathsf{O}(1)$ orientation of $\ket{\psi_3}$ drops for the same reason.

\begin{figure}[t]
    \centering
    \includegraphics[width=\linewidth]{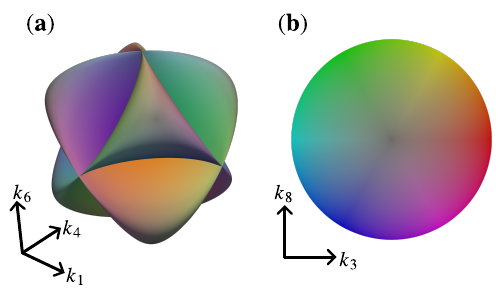}
    \caption{
    (a)~Projection of the nodal manifolds $M_1$ and $M_2$ in the symmetry class $\textrm{AI}$ [Eqs.~(\ref{eqn:AI-manifold-M1}) and~(\ref{eqn:AI-manifold-M1})] onto the space spanned by momenta $\{k_1,k_4,k_6\}$. 
    The manifolds are colored according to the values of the coordinates $\{k_3,k_8\}$ according to the legend in panel~(b).
    Each nodal manifold appears like a Roman surface in panel~(a), which corresponds to one of the standard representations of the real projective plane~\cite{Coffman:1996}.
    That the manifolds do not intersect inside $S^4 \subset \reals^5$ follows from the distinct coloring of any intersecting sheets.
    The linking of the two manifolds is not manifest in panel~(a); however, it is revealed by the nontrivial Stiefel-Whitney invariants as discussed in the main text.
    }
\label{fig:rp2_linking}
\end{figure}

Inserting the known eigenvalues together with the eigenframe in Eq.~(\ref{eqn:AI-eigenstate-rotation}) into Eq.~(\ref{eqn:spectral-decomp-AI}), and comparing it against the general decomposition in Eq.~(\ref{eqn:AI-3band-Ham-GM}), we can read off the five coefficients multiplying the real Gell-Mann matrices as 
\begin{subequations}
\label{eqn:AI-manifold-M1}
\begin{eqnarray}
k_1 & = & -\tfrac{\sqrt{3}}{2}\,\sin(2\phi)\,\sin^2\theta \\
k_3 & = & -\tfrac{\sqrt{3}}{2}\,\cos(2\phi)\sin^2\theta\\
k_4 & = & \tfrac{\sqrt{3}}{2}\,\sin\phi\,\sin(2\theta) \\
k_6 & = & -\tfrac{\sqrt{3}}{2}\,\cos\phi\,\sin(2\theta) \\
k_8 & = & -\tfrac{1}{4}\,\big[1 + 3\cos(2\theta)\big]
\end{eqnarray}
\end{subequations}
Since there are two free parameters, $M_1$ is a two-dimensional manifold, i.e., of codimension $\delta^\textrm{AI}_{(2)}=2$ on the enclosing $S^4$ as expected. 
To identify this manifold, recall the domain of the two parameters is $(\theta,\phi)\in[0,\tfrac{\pi}{2}]\times [0,2\pi]$, corresponding to the upper hemisphere of $S^2$ in the usual spherical coordinates.
In addition, the choices $\phi$ and $\phi+\pi$ for $\theta=\pi/2$ (i.e., antipodal points along the sphere's equator) differ only by a $\mathsf{O}(1)$ gauge transformation. 
With these constraints, we identify $M_1 \simeq \reals P^2$.
This result follows, in fact, directly from the specified freedom for the three eigenstates modulo the gauge ambiguity, i.e., $\mathsf{O}(3)/\mathsf{O}(2)\times \mathsf{O}(1) \simeq \reals P^2$, which corresponds to the classifying space of $3$-band class-$\textrm{AI}$ with a single energy gap. 
The projections of the nodal manifolds onto a few subsets of the momenta are shown in Fig.~\ref{fig:rp2_linking}.

The cell decomposition of the real projective plane $\reals P^2$ is well known, and consists of one cell in each dimension $D\in\{0,1,2\}$, where each of these cells acts as a generator of a homology group with $\ztwo$ coefficients~\cite{Hatcher_book}.
According to the discussion in Sec.~\ref{sec:general-section}, in the enclosing sphere $S^{4}$, one can anticipate the linking of the one-cell of $M_1$ and the two-cell of $M_2$ (equivalently, of the one-cell of $M_2$ with the two-cell of $M_1$). 
To demonstrate the linking, we show that both homology classes in dimensions one and two carry a nontrivial band invariant. 
We begin by calculating the Berry phase (first Stiefel-Whitney number) on a one-dimensional cycle of~$M_1$.

As a representative of the nontrivial one-cycle $\gamma_1 \subset M_1$, we take the path with fixed $\theta=\pi/2$ and with $\phi\in[0,\pi]$, where the $5$-momentum reduces to
\begin{equation}
(k_1,k_3,\ldots,k_8) \!=\! \big({-}\tfrac{\sqrt{3}}{2}\sin(2\phi),{-}\tfrac{\sqrt{3}}{2}\cos(2\phi),0,0,\tfrac{1}{2}\big),\!
\end{equation}
allowing us to compute the Berry phase of the degenerate states $\{\ket{\psi_1},\ket{\psi_2}\}$ as well as of the non-degenerate state $\ket{\psi_3}$ on the 1-cycle analytically.
Specifically, the Hamiltonian along the path as a function of $\phi\in[0,\pi]$ becomes
\begin{equation}
H^\textrm{AI}(\bs{k}(\phi))=\left(
\begin{array}{ccc}
\frac{1-3 \cos (2 \phi)}{2 \sqrt{3}} & -\tfrac{\sqrt{3}}{2} \sin (2\phi) & 0 \\
 -\frac{\sqrt{3}}{2} \sin (2\phi) & \frac{1+3 \cos (2 \phi)}{2 \sqrt{3}} & 0 \\
 0 & 0 & -\frac{1}{\sqrt{3}} \\
\end{array}
\right)
\end{equation}
whose eigensystem, also readable from Eq.~(\ref{eqn:AI-eigenframe}), is 
\begin{subequations}
\label{eqn:AI-S1-eigenstates}
\begin{eqnarray}
\ket{\psi_1}=(\cos\phi,\sin\phi,0)^\top \;\;\; &\qquad& \varepsilon_1 = -1/\sqrt{3} \quad \\
\ket{\psi_2}=(0,0,1)^\top \qquad \quad \;\;\;\,&\qquad & \varepsilon_2 = -1/\sqrt{3} \quad \\
\ket{\psi_3} = (\sin\phi,-\cos\phi,0)^\top &\qquad& \varepsilon_3 = 2/\sqrt{3}.\quad 
\end{eqnarray}
\end{subequations}
From the above, it is straightforward to show that both $\{\ket{\psi_1},\ket{\psi_2}\}$ and $\ket{\psi_3}$ carry a nontrivial Berry phase $\varphi_\textrm{B} = \pi$ on the $1$-cycle of $M_1$.
According to the general discussion in Sec.~\ref{sec:general-section}, the nontrivial band invariant signals nontrivial linking of the nodal manifolds $M_1$ and $M_2$, which, in turn, implies the existence of a threefold degeneracy inside the sphere~$S^4$.

While the presented conclusion has been derived for the elementary Hamiltonian in Eq.~(\ref{eqn:AI-3band-Ham-GM}), the robustness of the topological band invariants ensures that the same conclusions persist under continuous deformations of that elementary Hamiltonian to general models.

The same arguments can be repeated for the nodal manifold $M_2$ (in this case considering the non-degenerate eigenstate $\ket{\psi_1}$, which is first rotated by $\theta$ around the $y$-axis and subsequently by $\phi$ around the $x$-axis). 
We find a parameterization of $M_2$ with angles $\theta$ and $\phi$ (with the same domain and boundary conditions as before) as
\begin{subequations}
\label{eqn:AI-manifold-M2}
\begin{eqnarray}
k_1 & = & -\tfrac{\sqrt{3}}{2}\,\sin\phi\,\sin(2\theta) \\
k_3 & = & -\tfrac{\sqrt{3}}{4}\,\Big[(1 + \sin^2\phi)\,\cos(2\theta) + \cos^2\phi\Big] \\
k_4 & = & \tfrac{\sqrt{3}}{2}\,\cos\phi\,\sin(2\theta) \\
k_6 & = & \tfrac{\sqrt{3}}{2}\,\sin(2\phi)\,\sin^2\theta \\
k_8 & = & \tfrac{1}{8}\,\Big[(3\cos(2\phi)-1) - 6\cos^2\phi\,\cos(2\theta)\Big].
\end{eqnarray}
\end{subequations}
By repeating the earlier analysis, one finds that both the non-degenerate states $\ket{\psi_1}$ as well as the degenerate doublet $\{\ket{\psi_2},\ket{\psi_3}\}$ carry a nontrivial Berry phase $\varphi_\textrm{B}=\pi$ on the nontrivial $1$-cycle of $M_2$.
This analysis of band topology on $M_2$ provides equivalent evidence for the linking of $M_1$ and $M_2$, as well as for the presence of a threefold degeneracy inside $S^4$, as discussed in the former discussion of $M_1$.

Before proceeding with the discussion of the next symmetry class, we want to bring up three remarks related to the above analysis.
First, it is known that the integer topological invariant in Eq.~(\ref{eqn:pi1-S1-AI}) does not constitute the stable characterization of twofold band degeneracies in class AI; namely, the winding number reduces~\cite{Bzdusek:2017,Ahn:2018,Wu:2019} to the $\ztwo$-valued first Stiefel-Whitney class (equivalently:~to the quantized Berry phase) in models with three or more bands. 
Mathematically, this corresponds to replacing the classifying space of (single-gap) two-band insulators, $\mathsf{O}(2)/\mathsf{O}(1)\,{\times}\,\mathsf{O}(1) \simeq S^1$, with the classifying space of many-band insulators, $\mathsf{O}(n+\ell)/\mathsf{O}(n)\,{\times}\,\mathsf{O}(\ell)$ where $n+\ell \geq 3$.
The reduction of the fundamental group arises due to non-Abelian braiding with twofold degeneracies in adjacent energy gaps~\cite{Bouhon:2020}.
Such a reduction is not unique to the fundamental group; in particular, for the symmetry class $\textrm{AI}$ in two dimensions, the $\intg$-valued Euler class reduces to the $\ztwo$-valued second Stiefel-Whitney class due to nontrivial action of $\pi_1$ on $\pi_2$~\cite{Tiwari:2020}. 
Even more intricate interplay between band invariants on 1D, 2D, and 3D cycles of the Brillouin zone arises in multi-gap class AI in three dimensions~\cite{Lim:2023,Davoyan:2024,Jankowski:2024}.

Given the examples above, it appears conceivable that the $\intg$-valued winding number $\pi_4(S^4)$ that we use to characterize threefold band degeneracies in the minimal (i.e., three-band) models may similarly reduce to a stable invariant captured by a smaller group. 
However, in contrast to the mentioned cases in one to three dimensions, the study of such a topological reduction in the present context is significantly more challenging. 
The reason is the absence of a clean gap condition on the enclosing $S^4$. 
More specifically, since both band gaps close somewhere on the enclosing sphere (namely on $M_1$ and $M_2$, respectively), the topology that stabilizes the threefold degeneracy does not correspond to a homotopy group of any of the standard single-gap or multi-gap classifying spaces of Hamiltonians. 
We leave the study of the stability of the topological invariant of the threefold band degeneracy in class AI to a separate work rooted in homological constructions~\cite{Iliasov:2026}.

Second, note that to demonstrate the linking of $M_1$ and $M_2$ it is sufficient to find \emph{one} nontrivial cycle of either $M_1$ and $M_2$ that supports a nontrivial band invariant. 
The general description presented in Sec.~\ref{sec:general-section} does not specify the dimension of the nontrivial cycles whose nontrivial band invariant signals the presence of the multifold degeneracy inside the enclosing sphere. 
Generally, in this and the subsequent sections, we choose to consider the \emph{lowest}-dimensional cycle of dimension larger than $0$ (except for the classes $\textrm{BDI}$ and $\textrm{D}$, where a topological invariant can be readily constructed on $0$-cells), finding a nontrivial band invariant in each case.
That the lowest-dimensional cycle readily enables the detection of the multifold degeneracy is particularly convenient from the computational perspective. 
We therefore largely leave open the question of whether band invariants on higher-dimensional cycles of $M_1$ and $M_2$ (which are computationally more challenging to extract) can likewise be used for the diagnosis of the multifold degeneracies.
However, the case of symmetry class $\textrm{AI}$ is sufficiently simple to also admit an explicit cohomological as well as computational analysis of the topology on the $2$-cell, which we present below.

Let us use the symbol $E^{(1)}$ for the rank-one bundle spanned by the upper non-degenerate band ($E^{(2)}$ for the rank-two bundle spanned by the lower two degenerate bands) of the Hamiltonian (\ref{eqn:AI-3band-Ham-GM}) on $M_1 \simeq \reals P^2$. 
The cohomology ring on this base manifold is known~\cite{Hatcher_book} to take the form
\begin{equation}
H^*(\reals P^2,\ztwo) = \ztwo[a]/(a^3).
\end{equation}
The symbol on the right-hand side indicates sums of the form $w = \beta_0 + \beta_1 a + \beta_2 a^2$ with $\beta_i \in \ztwo = \{0,1\}$, where $a$ is the generator of $H^1(\reals P^2,\ztwo)$, $a^2 = a \smile a $ (with `$\smile$' the cup product) is the generator of $H^2(\reals P^2,\ztwo)$, and $a^3 = 0$ vanishes.
Owing to its rank-$1$ and nontrivial Berry phase, we know that the total Stiefel-Whitney class of the bundle $E^{(1)}$ is
\begin{subequations}
\begin{equation}
w(E^{(1)}) = 1 + a.   
\end{equation} 
Furthermore, since the total bundle $E^{(1)} \oplus E^{(2)} \simeq \reals P^2 \times \reals^3$ is trivial, we have 
\begin{equation}
w(E^{(1)} \oplus E^{(2)}) = 1.
\end{equation}
By inverting the Whitney product formula $w(E\oplus F) = w(E)\smile w(F)$, we find a unique solution for the topology of vector bundle $E^{(2)}$:
\begin{equation}
w(E^{(2)}) = (1+a)^{-1} = 1 + a + a^2.
\end{equation}
\end{subequations}
Reading off the degrees, we see that the bundle spanned by the two degenerate bands exhibits nontrivial first as well as nontrivial second Stiefel-Whitney invariant (2SW).
The first result has been checked with an explicit calculation below~Eq.~(\ref{eqn:AI-S1-eigenstates}).

To verify the second prediction, we study spectral flow of Wilson loop eigenvalues on a family of contractible paths on $M_1$. 
To facilitate the calculation of the Wilson loop eigenvalues,
we consider the family of loops on the $(\theta,\phi)$ parameterization of $M_1$ specified by $s_\theta : t \mapsto (\theta,2\pi t)$ with $t \in [0,1]$, which correspond to latitude circles at a fixed polar angle $\theta$. 
The subspace parallel-transported along these loops is spanned by the first two columns of Eq.~(\ref{eqn:AI-eigenframe}). 
The non-Abelian Berry connection (which we define as $\mathcal{A} = +\imi \Psi^\dagger d \Psi$) along $s_\theta$ is 
\begin{subequations}
\begin{equation}
A_\phi(\theta) = \sigma_y \cos \theta.     
\end{equation}    
The Wilson loop matrix is the path ordered product 
\begin{equation}
\label{eq:Wilson-matrix-definition}
W(\theta) = \mathcal{P} \exp \left( -\imi \int_0^{2\pi} A_\phi(\theta) d\phi \right).    
\end{equation} 
This gives 
\begin{equation}
W(\theta) = \exp( -\imi \sigma_y 2\pi \cos\theta)    
\end{equation} 
\end{subequations}
whose eigenvalues are $\lambda_\pm(\theta) = \exp(\pm 2\pi \imi \cos \theta)$ from which we can read off the phases $\nu_\pm(\theta) = \pm 2\pi \cos\theta$. 
We recover vanishing phases at $\theta = 0$ and at $\theta = \pi/2$. 
On the $[0,\pi/2]$ interval, which corresponds to covering $M_1$ once, the cosine function is monotonically decreasing; therefore, we find a single crossing of the Wilson phases at $\nu_\pm = \pi$ for the path at coordinate $\theta = \pi/3$. 
The single Wilson crossing at $\pi$ implies that the 2SW invariant is nontrivial~\cite{Ahn:2018}.

We finally briefly recast the above results about the topology of triple nodal points in light of the discussion in Sec.~\ref{sec:general-linking-to-invariants} and compare against certain results for the class $\textrm{AI}$ known in the literature. 
First, the claim that the nontrivial Berry phase on the $1$-cycle of $M_1$ implies its linking with $M_2$ translates to the fact that any disk inscribed to the $1$-cycle exhibits a gap closing where the real vector bundle of $M_1$ cannot be analytically continued. 
This gap closing is generically a point-like Dirac node.
We want to highlight how our analysis relates to the well-established fact that such point nodes act as monopole sources of Berry phase $\varphi_\textrm{B} = \pi$.
To that end, note that the quantized Berry phase remains well-defined under continuous deformations of the base; in particular, we can consider a loop in the parameter space obtained by a small distortion of the $1$-cycle.
Then, the band degeneracy constituting $M_1$ generally splits, and the nontrivial Berry phase is inherited by one of the two energetically split bands. 
In the spirit of Fig.~\ref{fig:W_variation}, one can consider shrinking of this loop to the neighborhood of the Dirac node, allowing one to establish the Dirac node as the source of the Berry phase quantum.

In a similar spirit, we relate our analysis to the established result that nodal-line rings in 3D that carry a nontrivial 2SW invariant on the enclosing sphere $S^2 \subset \reals^3$~\cite{Fang:2015} are linked with nodal lines in adjacent energy gap~\cite{Ahn:2018}.
Here, a subtle difference emerges, namely that the $S^2$ used to enclose nodal-line rings in 3D vs.~the $2$-cell $\reals P^2 \subset M_1$ that arises in our present analysis are different manifolds. 
Nevertheless, the two cases exhibit analogous phenomenology in that the rank-$2$ bundle with nontrivial 2SW invariant (and in the case of $\reals P^2$ with the additional derived property that the first Stiefel-Whitney class is nontrivial too) does not split as a product of two rank-$1$ vector bundles.
In our present context, this implies that smooth deformations of the base from the initial base $M_1$ cannot split the degeneracy of the bands $\{\ket{\psi_1},\ket{\psi_2}\}$ completely; rather, they remain degenerate somewhere on the two-dimensional base.
For codimensional reasons, this degeneracy is generically point-like on any of these bases.
Now, consider a $3$-cell\footnote{This $3$-cell is mathematically interpreted as a $3$-chain with $\ztwo$ coefficients. It does not correspond to a smooth manifold with a boundary because $\reals P^2$ is not cobordant with the empty set.} inscribed into $M_1 \subset S^4$.
On one hand, the established linking of $M_1$ with $M_2$ implies that this $3$-cell intersects with $M_2$; for codimensional reasons, this intersection is generically one-dimensional, i.e., a nodal \emph{ring}.
On the other hand, continuous deformation of the base of the rank-2 bundle from $M_1 \simeq \reals P^2$ inside the $3$-cell leave behind a line of band degeneracies of $M_1$ (thus, any deformation of $M_1$ inside $S^4$ necessarily intersects $M_1$).
We observe a correspondence where $\reals P^2$ that (\emph{i})~encloses a nodal-line ring (part of $M_2$) and that (\emph{ii})~carries a nontrivial 2SW invariant (of the two bands that extend to $M_1$) exhibits a nodal point (part of $M_1$). 
This observation constitutes a direct analog of the linking phenomenology established~\cite{Ahn:2018} for spheres enclosing nodal rings~in~3D.

In the next sections, devoted to other symmetry classes, we follow the same steps as described above for the symmetry class $\textrm{AI}$ whenever possible. 
If the calculation of band invariants on $M_1$ and $M_2$ is not feasible, we restrict ourselves only to describing the structure of the nodal manifolds.

\section{Symmetry class \texorpdfstring{$\textrm{A}$}{A}}
\label{sec:class-A}

The unitary class (labeled $\mathrm{A}$) captures Hamiltonians without time-reversal symmetry.  
The Hamiltonian is an arbitrary Hermitian matrix, and the eigenstates of an $N$-state Hamiltonian can be arranged as columns in a unitary matrix $\Psi \in \mathsf{U}(N)$.
Our discussion of the symmetry class $\textrm{A}$ parallels the discussion in the previous section. First, we review the codimension counting by von Neumann and Wigner, then we construct the nodal manifolds $M_1$ and $M_2$ as two non-intersecting copies of $\cmplx P^2$ on the enclosing sphere $S^7$, and finally we relate the threefold degeneracy inside the enclosing sphere to Chern numbers on cycles of the nodal manifolds.

To understand the dimension of the group $\mathsf{U}(n)$, first note that this is a $n \times n$ matrix with complex entries, implying $2n^2$ parameters.
The normalization, which corresponds to a real-valued product of the state with itself, implies $n$ constraints on these parameters, i.e., the same as for the orthogonal class.
In contrast, owing to the complex-valued inner product of distinct eigenstates, orthogonality of distinct pairs of eigenstates implies $2\times \tfrac{1}{2}n(n-1)$ constraints, i.e., twice as many as in the orthogonal class.
Therefore, we have
\begin{equation}
\dim[\mathsf{U}(n)]=2n^2 - n - n(n-1) = n^2.    
\end{equation}
The von Neumann-Wigner argument specifies the dimension of the space of class-$\mathrm{A}$ Hamiltonians with $N = \sum_{j=1}^k n_k$ bands~as
\begin{eqnarray}
d^{\textrm{A}}_{(n_1,n_2,\ldots,n_k)} &=& k + \dim [\mathsf{U}(N)] - \sum_{j=1}^k \dim [\mathsf{U}(n_j)] \nonumber \\
&=& k + N^2 - \sum_{j=1}^k n_k^2.
\end{eqnarray}
From here, we define 
\begin{equation}
\label{eqn:A-codim}
\delta^\textrm{A}_{(n)} = d^\textrm{A}_{(1,1,1,\ldots)} - d^\textrm{A}_{(n,1,1,\ldots)} = n^2-1
\end{equation}
as the codimension of forming an $n$-fold degeneracy in an otherwise non-degenerate spectrum.
A $2$-fold degeneracy has codimension $\delta_{(2)}^\textrm{A}=3$, corresponding to the three Pauli matrices. 
The codimension of a $3$-fold degeneracy is $\delta_{(3)}^\textrm{A}=8$, corresponding to the eight Gell-Mann matrices. 
More generally, for $n$-fold degeneracy, we count dimension of the space of traceless Hermitian $n \times n$ matrices
This equals $n-1$ (for the real diagonal terms) plus $2\times \tfrac{1}{2}n(n-1)$ (for the complex off-diagonal terms), which adds up to $n^2 - 1$, matching the codimension determined in Eq.~(\ref{eqn:A-codim}).

We next consider a three-band Hamiltonian that describes a threefold degeneracy in class $\mathrm{A}$.
Similarly to the three-band Hamiltonian in class $\textrm{AI}$, Eq.~(\ref{eqn:AI-3band-Ham-GM}), the class-A analogue is a linear combination of Gell-Mann matrices; however, due to the lack time-reversal symmetry, all eight Gell-Mann matrices listed in Eq.~(\ref{eqn:GellMann}) contribute. 
Therefore,
\begin{equation}
    H^\textrm{A} = \sum_{i = 1}^8 k_i \Lambda_i
    \label{eqn:A-3band-Ham-GM}
\end{equation}
with $8$-component momentum vector $\bs{k} = (k_1,\ldots,k_8)$.
Since Eqs.~(\ref{eqn:gell-mann-squared}) still apply, the nodal manifolds $M_1$ and $M_2$ on an enclosing sphere with radius $\kappa = 1$ are again defined by Eqs.~(\ref{eqn:AI-M1-M2-spectrum}), with the key difference that, in the present case, this sphere is seven-dimensional. 
In the following discussion, we focus on the particular case of the nodal manifold $M_1$, with the discussion of $M_2$ being completely analogous.

Let us first describe the geometry of $M_1$ on general grounds. For that purpose, one needs to consider not only the eigenvalues but also the eigenstates of the introduced three-band Hamiltonian. 
On $M_1$, since the eigenvalues are fixed by Eqs.~(\ref{eqn:AI-M1-M2-spectrum}), the Hamiltonian is fully specified by the third (non-degenerate) eigenstate $\ket{\psi_3}$, which is now a vector in $\mathbb{C}^3$. 
Since $\ket{\psi_3}$ is normalized and is defined up to a $\mathsf{U}(1)$ gauge freedom, it describes a point in $S^5/\mathsf{U}(1)$, which corresponds to a standard identification of $\mathbb{C}\mathsf{P}^2$. 
This allows us to conclude that $M_1 \simeq \mathbb{C}\mathsf{P}^2$. 
This result is consistent with $M_1$ being the classifying space of three-band class-$\textrm{A}$ Hamiltonians with one energy gap, $M_1 \simeq \mathsf{U}(3)/\mathsf{U}(2)\times \mathsf{U}(1) \simeq \cmplx P^2$.

Next, we show how one can parameterize $M_1$ with four angles, allowing us to explicitly verify the linking with $M_2$ in terms of Chern numbers on nontrivial cycles. 
Note that a general \emph{complex} three-component vector is obtained [up to an unimportant $\mathsf{U}(1)$ phase] from the general \emph{real} three-component $\ket{\psi_3}$ in Eqs.~(\ref{eqn:AI-eigenstates}) by including two relative phases between the individual components. 
Therefore, we construct the eigenframe as
\begin{subequations}
\begin{equation}
\label{eqn:A-eigenstate-rotation}
\Psi 
= \mathcal{D}(\alpha,\beta)\mathcal{R}_z(\phi)\mathcal{R}_x(\theta)\Psi_0,
\end{equation}
where $\mathcal{R}_z(\phi)$ and $\mathcal{R}_x(\theta)$ are the two orthogonal rotations present in Eq.~(\ref{eqn:AI-eigenstate-rotation}) while $\mathcal{D}(\alpha,\beta)=\diag(e^{\imi \alpha},e^{\imi \beta},1)$ is a diagonal matrix with arbitrary complex phases $\alpha$ and $\beta$.
The resulting eigenframe on $M_1$ takes the form
\begin{equation}
\label{eqn:A-eigenframe}
\Psi = \left(
\begin{array}{ccc}
 e^{i \alpha } \cos \phi  & \,-e^{i \alpha } \cos \theta \sin \phi  & e^{i \alpha } \sin \theta 
   \sin \phi  \\
 e^{i \beta } \sin \phi  & e^{i \beta } \cos \theta  \cos \phi  & \,-e^{i \beta } \sin \theta  \cos
   \phi  \\
 0 & \sin \theta  & \cos \theta  \\
\end{array}
\right).    
\end{equation}
\end{subequations}
Note that, due to the inclusion of the complex phases $\alpha$ and $\beta$ in range $[0,2\pi]$ [and in contradistinction to Eq.~(\ref{eqn:AI-eigenframe})], the range of $\phi$ must be reduced to $[0,\tfrac{\pi}{2}]$ to avoid double counting (while the range $\theta \in [0,\tfrac{\pi}{2}]$ remains unaffected). 
The reason is that the increase $\beta \mapsto \beta + \pi$ is equivalent\footnote{By `equivalence' we here mean that, up to a gauge transformation, we obtain the same three eigenstates.} to the `reflection' $\phi \mapsto \pi - \phi$, while the simultaneous increase $(\alpha,\beta) \mapsto (\alpha + \pi,\beta+\pi)$ is equivalent to the `translation' $\phi\mapsto \pi+\phi$.
In combination, the freedom of the phases $(\alpha,\beta)$ therefore implies quartering of the domain of $\phi$.

From $\Psi$, one extracts the three eigenstates as the individual columns.
The decomposition of $H^\textrm{A}=\Psi \mathcal{E} \Psi^\dagger$ into the Gell-Mann basis following Eq.~(\ref{eqn:A-3band-Ham-GM}) then gives an explicit parameterization of $\cmplx P^2 \subset S^7 \subset \reals^8$ as 
\begin{subequations}
\begin{eqnarray}
k_1 &=& -\tfrac{\sqrt{3}}{2}  \sin ^2\theta \, \sin (2 \phi ) \, \cos (\alpha -\beta )\\
k_2 &=& \tfrac{\sqrt{3}}{2}  \sin ^2\theta \, \sin (2 \phi ) \, \sin (\alpha -\beta ) \\
k_3 &=& -\tfrac{\sqrt{3}}{2}  \sin ^2\theta \, \cos (2 \phi )\\
k_4 &=& \tfrac{\sqrt{3}}{2}  \cos \alpha \, \sin (2 \theta ) \, \sin \phi \\
k_5 &=& -\tfrac{\sqrt{3}}{2}  \sin \alpha \, \sin (2 \theta ) \,\sin \phi \\
k_6 &=& -\tfrac{\sqrt{3}}{2}  \cos \beta \, \sin (2 \theta ) \, \cos \phi \\
k_7 &=& \tfrac{\sqrt{3}}{2}  \sin \beta \, \sin (2 \theta ) \, \cos \phi \\
k_8 &=& -\tfrac{1}{4} \left[1+3 \cos (2 \theta )\right].
\end{eqnarray}
\end{subequations}
The manifold is covered once if we restrict the parameters to the ranges $\alpha,\beta\in[0,2\pi]$ and $\theta,\phi\in[0,\tfrac{\pi}{2}]$.

To verify the linking of $M_1$ and $M_2$ we need to compute the Chern number of either band subspace on the nontrivial 2-cycle $\cmplx P^1 \subset \cmplx P^2$. 
To that end, recall that $\cmplx P^1 \simeq S^2$ can be interpreted as the Bloch sphere.
Furthermore, observe that we can recover the two eigenstates of the spinor Hamiltonian 
\begin{subequations}
\begin{equation}
\label{eqn:spinor-1/2-Ham}
H_{1/2}=\bs{n}\cdot \bs{\sigma} 
\end{equation} 
with a 2-spherical parameterization of the unit vector $\bs{n}$ on the second two components of $\ket{\psi_2}$ and $\ket{\psi_3}$ after setting $\phi = 0$ and $\alpha = 0$ (while keeping $\theta$ and $\beta$ as tunable parameters).
Specifically, if we parameterize 
\begin{equation}
\bs{n}=\big(\sin(2\theta)\cos\beta,-\sin(2\theta)\sin\beta,\cos(2\theta)\big)    
\end{equation}
with $\theta\in[0,\tfrac{\pi}{2}]$ and $\beta\in[0,2\pi]$, then the positive (`$+$') and the negative (`$-$') eigenstate of (\ref{eqn:spinor-1/2-Ham}) take the form
\begin{eqnarray}
\ket{\psi_+} &=& (e^{\imi\beta}\cos\theta, \sin\theta)^\top \\
\ket{\psi_-} &=& (-e^{\imi\beta}\sin\theta, \cos\theta)^\top .
\end{eqnarray}
\end{subequations}
On the other hand, setting $\phi = 0 = \alpha$ results in a two-parameter family of eigenstates 
\begin{subequations}
\label{eqn:A-S2-eigenstates}
\begin{eqnarray}
\ket{\psi_1}=(1,0,0)^\top \qquad \qquad \;\;\; \,&\quad \;\;& \varepsilon_1 = -1/\sqrt{3} \quad \\
\ket{\psi_2}=(0,e^{\imi\beta}\cos\theta,\sin\theta)^\top \;\;\, &\quad & \varepsilon_2 = -1/\sqrt{3} \quad \;\;\\
\ket{\psi_3} = (0,-e^{\imi\beta}\sin\theta,\cos\theta)^\top &\quad \;\;& \varepsilon_3 = 2/\sqrt{3}.\quad 
\end{eqnarray}
\end{subequations}
Owing to the interpretation of $\ket{\psi_2}$ and $\ket{\psi_3}$ in terms of the spinor Hamiltonian~(\ref{eqn:spinor-1/2-Ham}), we immediately recognize that both band subsets $\{\ket{\psi_1},\ket{\psi_2}\}$ and $\{\ket{\psi_3}\}$ carry a nontrivial Chern number $\abs{C}=1$ (of opposite relative sign) on the nontrivial $2$-cycle of $M_1$. 
According to the general discussion presented in Sec.~\ref{sec:general-section}, this band invariant signals the existence of the threefold band degeneracy inside the enclosing sphere $S^7$.

Similarly to the discussion for the class $\textrm{AI}$, we can deduce that the nodal manifolds carry a nontrivial \emph{second} Chern number as well. 
As in the case of the $\textrm{AI}$ class, we denote the rank-one bundle spanned by the upper non-degenerate band as $E^{(1)}$ and use $E^{(2)}$ for the rank-two bundle spanned by the lower two degenerate bands of the Hamiltonian (\ref{eqn:A-3band-Ham-GM}) on the base space $M_1 \simeq \mathbb{C} P^2$. 
The relevant cohomology ring on $\mathbb{C} P^2$ takes a similar form as well, but now, due to the involvement of \emph{complex} vector bundles, we can take it with integer coefficients~\cite{Hatcher_book},
\begin{equation}
H^*(\mathbb{C} P^2,\intg) = \intg[a]/(a^3).
\end{equation}
The symbol on the right-hand side indicates sums of the form $c = \beta_0 + \beta_1 a + \beta_2 a^2$ with $\beta_i \in \intg$, where $a$ is the generator of $H^2(\mathbb{C} P^2,\intg)$, $a^2 = a \smile a $ is the generator of $H^4(\mathbb{C} P^2,\intg)$, and $a^3 = 0$ vanishes.

Let us choose the sign of the Chern number of the bundle $E^{(1)}$ being positive (this is equivalent to choosing an appropriate orientation on the corresponding $2$-cycle). 
Since the bundle $E^{(1)}$ has Chern number and complex rank equal to $1$, its total Chern class is
\begin{subequations}
\begin{equation}
    c(E^{(1)})=1+a.
\end{equation}
As in the real case, the total bundle $E^{(1)} \oplus E^{(2)} \simeq \mathbb{C} P^2 \times \mathbb{C}^3$ is trivial; therefore, we have 
\begin{equation}
c(E^{(1)} \oplus E^{(2)}) = 1.
\end{equation}
By inverting the Whitney sum formula for the total Chern classes, $c(E\oplus F) = c(E)\smile c(F)$, we find the unique solution for the Chern class of $E^{(2)}$,
\begin{equation}
    c(E^{(2)})=1-a+a^2.
\end{equation}
\end{subequations}
This formula confirms that first Chern numbers of bundles $E^{(1)}$ and $E^{(2)}$ have different signs, and it demonstrates that bundle $E^{(2)}$ also has second Chern number equal to $1$. 
This indicates the linking of $4$-cycle in the nodal manifold $M_1$ with the $2$-cycle in the nodal manifold $M_2$.
The nontrivial value of the second Chern number of the degenerate states $\{\ket{\psi_1},\ket{\psi_2}\}$ on $M_1 \simeq \cmplx P^2$ can be verified through an explicit analytical calculation. 
We include the derivation in the Supplementary code and data~\cite{supplementary_code_data}.

Let us conclude with recasting the derived characterization of the triple-point topology in terms of Sec.~\ref{sec:general-linking-to-invariants} and relate it to result for the class $\textrm{A}$ established in earlier works.
First, the fact that nontrivial first Chern number on the $2$-cycle $\cmplx P^1 \simeq S^2$ of $M_1$ implies linking with $M_2$ translates to the fact that any $3$-cell (i.e., `ball') inscribed to the $2$-cycle exhibits a gap closing where the complex vector bundle of $M_1$ cannot be analytically continued. 
This gap closing is generically a point-like Weyl node.
The rank-$2$ vector bundle on $M_1$ can be split into two rank-$1$ bundles by continuous deformations of the base, with the first Chern number carried by one of the two energetically split bands. 
By shrinking the $2$-cycle tightly around the Weyl node, we can identify it as the source of one quantum of the Chern topology.
Second, it has been established that two-dimensional Weyl nodal surfaces in 5D that carry a nontrivial second Chern number on the enclosing sphere $S^4 \subset \reals^5$ are linked with Weyl nodal surfaces in the adjacent energy gap~\cite{Lian:2016}.
By adapting the reasoning presented at the end of Sec.~\ref{sec:class-AI} for the class $\textrm{AI}$, one can establish that an equivalent phenomenology arises in our present context, albeit for a different choice of the enclosing manifold. 
Namely, in the present context, we find that $\cmplx P^2$ that (\emph{i})~encloses a nodal surface (part of $M_2$) and that (\emph{ii})~carries a nontrivial second Chern number (of the two bands that extend to $M_1$) exhibits a nodal line (part of $M_1$). 
The latter degeneracy extends to a nodal surface linking with nodal surface in the adjacent energy~gap.

\section{Symmetry class \texorpdfstring{$\textrm{AII}$}{AII}}
\label{sec:class-AII}

The symplectic class (labeled $\mathrm{AII}$) captures time-reversal-symmetric models with $\mathcal{T}^2 = -\mathbb{1}$, as expected for fermions with half-integer spin.
We represent $\mathcal{T} = - \imi \sigma_y \mathcal{K}$, where $\mathcal{K}$ is the complex conjugation operator, and $\{\sigma_j\}_{j=0,x,y,z}$ are the Pauli matrices in the spin degree of freedom, including the identity matrix.
If there are other degrees of freedom besides spin, we will express the time-reversal operator in the block-diagonal structure
\begin{equation}
\label{eqn:spinful-TRS-blocks}
\mathcal{T} = \mathbb{1}_N\otimes (- \imi \sigma_y \mathcal{K})=\left(\begin{array}{cc|cc|c}
0 & -1 & 0 & 0 & \cdots \\
1 & 0 & 0 & 0 & \cdots \\ \hline
0 & 0 & 0 & -1 & \cdots \\
0 & 0 & 1 & 0 & \cdots \\ \hline
\vdots & \vdots & \vdots & \vdots & \ddots
\end{array}\right) \mathcal{K},    
\end{equation}
where the first matrix acts on additional degrees of freedom.
We also introduce the symbol $\Omega = \mathbb{1}_N\otimes \imi \sigma_y$ and call this matrix the \emph{symplectic form}.

Recall that spinful time-reversal symmetry is associated with Kramers degeneracy, i.e., if the Hamiltonian obeys $\mathcal{T} H \mathcal{T}^{-1} = H$ and $H\ket{u^\alpha} = \varepsilon^\alpha \ket{u^\alpha}$, then $\mathcal{T} \ket{u^\alpha}$ is another (orthogonal) eigenstate of $H$ with the same energy $\varepsilon^\alpha$. 
Specifically, if we decompose the eigenstate into the basis as
\begin{subequations}
\begin{equation}
\ket{u^\alpha} = \left(\begin{array}{cc|cc|c}
v^\alpha_1 & w^\alpha_1 & v^\alpha_2 & w^\alpha_2 & \cdots
\end{array}\right)^\top
\end{equation}
then the Kramers partner reads
\begin{equation}
\mathcal{T}\ket{u^\alpha} = \left(\begin{array}{cc|cc|c}
-\bar{w}^{\alpha}_1 & \bar{v}^{\alpha}_1 & -\bar{w}^{\alpha}_2 & \bar{v}^{\alpha}_2 & \cdots
\end{array}\right)^\top    
\end{equation}
\end{subequations}
where the horizontal bar above the amplitudes indicates complex conjugation.

Let the Hamiltonian matrix $H$ be of dimension $2N \times  2N$. 
Then, collecting all the eigenstates as columns into a $2N \times 2N$ matrix, with the Kramers partners grouped together, we obtain
\begin{equation}
\label{eqn:symplectic-eigenmatrix}
\Psi = \left(\begin{array}{cc|cc|c}
v_1^1 & -\bar{w}_1^1 & v_1^2 & -\bar{w}_1^2 & \cdots \\[2pt]
w_1^1 & \bar{v}_1^1 & w_1^2 & \bar{v}_1^2 & \cdots \\[2pt] \hline
v_2^1 & -\bar{w}_2^1 & v_2^2 & -\bar{w}_2^2 & \cdots \\[2pt]
w_2^1 & \bar{v}_2^1 & w_2^2 & \bar{v}_2^2 & \cdots \\[2pt] \hline
\vdots & \vdots & \vdots & \vdots & \ddots
\end{array}\right)    
\end{equation}
with $n \times n$ blocks of dimension $2 \times 2$.
The orthogonality of the eigenstates implies that $\Psi$ is unitary, $\Psi \in \mathsf{U}(2N)$, which translates to conditions
\begin{subequations}
\begin{eqnarray}
\label{eqn:Sp-ortho-1}
\bar{v}_j^\alpha v_j^\beta + \bar{w}_j^\alpha w_j^\beta &=& \delta^{\alpha \beta} \\
\label{eqn:Sp-ortho-2}
- w_j^\alpha v_j^\beta + v_j^\alpha w_j^\beta &=& 0.
\end{eqnarray}
\end{subequations}
Using the above, a straightforward calculation further reveals that
\begin{equation}
\label{eqn:U-is-symplectic}
\Psi^\top \Omega \Psi = \Omega,
\end{equation}
meaning that $\Psi$ is also a complex symplectic matrix, $\Psi \in \textsf{Sp}(2N,\cmplx)$. 
Therefore, the matrix $\Psi$ of eigenstates lies in the intersection
\begin{equation}
\Psi \,\in\, \mathsf{U}(2N) \cap \mathsf{Sp}(2N,\cmplx) =: \mathsf{Sp}(N),
\end{equation}
which is called the compact symplectic group.

Before proceeding with the codimension and linking analysis, we show that the presence of spinful time-reversal allows us to elegantly reformulate the Hamiltonian matrix and the Kramers-degenerate eigenstates using quaternion numbers. 
Such a construction has also appeared in the discussion of higher-dimensional class AII topology in Ref.~\citenum{Hatsugai:2010}, and it will enable us to compactify the subsequent discussion.

Recall that quaternions ($\quats$) are numbers of the form
\begin{equation}
\label{eqn:general-quaternion}
\mathfrak{q} = a 1 + b \imi + c \imj + d \imk   
\end{equation}
where $a,b,c,d\in\reals$ are real coefficients, and $\imi^2 = \imj^2 = \imk^2 = \imi \cdot \imj \cdot \imk = -1$ are three imaginary units that anticommute with each other; for example, $\imi \cdot \imj = -\imj \cdot \imi$. 
We also introduce 
\begin{equation}
\label{eqn:Hermitian-conjugation-quaternions}
\mathfrak{q}^\dagger = a 1 - b \imi - c \imj - d \imk 
\end{equation}
and call it the \emph{Hermitian conjugate} of $\mathfrak{q}$.

A convenient representation of quaternions in terms of Pauli matrices reads
\begin{subequations}
\begin{equation}
1 \mapsto \mathbb{1}, \quad\! \imi \mapsto -\imi \sigma_x,\quad \!\imj \mapsto -\imi \sigma_y,\quad \! \textrm{and}\quad \! \imk \mapsto -\imi \sigma_z,     
\end{equation}
in which case Eq.~(\ref{eqn:general-quaternion}) becomes
\begin{equation}
\label{eqn:quat-matrix-rep}
\mathfrak{q} = \left(\begin{array}{cc}
a - d \imi & - c - b \imi \\
c - b \imi & a + d \imi
\end{array}\right).   
\end{equation}
\end{subequations}
Notice that the matrices $\mathbb{1}$ and $-\imi \sigma_y$ are real while $-\imi \sigma_x$ and $\imi \sigma_z$ are imaginary. 
Therefore, we define the \emph{complex conjugate} of the quaternion in Eq.~(\ref{eqn:general-quaternion}) as
\begin{equation}
\label{eqn:quats-complex-conj}
\mathfrak{q}^* = a 1 - b \imi + c \imj -d \imk.
\end{equation}
Crucially, observe that the matrix in Eq.~(\ref{eqn:quat-matrix-rep}) has exactly the same structure as the blocks of the eigenstate matrix in Eq.~(\ref{eqn:symplectic-eigenmatrix}), with $v^\alpha_j = a - d \imi$ and $w^\alpha_j = c - b \imi $. 
Therefore, we can rewrite $\Psi$ as an $N\times N$ matrix over $\quats$ as
\begin{subequations}
\begin{equation}
U = \left(\begin{array}{c|c|c}
\mathfrak{q}^1_1 & \mathfrak{q}^2_1 & \cdots \\[2pt] \hline
\mathfrak{q}^1_2 & \mathfrak{q}^2_2 & \cdots \\[2pt] \hline 
\vdots & \vdots & \ddots
\end{array}\right)   
\end{equation}
with the entries further decomposed as 
\begin{equation}
\mathfrak{q}^\alpha_j = a^\alpha_j 1 + b^\alpha_j \imi + c^\alpha_j \imj + d^\alpha_j \imk.  
\end{equation}
With such a substitution, the pair of Eqs.~(\ref{eqn:Sp-ortho-1}) and~(\ref{eqn:Sp-ortho-2}) can be expressed more compactly as 
\begin{equation}
\label{eqn:symplectic-unitarity}
(\mathfrak{q}_j^\alpha)^\dagger \cdot \mathfrak{q}_j^\beta = \delta^{\alpha \beta} 1,
\end{equation}
\end{subequations}
which at the level of the matrix translates to $\Psi \Psi^\dagger = \Psi^\dagger \Psi = \mathbb{1}$. 
This means that we can treat $\Psi$ as an $N \times N$ unitary matrix over quaternion (rather than complex) numbers. 
For this reason, $\mathsf{Sp}(N)$ is sometimes also called the hyperunitary group.

Note also that the symplectic form is expressed through quaternions as $\Omega = -\mathbb{1}\imj$, and that the transpose of a quaternion number is 
\begin{eqnarray}
(a 1 + b \imi + c \imj + d \imk)^\top &=& \left(\begin{array}{cc}
a - d \imi & c - b \imi \\
- c - b \imi & a + d  \imi
\end{array}\right) \nonumber \\
&=&  a 1 + b \imi - c \imj + d \imk
\label{eqn:quats-transpose}
\end{eqnarray}
with only the sign in front of $c\imj$ flipped. 
Note that $\mathfrak{q}^\top = (\mathfrak{q}^*)^\dagger$, as expected for matrices.
Due to the anticommutativity of the imaginary units, it further follows that 
\begin{subequations}
\begin{equation}
\mathfrak{q}^\top \cdot \imj = \imj \cdot \mathfrak{q}^\dagger.
\end{equation}
Therefore, the left-hand side of Eq.~(\ref{eqn:U-is-symplectic}) translates to
\begin{equation}
(\mathfrak{q}_j^\alpha)^\top \cdot \imj \cdot \mathfrak{q}_j^\beta = \imj \cdot (\mathfrak{q}_j^\alpha)^\dagger \cdot \mathfrak{q}_j^\beta \stackrel{\textrm{(\ref{eqn:symplectic-unitarity})}}{=} \delta^{\alpha\beta}\imj
\end{equation}
\end{subequations}
which reproduces the right-hand side of Eq.~(\ref{eqn:U-is-symplectic}). 
This means that any unitary matrix over quaternion numbers is automatically symplectic.
The conclusion of this analysis is that we can treat the matrix $\Psi$ of eigenstates of a spinful time-reversal-symmetric Hamiltonian $H$ as a unitary matrix over quaternion numbers, with the column
\begin{equation}
\mathfrak{u}^\alpha = \left(\begin{array}{c} 
\mathfrak{q}^\alpha_1 \\ \mathfrak{q}^\alpha_2 \\ 
\vdots 
\end{array}\right)    
\end{equation}
encoding two states that form a Kramers pair.

We next study how to translate into the quaternion language a spinful time-reversal-symmetric Hamiltonian $H$.
To that end, note that a general $2N \times 2N$ matrix can be expressed as 
\begin{subequations}
\begin{equation}
\label{eqn:block-decomp-Ham}
H = h_{\alpha\beta j} M_{\alpha \beta} \otimes \sigma_j    
\end{equation}
where $h_{\alpha\beta j} \in \cmplx$ and $M_{\alpha\beta}$ is an $N\times N$ matrix with components $(M_{\alpha\beta})_{\gamma\delta} = \delta_{\alpha\gamma} \delta_{\beta\delta}$.
Observe that the commutation with the time-reversal operator, $H \mathcal{T} = \mathcal{T} H$, imposes 
\begin{equation}
\label{eqn:symplectic-TRS-with-quats}
h_{\alpha \beta j} \sigma_j \cdot (\imi \sigma_y) = (\imi \sigma_y) \cdot (h_{\alpha \beta j} \sigma_j)^*.
\end{equation}
From here it follows that the coefficient $h_{\alpha\beta j}$ is real for $j=0$ and imaginary for $j\in\{x,y,z\}$. 
Therefore, the blocks of spinful time-reversal-symmetric Hamiltonians take a more specific form 
\begin{eqnarray}
h_{\alpha \beta j} \sigma_j = \mathfrak{h}_{\alpha \beta} &=& a_{\alpha \beta} \mathbb{1} - b_{\alpha \beta} \imi \sigma_x - c_{\alpha \beta} \imi \sigma_y - d_{\alpha \beta} \imi \sigma_z  \nonumber \\
&=:& a_{\alpha \beta} 1 + b_{\alpha \beta} \imi + c_{\alpha \beta} \imj + d_{\alpha \beta} \imk ,
\label{eqn:Ham-quat-decomp}
\end{eqnarray}
\end{subequations}
i.e., the Hamiltonian can be expressed as an $N \times N$ matrix over quaternion numbers. 
In addition, the Hermiticity $H = H^\dagger$ implies at the level of the decomposition in Eq.~(\ref{eqn:block-decomp-Ham}) that $h_{\alpha\beta j} = h_{\beta \alpha j}^*$, which at the quaternion level translates to 
\begin{equation}
\mathfrak{h}_{\alpha \beta} = \mathfrak{h}_{\beta \alpha}^\dagger. 
\label{eqn:Hermiticity-with-quaternions}
\end{equation}
For diagonal terms ($\alpha = \beta$), comparison of Eqs.~(\ref{eqn:general-quaternion}) and~(\ref{eqn:quats-complex-conj}) reveals that only the term ${\propto}\,1$ is allowed, i.e., that all the imaginary numbers are absent.

To understand the dimension of the group $\mathsf{Sp}(n)$, first note that this is an $n \times n$ matrix with quaternion entries, implying $4n^2$ parameters. 
The normalization of the columns, computed as the real-valued inner product of the columns with themselves, implies $n$ constraints on these parameters.
In addition, ensuring the orthogonality for each pair of distinct eigenstates corresponds to additional $4\times \tfrac{1}{2}n(n-1)$ constraints. 
Therefore
\begin{equation}
\dim[\mathsf{Sp}(n)] = 4n^2 - n - 2n(n-1) = n(2n+1).
\end{equation}
The von Neumann-Wigner argument specifies the dimension of the space of class-$\textrm{AII}$ Hamiltonians with $N= \sum_{j=1}^k n_j$ bands~as
\begin{eqnarray}
d^{\textrm{AII}}_{(n_1,n_2,\ldots,n_k)} &=& k + \dim [\mathsf{Sp}(N)] - \sum_{j=1}^k \dim [\mathsf{Sp}(n_j)] \nonumber \\
&=& k + N(2N+1) - \sum_{j=1}^k  n_k(2n_k+1).
\end{eqnarray}
From here, we define 
\begin{equation}
\label{eqn:AII-codim}
\delta^\textrm{AII}_{(n)} = d^\textrm{AII}_{(1,1,1,\ldots)} - d^\textrm{AII}_{(n,1,1,\ldots)} = (2n+1)(n-1)
\end{equation}
as the codimension of forming an $n$-fold degeneracy in an otherwise non-degenerate spectrum.

For $n=2$ (which, owing to the Kramers doubling, encodes a fourfold degeneracy), we obtain codimension $\delta^\textrm{AII}_{(2)}=5$.
This corresponds to a decomposition of an effective two-band Hamiltonian into the five Dirac matrices, which can be encoded in the quaternion language compactly as 
\begin{equation}
\label{eqn:Dirac-matrices-quaternionic}
\left(\begin{array}{cc}
1 & 0 \\
0 & -1
\end{array}\right),
\left(\begin{array}{cc}
0 & 1 \\
1 & 0
\end{array}\right),
\left(\begin{array}{cc}
0 & \imi\\
-\imi & 0
\end{array}\right),
\left(\begin{array}{cc}
0 & \imj\\
-\imj & 0
\end{array}\right)\;\textrm{and}\;
\left(\begin{array}{cc}
0 & \imk\\
-\imk & 0
\end{array}\right).
\end{equation} 
The twofold degeneracy is characterized by $\pi_4(S^4) = \intg$ that corresponds to the second Chern number~\cite{Hatsugai:2010}. 
To consider $n$-fold degeneracies for general $n \geq 2$, we should study the space of traceless Hermitian $n \times n$ matrices with quaternion numbers. 
Such traceless matrices are characterized by $n-1$ real terms on the diagonal and by $4\times\tfrac{1}{2}n(n-1)$ off-diagonal quaternion numbers. 
Taken together, we decompose an $n$-band quaternionic Hamiltonian into $(2n+1)(n-1)$ linearly independent matrices, matching the codimension computed~in~Eq.~(\ref{eqn:AII-codim}).

Let us now analyze in detail the case $n=3$ (which, owing to the Kramers doubling, encodes a sixfold degeneracy). 
To that end, we utilize the $14$ quaternionic Gell-Mann matrices, which comprise the two diagonal generators $\Lambda_3=\diag(1,-1,0)$ and $\Lambda_8=\tfrac{1}{\sqrt 3}\diag(1,1,-2)$ [identical to Eq.~(\ref{eqn:GellMann})], together with, for each of the three off-diagonal slots, the four Hermitian quaternion fillings $\mu\in\{1,\imi,\imj,\imk\}$ placed in the lower-left entry and with $\mu^\dagger$ in the upper-right entry [cf.~Eq.~(\ref{eqn:Hermiticity-with-quaternions})].
The eight matrices containing only entries ${\propto}\,1$ and ${\propto}\,\imi$ reproduce the complex Gell-Mann basis in Eq.~(\ref{eqn:GellMann}), while the six remaining matrices, built from $\mu\in\{\imj,\imk\}$, are
\begin{eqnarray}
\Lambda_{9} &=&
\left(\begin{array}{ccc} 0&-\imj&0\\ \imj&0&0\\ 0&0&0 \end{array}\right),\;\;
\Lambda_{10} =
\left(\begin{array}{ccc} 0&-\imk&0\\ \imk&0&0\\ 0&0&0 \end{array}\right),
\nonumber\\
\Lambda_{11} &=&
\left(\begin{array}{ccc} 0&0&-\imj\\ 0&0&0\\ \imj&0&0 \end{array}\right),\;\;
\Lambda_{12} =
\left(\begin{array}{ccc} 0&0&-\imk\\ 0&0&0\\ \imk&0&0 \end{array}\right),
\label{eqn:quat-GellMann}\\
\Lambda_{13} &=&
\left(\begin{array}{ccc} 0&0&0\\ 0&0&-\imj\\ 0&\imj&0 \end{array}\right),\;\;
\Lambda_{14} =
\left(\begin{array}{ccc} 0&0&0\\ 0&0&-\imk\\ 0&\imk&0 \end{array}\right).\nonumber
\end{eqnarray}
The trace property (\ref{eqn:gell-mann-squared}) is replaced with 
\begin{equation}
\tfrac{1}{2}\tr[\Lambda_i\Lambda_j+\Lambda_j \Lambda_i]=2\delta_{ij}. 
\end{equation}
However, this still ensures that the three-band Hamiltonian 
\begin{equation}
    H^\textrm{AII} = \sum_{i = 1}^{14} k_i \Lambda_i
    \label{eqn:AII-3band-Ham-GM}
\end{equation}
obeys
\begin{eqnarray}
\varepsilon_1^2 + \varepsilon_2^2 + \varepsilon_3^2 &=& \tr[(H^\textrm{AII})^2] \nonumber \\
&=& \nonumber \tfrac{1}{2}\sum_{i,j=1}^{14} k_i k_j \tr[\Lambda_i \Lambda_j + \Lambda_j \Lambda_i] \nonumber \\
&=& 2 \sum_{i=1}^{14} k_i^2 = 2
\end{eqnarray}
where, as usual, we set the radius of the enclosing $13$-sphere to $\kappa = 1$.
It follows that the nodal manifolds $M_1$ and $M_2$ can be identified within the $S^{13} \subset \reals^{14}$ by the conditions~(\ref{eqn:AI-M1-M2-spectrum}).

We proceed to characterize the nodal manifold $M_1$. Note that the Hamiltonian on $M_1$ is fully specified by the non-degenerate top eigenstate $\ket{\psi_3}$, which is now a vector in $\quats^3$. 
Being normalized and defined only up to a $\mathsf{Sp}(1) \cong \mathsf{SU}(2)\simeq S^3$ gauge, the eigenstates specifies a point in $S^{11}/\mathsf{Sp}(1) \simeq \quats P^2$, i.e., in the quaternionic projective plane.
This result is consistent with $M_1$ being the classifying space of three-band class-$\textrm{AII}$ Hamiltonians with one energy gap, $M_1 \simeq \mathsf{Sp}(3)/\mathsf{Sp}(2)\times \mathsf{Sp}(1) \simeq \quats P^2$.

To parameterize $M_1$ explicitly, we multiply the first two components of the real eigenstate in Eqs.~(\ref{eqn:AI-eigenstates}) with unit quaternions $\mathfrak{p},\mathfrak{q}\in\mathsf{Sp}(1)$, in direct analogy with the complex phases $e^{\imi\alpha},e^{\imi\beta}$ of Eq.~(\ref{eqn:A-eigenstate-rotation}).
We thus construct the eigenframe as
\begin{subequations}
\begin{equation}
\Psi=\mathcal{D}(\mathfrak{p},\mathfrak{q})\,\mathcal{R}_z(\phi)\,\mathcal{R}_x(\theta)\,\Psi_0,
\end{equation}
where $\mathcal{R}_z(\phi)$ and $\mathcal{R}_x(\theta)$ are the two orthogonal rotations present in Eq.~(\ref{eqn:AI-eigenstate-rotation}) while $\mathcal{D}(\mathfrak{p},\mathfrak{q})=\diag(\mathfrak{p},\mathfrak{q},1)$ is a diagonal matrix. 
The resulting eigenframe on $M_1$ takes the form
\begin{equation}
\label{eqn:AII-eigenframe}
\Psi = \left(
\begin{array}{ccc}
 \mathfrak{p} \cos \phi  & \,-\mathfrak{p}\cos \theta \sin \phi  & \mathfrak{p} \sin \theta 
   \sin \phi  \\
 \mathfrak{q} \sin \phi  & \mathfrak{q} \cos \theta  \cos \phi  & \,-\mathfrak{q} \sin \theta  \cos
   \phi  \\
 0 & \sin \theta  & \cos \theta  \\
\end{array}
\right).    
\end{equation}
\end{subequations}
In analogy with the discussion of Eq.~(\ref{eqn:A-eigenframe}), the possibility to set either $\mathfrak{p}$ or $\mathfrak{q}$ to $-1$ forces us to reduce the range of $\phi$ to $[0,\tfrac{\pi}{2}]$ (while keeping $\theta \in [0,\tfrac{\pi}{2}]$).
Utilizing the parameterization of unit quaternions with Hopf coordinates as
\begin{subequations}
\begin{eqnarray}
\mathfrak q(\chi,\alpha,\beta) &=&
\begin{pmatrix}
\cos\chi\, e^{\imi\alpha} & \sin\chi\, e^{\imi\beta} \\
-\sin\chi\, e^{-\imi\beta} & \cos\chi\, e^{-\imi\alpha}
\end{pmatrix}, \\
\mathfrak p(\rho,\gamma,\delta) &=&
\begin{pmatrix}
\cos\rho\, e^{\imi\gamma} & \sin\rho\, e^{\imi\delta} \\
-\sin\rho\, e^{-\imi\delta} & \cos\rho\, e^{-\imi\gamma}
\end{pmatrix},
\end{eqnarray}
\end{subequations}
where $\chi, \rho \in [0,\pi/2]$ and $\alpha,\beta, \gamma, \delta \in[0,2\pi]$, one can compare the eigensystem decomposition $H = \Psi \mathcal{E} \Psi^\dagger$ on $M_1$ against the Gell-Mann decomposition~(\ref{eqn:AII-3band-Ham-GM}) to find an explicit parameterization of $M_1 \subset S^{13} \subset \reals^{14}$ with eight angular coordinates.
This parameterization can be found in the Supplementary Code and Data~\cite{supplementary_code_data}.
An analogous parameterization can be established for $M_2$.

In analogy with the symmetry classes $\textrm{AI}$ and $\textrm{A}$ discussed in the previous sections, we next aim to investigate the linking of $M_1$ and $M_2$ in terms of topological band invariants on nontrivial cycles.
Recall that $\quats P^2$ can be decomposed into one cell in each dimension $D \in \{0,4,8\}$, each acting as a generator of a homology group with integer coefficients.
The nontrivial $4$-cycle $\quats P^1 \simeq S^4$ corresponds to setting $\phi = 0$ and $\mathfrak{p} =1$, while keeping $\theta$ and $\mathfrak{q}$ as tunable parameters, in which case the columns of~(\ref{eqn:AII-eigenframe}) give the eigenstates as
\begin{equation}
\label{eqn:AII-S4-eigenstates}
\ket{\psi_1}=\left(\begin{array}{c} 
1 \\ 0 \\ 0 
\end{array}\right),
\quad\!\!
\ket{\psi_2}=\left(\begin{array}{c} 
0 \\ \mathfrak{q}\cos\theta \\ \sin\theta 
\end{array}\right),
\quad\!\!
\ket{\psi_3}=\left(\begin{array}{c} 
0 \\ -\mathfrak{q}\sin\theta \\ \cos\theta 
\end{array}\right)
\end{equation}
with energies $\varepsilon_1 = \varepsilon_2 = -1/\sqrt{3}$ and $\varepsilon_3 = 2\sqrt{3}$. 
By mapping $\ket{\psi_2}$ and $\ket{\psi_3}$ onto eigenstates of the two-band problem, decomposed into Dirac matrices~(\ref{eqn:Dirac-matrices-quaternionic}), one can establish that the nontrivial $4$-cycle exhibits a nontrivial second Chern number~\cite{supplementary_code_data}. 
By virtue of the general analysis in Sec.~\ref{sec:general-section}, this band invariant ensures that $M_1$ and $M_2$ are linked. 
Owing to the robustness of topological band invariants, analogous characterization of triple points via linking applies to general three-band models extending the minimal Hamiltonian in Eq.~(\ref{eqn:AII-3band-Ham-GM}).

Finally, by taking the cohomology ring $H^*(\quats P^2,\intg) = \intg[a]/(a^3)$ with $a$ the generator of $H^4(\quats P^2,\intg)$, one can adapt the argument presented for the class $\textrm{A}$ to show that the rank-$2$ quaternionic bundle (which can also be seen as a rank-$4$ complex bundle) on the eight-dimensional nodal manifolds $M_1$ and $M_2$ also exhibits a nontrivial \emph{fourth} Chern number.
In the context of band theory, in analogy with the result for the class \textrm{AI}~\cite{Ahn:2018} and for the class $\textrm{A}$~\cite{Lian:2016}, we anticipate that nodal $4$-manifolds in $9D$ that carry a nontrivial fourth Chern number on the enclosing $S^8 \subset \reals^9$, are linked with nodal $4$ manifolds in adjacent energy gap.
However, the present context appears to offer a genuinely richer topological characterization, since the classifying space $\mathsf{Sp}(3)/\mathsf{Sp}(1)\times\mathsf{Sp}(2)$ also exhibits nontrivial topological invariants in dimensions $4 < D < 8$, including stable $\ztwo$ invariants in dimensions $D\in \{5,6\}$~\cite{Lundell:1992}.\footnote{Note that the mentioned stable $\ztwo$ invariants come from \emph{homotopy} considerations, while the (co)homology groups only give $\intg$ invariants in degrees $\mathrm{mod}\,4$.} 
We leave the investigation of the physical meaning of these further invariants, including their prospective relation to multifold band degeneracies and linking, open to future investigation.

\section{Symmetry class \texorpdfstring{$\textrm{BDI}$}{BDI}}
\label{sec:class-BDI}

From this section onward, we leave the realm of Wigner-Dyson symmetry classes and consider multifold band degeneracies of Hamiltonians with either particle-hole symmetry or chiral symmetry. 
Since these symmetries flip the sign of energy, they only influence the codimension and the topology of band nodes pinned to zero energy ($\varepsilon = 0$); in contrast, the characterization of band nodes at $\varepsilon \neq 0$ is unaffected by the particle-hole and chiral symmetry, reducing their treatment to one of the formerly addressed Wigner-Dyson class.
We begin with the case of the symmetry class $\textrm{BDI}$, which is characterized by spinless time-reversal symmetry ($\mathcal{T}^2 = +1$) and particle-hole symmetry ($\mathcal{C}^2 = +1$); the composition of the two gives the chiral symmetry $\mathcal{S}=\mathcal{TC}$.
We restrict our analysis to the usual `balanced' case, by which we mean the vanishing trace\footnote{In models with $\tr\mathcal{S}\neq 0$, the chiral imbalance (possibly interpretable as an unequal number of sites on two sublattices of a bipartite lattice) enforces the presence of flat zero-energy bands throughout the parameter space~\cite{Neupert:2012,Calugaru:2022}.} of the chiral operator: $\tr \mathcal{S} = 0$.

We first address how to adapt the von Neumann-Wigner counting to the (balanced) symmetry class $\textrm{BDI}$. 
The chiral symmetry $\mathcal{S} = \sigma_z\otimes \mathbb{1}_N$ allows us to write the Hermitian Hamiltonian with $2N$ energy bands in a block-off-diagonal form
\begin{equation}
\label{enq:Ham-BDI-off-blocks}
H = \left(\begin{array}{cc} 
\mathbb{0}  & h         \\
h^\top   & \mathbb{0}
\end{array}\right)    
\end{equation}
where $h \in \reals^{N \times N}$ owing to the spinless time-reversal $\mathcal{T}=\mathcal{K}$. 
The Hamiltonian $H$ has a $2n$-fold zero eigenvalue if the matrix $h$ has an $n$-fold zero eigenvalue.
Nonetheless, given the general doubling that arises in particle-hole-symmetric classes (cf.~footnotes to Table~\ref{tab:codims-WD+balanced}), we denote the codimension of forming such a $2n$-fold band degeneracy~as~$\delta^\textrm{BDI}_{(n)}$.

According to the rank-nullity theorem, 
\begin{equation}
\label{eqn:rank-nullity-theorem}
\rank h = N - \nullop h \stackrel{!}{=} N - n.     
\end{equation}
The dimension of the space of general real $N \times N$ matrices is $N^2$. 
Furthermore, the rank factorization reveals that the space of rank-$r$ real $N \times N$ matrices has dimension $2Nr - r^2$.
Specifically, a real $N\times N$ matrix $C$ with rank $r$ can be decomposed as $C=AB$, where $A$ is an $N\times r$ matrix and $B$ is an $r\times N$ matrix. 
The matrices $A$ and $B$ in total have $2Nr$ degrees of freedom.
However, the decomposition is not unique because one can multiply $A$ and $B$ by an invertible matrix as
\begin{equation}
\label{eqn:rank-factorization}
    C=AB=AUU^{-1}B \equiv \tilde{A}\tilde{B}.
\end{equation}
Since the space $\mathsf{GL}(r,\reals)$ of real invertible $r \times r$ matrices has dimension $r^2$, we conclude that the space of rank-$r$ real $N\times N$ matrices [and, by extension, of class-$\textrm{BDI}$ Hamiltonians with $2(N\,{-}\,r)$-fold band degeneracy at $\varepsilon=0$] has dimension 
\begin{subequations}
\begin{equation}
d_{(N,N-r)}^\textrm{BDI} = 2Nr-r^2, 
\end{equation}
where the first subscript indicates the dimension of the off-diagonal block $h$, and the second subscript counts the dimension of the null-space of $h$.
Substituting $r = N-n$, where $n$ is the number of zero eigenvalues of $h$, we find the alternative expression
\begin{equation}
d_{(N,n)}^\textrm{BDI} = N^2 - n^2.    
\end{equation}
\end{subequations}
From here, the codimension of forming a $2n$-fold degeneracy in nodal class BDI is
\begin{equation}
\delta_{(n)}^\textrm{BDI} = d^\textrm{BDI}_{(N,0)} - d^\textrm{BDI}_{(N,n)} = n^2.
\end{equation}
In particular, $\delta_{(1)}^\textrm{BDI} = 1$, meaning that an effective two-band model is expanded into a single Pauli matrix~\cite{Bzdusek:2017}.
This result implies the existence of topologically stable nodal surfaces in three-dimensional crystals in symmetry class $\textrm{BDI}$~\cite{Zhong:2016} characterized by a Pfaffian invariant $\pi_0(S^0) = \ztwo$; here, the `zero-dimensional sphere' corresponds to the space of normalized two-band Hamiltonians expanded into a single Pauli matrix.

In the following, we focus on the topological characterization of the multifold degeneracy of the next available order, i.e., for $n=2$, in which case the minimal Hamiltonian contains four energy bands.
In analogy with Eqs.~(\ref{eqn:AI-3band-Ham-GM}), (\ref{eqn:A-3band-Ham-GM}), and~(\ref{eqn:AII-3band-Ham-GM}), the minimal four-band Hamiltonian near such a fourfold band degeneracy can be expressed as 
\begin{equation}
\label{eqn:BDI-4band-Ham-Dirac}
    H^\textrm{BDI} = \boldsymbol{k} \cdot \boldsymbol{V} + \boldsymbol{q} \cdot \boldsymbol{W}, 
\end{equation} 
where $\boldsymbol{k} = (k_1,k_2)$ and $\boldsymbol{q} = (q_1,q_2)$ are coordinates in a four-dimensional parameter space, and 
\begin{subequations}
\label{eqn:BDI-basis-matrices}
\begin{align}
    \boldsymbol{V} &= (\sigma_x \otimes \mathbb{1}, -\sigma_y \otimes \tau_y) \\ 
    \boldsymbol{W} &= (\sigma_x \otimes \tau_z, -\sigma_x \otimes \tau_x)
\end{align}
are matrices satisfying 
\begin{eqnarray}
\{V_i,V_j \} = \{ W_i, W_j \} &=& 2 \delta_{ij} \\
\textrm{and}\quad [V_i,W_j] &=&0 .     
\end{eqnarray}
\end{subequations}
The spectrum of Hamiltonian~(\ref{eqn:BDI-4band-Ham-Dirac}) can be expressed as~\cite{Bzdusek:2017} 
\begin{equation}
\label{eq:BDI-spectrum}
    \varepsilon = \pm ||\boldsymbol{k}|| \pm ||\boldsymbol{q}||
\end{equation}
We order the eigenvalues of the Hamiltonian such that $\varepsilon_1 \leq \dots\leq \varepsilon_4$ and define the degeneracy loci $\mathcal{L}_\medcirc$ and $\mathcal{L}_\varnothing$ by adapting Eq.~(\ref{eqn:class-AI-loci}) to 
\begin{subequations}
\label{eqn:chiral-loci}
\begin{align}
    \mathcal{L}_\medcirc &= \{ (\boldsymbol{k},\boldsymbol{q}) \,|\, \varepsilon_2 = \varepsilon_3 = 0\} \\ 
    \mathcal{L}_\varnothing &= \{ (\boldsymbol{k},\boldsymbol{q}) \,|\, \varepsilon_1 = \varepsilon_2
    \, ~(\mathrm{and} ~\,\varepsilon_3 = \varepsilon_4
    )\} \label{eqn:class-BDI-loci-non-zero}
\end{align}
\end{subequations}
In Eq.~(\ref{eqn:class-BDI-loci-non-zero}), the property in parentheses follows from the first condition in combination with the chiral symmetry.

\begin{figure}
    \centering
    \includegraphics[width=\linewidth]{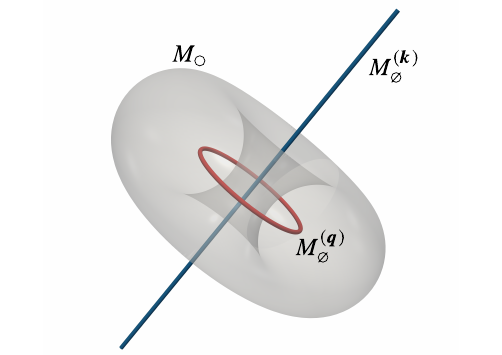}
    \caption{Stereographic projection of the nodal manifolds $M_\medcirc$ and $M_\varnothing$ in class $\textrm{BDI}$. The red ($M_\varnothing^{(\boldsymbol{q})}$) and blue ($M_\varnothing^{(\boldsymbol{k})}$) lines correspond to the non-intersecting subsets of $M_\varnothing$ that lie in the $\boldsymbol{q}=0$ and $\boldsymbol{k}=0$ planes, respectively. The gray torus represents the projection of the Clifford torus corresponding to $M_\medcirc$.}
    \label{fig:class-BDI-linking}
\end{figure}

We enclose the fourfold degeneracy located at $\boldsymbol{k} = \bs{0} = \boldsymbol{q}$  with a three-dimensional sphere $S^3\subset \reals^4$ of radius $\kappa = 1$ and investigate the nodal manifolds $M_\medcirc = \mathcal{L}_\medcirc \cap S^3$ and $M_\varnothing = \mathcal{L}_\varnothing \cap S^3$.
In the following, we discuss, in turn, both of these nodal manifolds, aiming to relate the presence of the fourfold degeneracy to their linking and band invariants. 
The winding of these manifolds can be visualized via stereographic projection of $S^3$ onto $\reals^3 \cup \{\infty\}$ as shown in Fig.~\ref{fig:class-BDI-linking}.

First, $M_\medcirc$ is the set of points $(\boldsymbol{k},\boldsymbol{q})$ satisfying the following set of equations
\begin{subequations}
\begin{align}
    & k_1 ^2 + k_2^2 + q_1^2 + q_2^2 = 1 \\ 
    & k_1^2+k_2^2 - q_1^2-q_2^2=0.
\end{align}
\end{subequations}
These equations define the product of two circles, $M_\medcirc \simeq S^1 \times S^1$ (the Clifford torus); both circles have radius $1/\sqrt{2}$, and they lie in the $(k_1,k_2)$ and the $(q_1,q_2)$ subspace, respectively. 
The manifold can be explicitly parameterized as 
\begin{equation}
(\bs{k},\bs{q}) = \tfrac{1}{\sqrt{2}} (\cos\theta,\sin\theta, \cos\phi,\sin\phi)
\end{equation}
with $\theta,\phi\in[0,2\pi]$. We argue in Appendix~\ref{app:class-space-w-zero-states} that, more generally, the classifying space of class-$\textrm{BDI}$ Hamiltonians with $2N$ bands and $2n$ zero-energy states is\footnote{The distinction between the factors inside vs.~outside of the square brackets is clarified in Appendix~\ref{app:class-space-w-zero-states}.}
\begin{equation}
\mathfrak{X}^\textrm{BDI}_{(n,N-n)} = \frac{\mathsf{O}(N)\times\mathsf{O}(N)}{[\mathsf{O}(n)\times \mathsf{O}(n)]\times \mathsf{O}(N-n)},    
\end{equation}
of which the special case $\mathfrak{X}^\textrm{BDI}_{(1,1)}$ reproduces the nodal manifold~$M_\medcirc$ as described above.

Being a torus, the nodal manifold has nontrivial cycles in dimensions $D\in\{0,1,2\}$.
To probe the linking of the nodal manifold $M_\medcirc$ with $M_\varnothing$, we first inspect topological invariants on the nontrivial $1$-cycles.
To that end, we compute the first Stiefel-Whitney class (i.e., the quantized Berry phase) of one of the non-degenerate bands. 
We find the eigensystem
\begin{subequations}
\label{eqn:BDI-zero-eigensystem}
\begin{eqnarray}
\label{eqn:BDI-1cycle-Berry}
\ket{u_-} &=& \tfrac{1}{\sqrt{2}}\left( \cos \alpha, -\sin \alpha, - \cos \beta, -\sin \beta \right)^\top \\
\ket{u_0^+} &=& \left( \sin \alpha, \cos \alpha, 0, 0 \right)^\top  \\
\ket{u_0^-} &=& \left( 0,0, - \sin \beta, \cos \beta \right)^\top \\
\ket{u_+} &=& \tfrac{1}{\sqrt{2}}\left( \cos \alpha, -\sin \alpha, \cos \beta, \sin \beta \right)^\top,
\end{eqnarray}
\end{subequations}
where the subscript $s\in\{-,0,+\}$ indicates the sign of energy, $\varepsilon_s = s/\sqrt{2}$, the superscript of the zero-energy states indicates their chiral eigenvalue, and we introduced angles $\alpha=(\theta+\phi)/2$ and $\beta=(\theta-\phi)/2$.
The nontrivial $1$-cycles of $M_1$ are obtained by fixing either $\theta$ or $\phi$ and allowing the other parameter to increase in range $[0,2\pi]$. 
Specifically, we consider the closed loops $s_\theta$ with fixed $\theta = 0$ and $s_\phi$ with fixed $\phi = 0$. 
Along $s_\theta$, as we vary $\phi$, each eigenstate acquires a minus sign, which follows from observing that the eigenstate is a function of halved angles. 
As a consequence, $\varphi_{\textrm{B},\theta} = \pi$ for each state. 
Identical reasoning can be applied to derive that $\varphi_{\textrm{B},\phi} = \pi$ for each eigenstate.
Let us also point out that if one treats the zero-energy subspace on $M_\medcirc$ as a single rank-$2$ bundle, then it is characterized by trivial Berry phases.
Nevertheless, the presence of chiral symmetry allows us to define a \emph{chirally resolved Berry phase}, i.e., the Berry phase carried by zero-energy states of positive chirality, as a robust topological band invariant that detects the linking of $M_\circ$ with $M_\varnothing$.

In contrast to the class $\textrm{AI}$, we find no topological band invariant on the nontrivial $2$-cycle of the nodal manifold $M_\medcirc$ in the class $\textrm{BDI}$. 
On one hand, the bundles spanned by the positive resp.~by the negative energy state have rank-$1$, and therefore cannot define a two-dimensional invariant. 
On the other hand, because the rank-$2$ bundle spanned by the zero-energy eigenstates in Eq.~(\ref{eqn:BDI-zero-eigensystem}) has vanishing Berry phases, it can be assigned a well-defined integer-valued Euler class~\cite{Bouhon:2020}.
However, since the two eigenstates point along different directions in the Hilbert space, their Euler curvature
\begin{equation}
\Eu(\theta,\phi) = \braket{\partial_\theta u_0^+}{\partial_\phi u_0^-} - \braket{\partial_\theta u_0^+}{\partial_\phi u_0^-} = 0
\end{equation}
vanishes. 
This integrates to a vanishing Euler class, and its modulo-$2$ reduction implies trivial 2SW class.

We next turn to the analysis of the nodal manifold $M_\varnothing$. 
It is specified as the union of two non-intersecting circles given by $||\boldsymbol{k}|| = 0$ and $|| \boldsymbol{q}|| =0$, respectively. 
That is,
\begin{subequations}
\begin{gather}
k_1 ^2 + k_2^2 + q_1^2 + q_2^2 = 1 \;\;\; \textrm{and} \\
k_1^2 + k_2^2 =0 \;\;\; \mathrm{or} \;\;\; q_1^2 + q_2^2 =0.
\end{gather}
\end{subequations}
Taking $k_1^2 + k_2^2=0$ results in $q_1^2+q_2^2=1$, 
defining one of the circles. 
The other circle is defined by $q_1^2 + q_2^2 = 0$. 
The two circles correspond to momenta $(\bs{k},\bs{q})$
$(\cos \phi,\sin \phi, 0, 0)$ and $(0, 0,\cos \phi,\sin \phi)$ with $\phi \in [0, 2\pi]$.
The two circles form a Hopf link inside the enclosing $S^3$ as illustrated by Fig.~\ref{fig:class-BDI-linking}. Note that the manifold $ M_\varnothing = S^1 \sqcup S^1$ can also be interpreted as
the standard classifying space of class-$\textrm{BDI}$ $4$-band Hamiltonians, $M_\varnothing\simeq \mathsf{O}(2)$.

We inspect topological invariants on $M_\varnothing$. Here, we find that $0$-dimensional invariants computed on the two circles already indicate that $M_\varnothing$ and $M_\medcirc$ are linked. 
Since the Hamiltonian admits a block-off-diagonal form with real entries, one can compare the sign of the determinant of the off-diagonal on the two components of $M_\varnothing$. 
The sign of the determinant of this matrix cannot change without closing the energy gap at $\varepsilon = 0$, which corresponds to passing across the nodal manifold $M_\medcirc$. 
For the circle $(\cos \phi, \sin \phi, 0, 0)$ we find $\mathrm{sign} \det h = +1$, while for the circle $(0,0,\cos \phi, \sin \phi)$) we have $\mathrm{sign} \det h = -1$. 
We therefore conclude that one of the circles is inside $M_\medcirc$, while the other is on the outside, which we interpret as linking of the two manifolds.

Let us point out that the sign of the determinant in the present calculation can be identified with the Pfaffian invariant ($\sign [\Pf]$) of the Hamiltonian in an appropriately adapted choice of basis. 
Specifically, by rotating the Hilbert space to a basis where $\mathcal{C} = \mathcal{K}$ (as adapted in the symmetry class $\textrm{D}$ in a later section) reveals that the Hamiltonian can, in fact, be rotated into an even-dimensional antisymmetric matrix, for which the Pfaffian can be defined as the signed square root of the determinant. 
It can be shown~\cite{Bzdusek:2017} that the determinant of the off-diagonal block $h$ in Eq.~(\ref{enq:Ham-BDI-off-blocks}) is identical to the Pfaffian of the antisymmetric Hamiltonian matrix. 
For this reason, we refer to the $0$-dimensional invariant in the class $\textrm{BDI}$ as the Pfaffian sign invariant.

In the case of $M_\varnothing$, we can additionally formulate a band invariant on the nontrivial $1$-cycles, i.e., along each of the two circles. 
Taking for concreteness the circle $(\cos \phi, \sin \phi, 0, 0)$, we find that the off-diagonal block takes the form
\begin{equation}
\label{eqn:BDI-1D-winding}
h = \left(\begin{array}{cc} 
\cos \phi & \sin\phi \\
-\sin \phi & \cos \phi
\end{array}\right) \in \mathsf{O}(2).   
\end{equation}
As $\phi$ grows in range $[0,2\pi]$, the Hamiltonian clearly traces a nontrivial loop in $\mathsf{O}(2)$.
In combination with our earlier results, we conclude that the fourfold degeneracy inside the enclosing $S^3$ implies that the $1$-cycles of $M_\medcirc$ are linked with $1$-cycles if $M_\varnothing$, and that the $2$-cycle of $M_\medcirc$ is linked with the $0$-cycle of $M_\varnothing$.
Owing to the robustness of the topological invariants, this linking property persists in general four-band models extending the minimal Hamiltonian in Eq.~(\ref{eqn:BDI-4band-Ham-Dirac}).

Let us finally relate the calculation in the previous paragraph to the result of Ref.~\citenum{Kim:2021} that class-$\textrm{BDI}$ nodal rings in 2D that carry a nontrivial winding number $\pi_1[\mathsf{O}(n)]=\ztwo$ on the enclosing circle necessarily contain a Dirac point in adjacent energy gaps. 
In the context of our analysis of the multifold degeneracy, consider inscribing a disk to either $1$-cycle of $M_\varnothing$.
As visible in Fig.~\ref{fig:class-BDI-linking}, any such disk intersects $M_\medcirc$ along a ring (corresponding to a zero-energy nodal ring) while also intersecting $M_\varnothing$ at a point (corresponding to the enclosed Dirac points in adjacent energy gaps).
The winding number in Eq.~(\ref{eqn:BDI-1D-winding}) can be computed as the base of the integration shrinks from the $1$-cycle in $M_\varnothing$ to any other loop that encloses the nodal ring. 
Therefore, our topological characterization of fourfold degeneracy in class $\textrm{BDI}$ exactly reproduces the finding of Ref.~\citenum{Kim:2021}.

\section{Symmetry class \texorpdfstring{$\textrm{AIII}$}{AIII}}
\label{sec:class-AIII}

Similarly to the class $\textrm{BDI}$, the condition $\tr \mathcal{S} = 0$ in the symmetry class $\textrm{AIII}$ allows us to write the Hermitian Hamiltonian with $2N$ energy bands in a block-off-diagonal form
\begin{equation}
H = \left(\begin{array}{cc} 
\mathbb{0}  & h         \\
h^\dagger   & \mathbb{0}
\end{array}\right)    
\label{eqn:block-off-diag-Ham}
\end{equation}
where $h \in \cmplx^{N \times N}$ due to the absence of time-reversal symmetry. 
The Hamiltonian $H$ has a $2n$-fold zero eigenvalue if the matrix $h$ has an $n$-fold zero eigenvalue.
The rank-nullity theorem in Eq.~(\ref{eqn:rank-nullity-theorem}) as well as the rank factorization and its redundancy in Eq.~(\ref{eqn:rank-factorization}) still apply; the sole difference is that we presently have to replace real matrix elements by complex ones, which results in doubling all the dimensions in the considerations carried formerly for the class $\textrm{BDI}$. 
Therefore, we find 
\begin{subequations}
\begin{equation}
d_{(N,n)}^\textrm{AIII} = 2(N^2 - n^2)    
\end{equation}
for the space of class-$\textrm{AIII}$ Hamiltonians with $2n$-fold band degeneracy at $\varepsilon = 0$, and
\begin{equation}
\delta_{(n)}^\textrm{AIII} = 2n^2.    
\end{equation}
\end{subequations}
for the codimension of forming a $2n$-fold degeneracy at $\varepsilon=0$.
In particular, $\delta^\textrm{AIII}_{(1)} = 2$, meaning that an effective two-band model is expanded into two Pauli matrices, which, in two dimensions, results in stable point nodes protected by the winding number $\pi_1(S^1) = \intg$.

We further focus on the topological characterization of the multifold degeneracy with $n=2$. Therefore, we consider four-band Hamiltonians in the symmetry class AIII. 
We represent the chiral symmetry $\mathcal{S} = \sigma_z \otimes \mathbb{1}_\tau$; then, the chiral-symmetric Hamiltonians can be conveniently expressed as 
\begin{equation}
\label{eqn:AIII-4band-Ham-Dirac}
H^\textrm{AIII} = \bs{a} \cdot \bs{V} + \bs{b} \cdot \bs{W}  
\end{equation}
where $\bs{a} = (a_1,a_2,a_3,a_4)$ and $\bs{b}=(b_1,b_2,b_3,b_4)$ are vectors parameterizing the $\reals^8$ momentum space, and
\begin{subequations}
\begin{eqnarray}
\bs{V} &=& (\sigma_x\otimes \mathbb{1}_\tau, -\sigma_y\otimes \tau_x,-\sigma_y\otimes \tau_y,-\sigma_y\otimes \tau_z)  \\  
\bs{W} &=& (\sigma_y\otimes \mathbb{1}_\tau, \sigma_x\otimes \tau_x,\sigma_x\otimes \tau_y,\sigma_x\otimes \tau_z)
\end{eqnarray}
are collections of Dirac matrices obeying\footnote{Interestingly, for classes with chiral symmetry $\mathcal{S}$, the existence of the collection $\bs{W}$ of anticommuting matrices follows from the existence of the complementary collection $\bs{V}$; namely, $\bs{W}$ can be defined as $\bs{W}=\mathcal{S}\bs{V}$. 
For example, let us denote $\alpha,\beta$ to be chirally symmetric matrices in $\bs{V}$, such that $\alpha\beta=-\beta\alpha$. Then we have $(\mathcal{S}\alpha)(\mathcal{S}\beta)=-\mathcal{S}\alpha\beta\mathcal{S}=\mathcal{S}\beta\alpha\mathcal{S}=-(\mathcal{S}\beta)(\mathcal{S}\alpha)$.}
\begin{eqnarray}
\{V_a,V_b\} = \{W_a,W_b\} &=& 2\delta_{ab} \\
\textrm{and}\quad [V_a,W_b] &=& 2\imi \delta_{ab}\mathcal{S} 
\end{eqnarray} 
\end{subequations}
Furthermore, we find for $a\neq b$
\begin{subequations}
\begin{equation}
\{V_a,W_b\}=2G_{ab}=-2G_{ba}
\end{equation}
with 
\begin{eqnarray}
G_{12}=\unit_\sigma\otimes \tau_x &\qquad& G_{23} = -\sigma_z \otimes \tau_z \\ 
G_{13}=\unit_\sigma \otimes \tau_y &\qquad& G_{24} = \sigma_z \otimes \tau_y \\
G_{14}=\unit_\sigma\otimes \tau_z &\qquad & G_{34} = -\sigma_z \otimes \tau_x,
\end{eqnarray}
\end{subequations}
corresponding to two sets of three anticommuting Dirac matrices. 
By squaring the Hamiltonian twice, the algebra of the matrices $V_a$, $W_b$, and $G_{ab}$ allows us to derive the spectrum~as
\begin{subequations}
\begin{equation}
\varepsilon = \pm\sqrt{\bs{a}^2 + \bs{b}^2 \pm 2\lVert \bs{a} \wedge \bs{b}\rVert}    
\end{equation}
where the $2$-vector $\bs{a} \wedge \bs{b}$ is constructed using the wedge product. 
It has six independent components 
\begin{equation}
(\bs{a} \wedge \bs{b})_{ab} = a_a b_b - b_a a_b = -(\bs{a} \wedge \bs{b})_{ba},
\end{equation}
\end{subequations}
and with the norm of the $2$-vector we mean the norm of the vector of its independent components.

We consider the Hamiltonian on the enclosing sphere 
\begin{equation}
\label{eqn:AIII-S}
S^7 = \{ (\bs{a},\bs{b})\,|\,\bs{a}^2 + \bs{b}^2 = 1\} \subset \reals^8.
\end{equation}
On this sphere, $\tr[H^2] = \sum_{j=1}^4 \varepsilon_j^2 = 4$ is constant.
We define the submanifolds $M_\medcirc \subset S^7$ and $M_\varnothing \subset S^7$ that correspond to the locus of twofold degeneracy at $\varepsilon=0$ and at $\varepsilon\neq 0$, respectively, in analogy with the discussion in Sec.~\ref{sec:class-BDI}.

The case $M_\medcirc$ corresponds to requiring $\bs{a}^2 + \bs{b}^2 = 2\lVert \bs{a} \wedge \bs{b}\rVert$, which translates to the simultaneous conditions $\lVert \bs{a} \rVert = \lVert \bs{b} \rVert$ and $\bs{a}\perp \bs{b}$.
The first condition, in combination with Eq.~(\ref{eqn:AIII-S}), fixes the magnitude of both four-component vectors to 
$\lVert \bs{a}\rVert = 1/\sqrt{2} = \lVert \bs{b} \rVert$.
We therefore set $\bs{a} = \bs{n}/\sqrt{2}$ with $\bs{n}\in S^3$ a unit vector; then, $\bs{b}$ is a vector in the tangent space $\mathrm{T}_{\bs{n}}S^3$ with a fixed norm. 
Therefore, we recognize $M_\medcirc$ as the unit tangent bundle $\mathrm{UT}S^3$. 
It is well-known that $S^3$ is parallelizable; therefore, 
\begin{equation}
\label{eqn:AIII-M-zero}
M_\medcirc \simeq S^3 \times S^2.
\end{equation}
There are several equivalent identifications of this manifold. First, it can be expressed as the Stiefel manifold $\mathsf{SO}(4)/\mathsf{SO}(2)$. 
Second, we can use the general form of classifying spaces of the class AIII with zero energy values, which gives $M_\medcirc \simeq \mathsf{U}(2)\times \mathsf{U}(2)/[\mathsf{U}(1)\times \mathsf{U}(1)]\times \mathsf{U}(1)$ (see Appendix~\ref{app:class-space-w-zero-states}).

The nodal manifold $M_\medcirc$ is constructed from one cell in dimensions $D\in\{0,2,3,5\}$, and each acts as a generator of a homology group with $\intg$ coefficients.
In the following, we narrow our attention to the nontrivial $2$-cycle of $M_\medcirc$, finding that it supports a non-zero first Chern number.
To reveal this result, we fix $\boldsymbol{a} = (1, 0, 0, 0)/\sqrt{2}$ which determines a point in the $S^3$ component. 
To probe the $S^2$ component of the nodal manifold, we let $\boldsymbol{b} = (0, \sin\theta \cos \phi, \sin \theta \sin \phi, \cos \theta)/\sqrt{2}$ with $\theta \in [0, \pi]$ and $\phi \in [0, 2 \pi]$ the usual spherical coordinates. 
The zero-energy states are then expressed as 
\begin{subequations}
\label{eqn:AII-first-Chern}
\begin{eqnarray}
\ket{u_0^+} &=& \left( -e^{-\imi\phi} \sin\tfrac{\theta}{2}, \cos \tfrac{\theta}{2}, 0, 0 \right)^\top,  \\
\ket{u_0^-} &=& \left( 0,0, -e^{-\imi\phi} \sin\tfrac{\theta}{2}, \cos \tfrac{\theta}{2} \right)^\top,  
\end{eqnarray}    
\end{subequations}
where the superscript indicate the eigenvalue of $\mathcal{S}$.
Observe that the non-zero components of each eigenstate match exactly the negative-energy eigenstate of the spinor Hamiltonian $\bs{n}\cdot\bs{\sigma}$ [cf.~Eq.~(\ref{eqn:spinor-1/2-Ham})].
Therefore, the two states carry the same Chern number $\abs{C} = 1$ on the $2$-cycle.
In analogy with the discussion below Eq.~(\ref{eqn:BDI-zero-eigensystem}) for the class $\textrm{BDI}$, one may call these the \emph{chirally resolved Chern number} of the zero-energy subspace.
However, in contrast to the class $\textrm{BDI}$, the total zero-energy subspace here carries a non-vanishing total Chern number $\abs{C}=2$. 
The nontrivial band invariant implies linking of $M_\medcirc$ with $M_\varnothing$.

The nodal manifold $M_\varnothing$ is determined by requiring $ \lVert \bs{a} \wedge \bs{b}\rVert =0$, which translates to the parallelity $\bs{a}\parallel \bs{b}$. 
We therefore write $\bs{a} = a \bs{n}$ and $\bs{b} = b \bs{n}$ with $\bs{n} \in S^3$ a unit vector. 
It follows from Eq.~(\ref{eqn:AIII-S}) that we can set $a= \cos\chi$ and $b = \sin\chi$ with $\chi \in [0,2\pi]\simeq S^1$. 
Using further the $3$-spherical coordinates $\phi\in[0,2\pi]$, $\theta\in[0,\pi]$ and $\psi\in[0,\pi]$ to characterize the unit vector, we can parameterize $M_\varnothing$ explicitly as  
\begin{eqnarray}
\label{eqn:class-AIII-non-zero-parametrization}
a_1 = \cos\chi\sin\psi\sin\theta\cos\phi &\quad& b_1 = \sin\chi\sin\psi\sin\theta\cos\phi \nonumber \\
a_2 = \cos\chi\sin\psi\sin\theta\sin\phi&\quad & b_2 = \sin\chi\sin\psi\sin\theta\sin\phi \nonumber \\
a_3 = \cos\chi\sin\psi\cos\theta &\quad & b_3 = \sin\chi\sin\psi\cos\theta \nonumber \\
a_4 = \cos\chi\cos\psi &\quad& b_4 = \sin\chi\cos\psi 
\end{eqnarray}
However, there is a redundancy in this description; namely, changing $(\chi,\bs{n})\mapsto (\chi+\pi,-\bs{n})$ reproduces the same point $(\bs{a},\bs{b})$. 
Therefore, we conclude that
\begin{equation}
\label{eqn:AIII-M-non-zero}
M_\varnothing \simeq (S^1 \times S^3)/\ztwo    
\end{equation}
where $\ztwo$ acts as a simultaneous antipodal map on $S^1$ and $S^3$. 
The nodal manifold $M_\varnothing$ can equivalently be identified as the space of class-$\textrm{AIII}$ $4$-band Hamiltonians with a single gap at half-filling, which is the Lie group~$\mathsf{U}(2)$.

To investigate the linking of $M_\varnothing$ with $M_\medcirc$, we first narrow our attention to the nontrivial $1$-cycle of $M_\varnothing$.
To this end, we consider the closed path specified by $\chi\in[0,2\pi]$ while keeping the other angles fixed to $\psi = \theta = \phi = 0$. 
Note that, owing to the antipodal identification in Eq.~(\ref{eqn:AIII-M-non-zero}), this path covers the nontrivial $1$-cycle \emph{twice}.
With this choice of coordinates, the Hamiltonian simplifies to $H= a_4 v_4 + b_4 w_4$ with $a_4 = \cos\chi$ and $b_4 = \sin\chi$.
Explicitly,
\begin{subequations}
\label{eqn:AIII-1D-winding}
\begin{equation}
    H = a_4 (-\sigma_y \otimes \tau_z) + b_4(\sigma_x \otimes \tau_z).
\end{equation}
This corresponds to the block-off-diagonal form~(\ref{eqn:block-off-diag-Ham}) 
with 
\begin{equation}
    h(\chi) = \begin{pmatrix}
        \imi e^{-\imi\chi} & 0 \\ 
        0 & -\imi e^{-\imi \chi}
    \end{pmatrix}.
\end{equation}
The winding number on the closed path is computed as
\begin{equation}
    2\times\nu = \frac{1}{2\pi \imi} \int_0^{2\pi} d\chi ~ \partial_\chi \log \det q(\chi)  = -2. 
\end{equation}
\end{subequations}
Accounting for the fact that the path covers the nontrivial $1$-cycle twice, we find that the $1$-cycle of $M_\varnothing$ is characterized~by~$\abs{\nu}=1$.

We furthermore establish that the 3D winding number over the nontrivial 3-cycle of $M_\varnothing$ is also nontrivial. 
The 3-cycle is specified by fixing $\chi=0$ which sets $\boldsymbol{b} = 0$ and $\boldsymbol{a} \in S^3$. 
On this submanifold, the off-diagonal block of the Hamiltonian takes the form 
\begin{equation*}
    h(\boldsymbol{a}) = a_1 \mathbb{1} - \imi \sum_{j=1}^3 a_{j+1} \sigma_{j}, \quad ||\boldsymbol{a}|| = 1.
\end{equation*}
It is straightforward to verify that $h^\dagger h = \mathbb{1}$ and $\text{det} h = 1$ implying that $h \in \mathsf{SU}(2)$ for all $\boldsymbol{a} \in S^3$. 
Furthermore, the map $h: S^3 \rightarrow \mathsf{SU}(2)$ corresponds to the unit quaternion representation of $\mathsf{SU}(2)$ and hence defines a diffeomorphism $S^3\simeq\mathsf{SU}(2)$ and as such has winding number $|\nu_3| = 1$. 
We confirm this in~\cite{supplementary_code_data} via evaluating the winding number 
\begin{equation}
    \nu_3[h] = \frac{1}{24 \pi^2} \int_{S^3}   ~\text{Tr} \left[(h^{-1} dh)^{\wedge3} \right]
\end{equation}
explicitly.

To summarize our discussion of the class $\textrm{AIII}$, we find that the presence of a fourfold degeneracy inside the enclosing $S^7$ is manifested in the linking of a $2$-cycle in $M_\medcirc$ with a $4$-cycle in $M_\varnothing$ (revealed by the first Chern number on the $2$-cycle) as well as in the linking of a $5$-cycle in $M_\medcirc$ with a $1$-cycle in $M_\varnothing$ (revealed by the winding number on the $1$-cycle), and by the linking of the $3$-cycle in $M_\medcirc$ with the $3$-cycle in $M_\varnothing$ (revealed by the 3D winding number on the $3$-cycle in $M_\varnothing$).
By virtue of the general discussion in Sec.~\ref{sec:general-section}, these linking properties persist in general four-band models extending the minimal Hamiltonian in Eq.~(\ref{eqn:AIII-4band-Ham-Dirac}).

\section{Symmetry class \texorpdfstring{$\textrm{CII}$}{CII}}\label{sec:class-CII}.

The symmetry class $\textrm{CII}$ is characterized by the presence of spinful time-reversal symmetry ($\mathcal{T}^2 = -1$) and particle-hole symmetry ($\mathcal{C}^2 = -1$); the composition of the two gives the chiral symmetry ($\mathcal{S}=\mathcal{T}\mathcal{C}$).
We represent
\begin{subequations}
\begin{eqnarray}
\mathcal{T} &=& \mathbb{1}_\sigma\otimes  \mathbb{1}_N\otimes (-\imi \tau_y) \mathcal{K} \\    
\mathcal{C} &=& \sigma_z \otimes \mathbb{1}_N\otimes (+\imi \tau_y) \mathcal{K} \\
\mathcal{S} &=& \sigma_z \otimes \mathbb{1}_N\otimes \mathbb{1}_\tau ,
\end{eqnarray}
\end{subequations}
where the second matrix indicates any additional degrees of freedom and specifies the Hamiltonian size as $4N \times 4N$. 
The spinful time-reversal symmetry allows for rewriting the Hamiltonian as a $2N \times 2N$ matrix over quaternions while dropping the Pauli matrices $\tau$, as presented for the class $\textrm{AII}$ in Sec.~\ref{sec:class-AII}. 
Then, the presence of chiral symmetry allows us to write the Hamiltonian in the block off-diagonal form
\begin{equation}
\label{eq:CII_H_blockdiag}
H = \left(\begin{array}{cc} 
\mathbb{0}  & \mathfrak{h}         \\
\mathfrak{h}^\dagger   & \mathbb{0}
\end{array}\right),    
\end{equation}
where $\mathfrak{h}$ is an $N \times N$ matrix over the quaternions ($\mathbb{H}$).

The rank factorization, used to determine the codimension of zero-energy band nodes in the real (Sec.~\ref{sec:class-BDI}) and complex (Sec.~\ref{sec:class-AIII}) chiral symmetry classes, also applies to quaternionic matrices.
The only point that requires care is that, in the present case, `rank' is understood as the quaternionic dimension of the image of the corresponding right-linear operator.
This amounts to the number of right-linearly independent columns (and agrees with the number of left-linearly independent rows~\cite{Zhang:1997}) of the matrix with quaternion components.
Thus, the Hamiltonian $H$ has a $4n$-fold zero-energy degeneracy iff the quaternionic matrix $\mathfrak h$ has an $n$-dimensional kernel over $\mathbb H$. Applying the rank-factorization argument together with the rank-nullity theorem, and accounting for the non-uniqueness of the decomposition, we find that the codimension of forming a $4n$-fold degeneracy at $\varepsilon=0$ in the symmetry class $\textrm{CII}$ as
\begin{equation}
\delta_{(n)}^\textrm{CII} = 4n^2,    
\end{equation}
where the factor $4$ arises from the quaternionic quadrupling of the dimensions compared with the real BDI class.

Note that, in agreement with the footnotes in Table~\ref{tab:codims-WD+balanced}, $\delta^\textrm{CII}_{(n)}$ encodes the formation of a $4n$-fold degeneracy. 
This notation is motivated by the fact that the chiral symmetry together with the Kramers' degeneracy imply that zero-energy states arise in multiples of four.
In particular, $\delta^\textrm{CII}_{(1)} = 4$, meaning that an effective four-band model is expanded into four Dirac matrices, which in four dimensions results in a stable fourfold degenerate nodal point protected by the winding number $\pi_3(S^3) = \intg$.

We further focus on the topological characterization of the multifold degeneracy with $n=2$.
The minimal model requires sixteen matrices in $\cmplx^{8 \times 8}$ (over $\quats^{4 \times 4}$) and it can be expressed similarly to the case of class as 
\begin{equation}
\label{eqn:CII-8band-Ham-Dirac}
H^\textrm{CII} = \sum^{16}_{j=1}k_j V_j,  
\end{equation}
where $\bs{k}={k_1,\ldots, k_{16}}$ is a vector of $16$ real components, i.e., coordinates of the momentum space $\reals^{16}$, and $V_j$ is an appropriate basis in the matrix space.

We continue with the derivation of the spectrum of the Hamiltonian~(\ref{eqn:CII-8band-Ham-Dirac}).
Using the quaternion decomposition as in Eq.~\eqref{eq:CII_H_blockdiag}, we can split the vector $\bs{k}$ into $2$ vectors, each with $2$ quaternionic components, as $\bs{\frak{u}}=(\frak{u}_1,\frak{u}_2)$ and $\bs{\frak{v}}=(\frak{v}_1,\frak{v}_2)$. We can choose the basis in such a way that the off-diagonal matrices $\mathfrak{h}$ and $\mathfrak{h}^\dagger$ in $H^\textrm{CII}$ become
\begin{subequations}
\begin{equation}
\mathfrak{h} = \left(\begin{array}{cc} 
\frak{u}_1  & \frak{v}_1    \\
\frak{u}_2   & \frak{v}_2
\end{array}\right),\quad\quad
\mathfrak{h}^{\dagger} = \left(\begin{array}{cc} 
\frak{u}^{\dagger}_1  & \frak{u}^{\dagger}_2  \\
\frak{v}^{\dagger}_1   & \frak{v}^{\dagger}_2
\end{array}\right),
\end{equation}
where $\mathfrak{x}^\dagger$ is the Hermitian conjugate of $\mathfrak{x}$ as defined in Eq.~(\ref{eqn:Hermitian-conjugation-quaternions}).
The square of the Hamiltonian $H^{\textrm{CII}}$ is a block diagonal matrix
\begin{equation}
(H^{\textrm{CII}})^2 = \left(\begin{array}{cc} 
\mathfrak{h}\mathfrak{h}^{\dagger}  & \mathbb{0}  \\
\mathbb{0}   & \mathfrak{h}^{\dagger}\mathfrak{h}
\end{array}\right).    
\end{equation}
Therefore, the eigenvalues of $H^{\textrm{CII}}$ are given by the square roots of the eigenvalues of $\mathfrak{h}\mathfrak{h}^\dagger$ and $\mathfrak{h}^\dagger \mathfrak{h}$; the latter two spectra are known to coincide.\footnote{The isospectrality of $\mathfrak{h}\mathfrak{h}^\dagger$ and $\mathfrak{h}^\dagger\mathfrak{h}$ follows from the equality of their characteristic polynomials, which is the consequence of the Weinstein-Aronszajn identity $\mathrm{det}(1+AB)=\mathrm{det}(1+BA)$ holding for any matrices with complex components~\cite{Pozrikidis:2014}. To apply this identity to matrices over quaternions, we can treat quaternionic matrices as larger matrices over complex numbers.}
Both positive and negative square roots are realized due to the chiral symmetry. 
It is convenient to analyze the block $\mathfrak{h}^{\dagger}\mathfrak{h}$, which takes the form of a Gram matrix over quaternions,
\begin{equation}
\mathfrak{h}^{\dagger}\mathfrak{h} = \left(\begin{array}{cc} 
\lVert \bs{\frak{u}}\rVert^2  & \langle\bs{\frak{u}}|\bs{\frak{v}}\rangle  \\
\langle\bs{\frak{v}}|\bs{\frak{u}}\rangle   & \lVert \bs{\frak{v}}\rVert^2
\end{array}\right),
\end{equation}
\end{subequations}
where we utilized the inner product $\langle \bs{\mathfrak{x}}|\bs{\mathfrak{y}}\rangle=\sum_j \mathfrak{x}^\dagger_j \mathfrak{y}_j$ and the norm $\lVert \bs{\mathfrak{x}}\rVert=\sqrt{\langle \bs{\mathfrak{x}}|\bs{\mathfrak{x}}\rangle}$.
Since the four components of the matrix $\mathfrak{h}^{\dagger}\mathfrak{h}$ commute with each other, the characteristic polynomial can be expressed as if the entries were real. 
Thus, we find
\begin{equation}
\varepsilon \!=\! \pm \frac{1}{\sqrt{2}}\sqrt{\lVert \bs{\frak{u}}\rVert^2 \!+\! \lVert \bs{\frak{v}}\rVert^2 \!\pm\! \sqrt{(\lVert \bs{\frak{u}}\rVert^2 \!- \! \lVert \bs{\frak{v}}\rVert^2)^2 \!+\! 4|\langle\bs{\frak{u}} | \bs{\frak{v}}\rangle|^2 }},
\end{equation}
as the eigenvalues of the minimal Hamiltonian~(\ref{eqn:CII-8band-Ham-Dirac}).

We proceed to study the nodal manifolds $M_\medcirc$ and $M_\varnothing$. Therefore,
we enclose the multifold nodal point at $\boldsymbol{k}=\boldsymbol{0}$ with a sphere $S^{15}$
\begin{equation}
\label{eqn:CII-S}
S^{15} = \{ (\bs{\frak{u}},\bs{\frak{v}})\,|\,\lVert \bs{\frak{u}}\rVert^2 + \lVert \bs{\frak{v}}\rVert^2 = 1\} \subset \reals^{16}.
\end{equation}
On this sphere, the eigenvalues obey $\tr[H^2] =\sum_{j=1}^8 \varepsilon_j^2 = 4$, where we sum over all eight eigenvalues (i.e., explicitly including both members of each Kramers pair).
The nodal manifolds $M_\medcirc\subset S^{15}$ and $M_\varnothing\subset S^{15}$ are defined in analogy with Sec.~\ref{sec:class-BDI} and~\ref{sec:class-AIII}, and correspond to the loci of a band degeneracy at $\varepsilon=0$ and at $\varepsilon\neq 0$, respectively.

The case $M_\medcirc$ corresponds to the conditions
\begin{subequations}
\begin{eqnarray}
&\lVert \bs{\frak{u}}\rVert^2 + \lVert \bs{\frak{v}}\rVert^2 =1\\
&|\langle\bs{\frak{u}} | \bs{\frak{v}}\rangle|=\lVert \bs{\frak{u}}\rVert \lVert \bs{\frak{v}}\rVert.\label{eq:quaternion_parallel}
\end{eqnarray}
\end{subequations}
Equation~\eqref{eq:quaternion_parallel} is satisfied iff $\bs{\frak{v}}=\bs{\frak{u}}\mathfrak{q}$ for some $\mathfrak{q}\in \mathbb{H}$; equivalently, $\bs{\frak{u}}$ and $\bs{\frak{v}}$ lie in the same quaternionic line $\ell$. 
Let us choose a unit vector  $\bs{\mathfrak{w}}=(\mathfrak{w}_1,\mathfrak{w}_2)\in S^7$ spanning $\ell$. 
Then there exist unique quaternions $\frak{a},\frak{b}\in \mathbb{H}$ such that $\bs{\frak{u}}=\bs{\mathfrak{w}}\frak{a}$ and $\bs{\frak{v}}=\bs{\mathfrak{w}}\frak{b}$. The norm condition becomes
\begin{equation}
    \lVert \bs{\frak{u}}\rVert^2 + \lVert \bs{\frak{v}}\rVert^2=|\frak{a}|^2+|\frak{b}|^2=1,
\end{equation}
which defines a sphere $S^7$ in the space of pairs $(\frak{a},\frak{b})\in \mathbb{H}^2$. Hence a pair of vectors $(\bs{\frak{u}},\bs{\frak{v}})=(\bs{\mathfrak{w}}\frak{a},\bs{\mathfrak{w}}\frak{b})$ lies in $S^7\times S^7$, where the first sphere corresponds to the space of unit vectors $\bs{\mathfrak{w}}$ and the second sphere corresponds to the space of pairs $(\frak{a},\frak{b})$. 
Next, observe that $\bs{\mathfrak{w}}\frak{p}$ defines the same quaternionic line as $\bs{\mathfrak{w}}$, iff $\frak{p}$ is a unit quaternion $\frak{p}\in \mathsf{Sp}(1)$. 
Since the values of $\frak{a}$ and $\frak{b}$ are unique for given $\bs{\mathfrak{w}}$, the choice of $\frak{p}$ is the only gauge freedom that is available when identifying the pair $(\bs{\frak{u}},\bs{\frak{v}})$ in $S^7\times S^7$. 
The action of the gauge group is defined by the requirement that the pair of vectors $(\bs{\frak{u}},\bs{\frak{v}})$ does not change after multiplying $\bs{\mathfrak{w}}$ by a unit quaternion $\frak{p}$. 
This can be achieved if one simultaneously multiplies the pair $(\frak{a},\frak{b})$ by $\frak{p}^{-1}$. 
More precisely, the gauge group $\mathsf{Sp}(1)$ acts on both spheres as $\frak{p}(\bs{\mathfrak{w}},(\frak{a},\frak{b}))=(\bs{\mathfrak{w}}\frak{p},(\frak{p}^{-1}\frak{a},\frak{p}^{-1}\frak{b}))$. 
Therefore, we can conclude that $M_\medcirc\simeq (S^7\times S^7)/\mathsf{Sp}(1)$, where the quotient $\textsf{Sp}(1) \simeq S^3$ acts simultaneously on both $7$-spheres following the quaternionic Hopf map.

It is also possible to reproduce this result from the general formula for classifying spaces with zero energies (see Appendix~\ref{app:class-space-w-zero-states}).
The classifying space for class-$\textrm{CII}$ Hamiltonians with $4$ zero energy states and with two states at both the positive and the negative energy is determined as $\mathsf{Sp}(2) \times \mathsf{Sp}(2)/\left[\mathsf{Sp}(1)\times \mathsf{Sp}(1)\right] \times \mathsf{Sp}(1)$, where each $\mathsf{Sp}(1)$ inside the brackets acts respectively on one copy of $\mathsf{Sp}(2)$, while the $\mathsf{Sp}(1)$ outside the brackets acts diagonally on both copies of $\mathsf{Sp}(2)$. 
By noticing that $\mathsf{Sp}(2)/\mathsf{Sp}(1)\simeq S^7$, we obtain the same space, $M_\medcirc\simeq (S^7\times S^7)/\mathsf{Sp}(1)$.

The nodal manifold $M_\medcirc$ as determined above is $11$-dimensional. Being one of the highest-dimensional nodal manifolds encountered in our present analysis, its explicit characterization is prohibitively difficult. 
Although we are currently not able to verify this statement explicitly, we anticipate that the nodal manifold contains nontrivial cycles in dimensions $D\in\{0,4,7,11\}$, with the $4$-cycle originating from the quaternionic Hopf map $S^7/S^3\simeq S^4$. 
Furthermore, we conjecture that the presence of the multifold degeneracy inside the enclosing $S^{15}$ enforces the $4$-cycle to exhibit a nontrivial value of the (chirally resolved) second Chern number, as this provides a natural quaternionic analog of the first Chern number found for the $2$-cycle of $M_\medcirc$ in the complex chiral class $\textrm{AIII}$ in Eq.~(\ref{eqn:AII-first-Chern}) and of the first Stiefel-Whitney class found for the $1$-cycle of $M_\medcirc$ in the real chiral class $\textrm{BDI}$ in Eq.~(\ref{eqn:BDI-zero-eigensystem}). 
Owing to the lack of explicit parameterization of $M_\medcirc$ in the present quaternionic class, we cannot verify this anticipated result explicitly.

On $M_\varnothing$, the vectors $\bs{\frak{u}}$ and $\bs{\frak{v}}$ satisfy the following set of equations:
\begin{subequations}
\begin{eqnarray}
&\lVert \bs{\frak{u}}\rVert^2 + \lVert \bs{\frak{v}}\rVert^2 =1\\
&\lVert \bs{\frak{u}}\rVert^2 - \lVert \bs{\frak{v}}\rVert^2 =0\\
&\langle\bs{\frak{u}} | \bs{\frak{v}}\rangle=0.
\end{eqnarray}
\end{subequations}
These equations fix the norms of vectors $\bs{\frak{u}}$ and $\bs{\frak{v}}$ and require their orthogonality. 
In other words, $M_\varnothing$ is the space of orthonormal quaternionic frames in $\mathbb{H}^2$, which is the quaternionic unitary group $\mathsf{Sp}(2)$. 
This is the standard classifying space of class-$\textrm{CII}$ Hamiltonians with $4$ occupied and $4$ unoccupied bands. 
The space $\mathsf{Sp}(2)$ is known to be constructed with one cell in each dimension $D\in\{0,3,7,10\}$, each acting as a generator of the homology group with $\intg$ coefficients.

While unable to provide explicit verification, we conjecture that the presence of the multifold degeneracy inside the enclosing $S^{15}$ is manifested in the quaternionic chiral class $\textrm{CII}$ by a nontrivial three-dimensional winding number of the block $\mathfrak{h}$ on the $3$-cycle of $M_\varnothing$. 
The motivation is that this invariant provides a natural quaternionic ($\mathsf{Sp}(1)\simeq S^3$) analog of the one-dimensional winding number found for the $1$-cycle of $M_\varnothing$ in the complex ($\mathsf{U}(1)\simeq S^1$) chiral class $\textrm{AIII}$ in Eq.~(\ref{eqn:AIII-1D-winding}) and of the Pfaffian invariant found on the $0$-cycle of $M_\varnothing$ in the real ($\mathsf{O}(1)\simeq S^0$) chiral class $\textrm{BDI}$.
Observe also that the anticipated homology of $M_\medcirc$ is compatible with nontrivial cycles determined for $M_\varnothing$; namely, the $4$-cycle of $M_\medcirc$ can link with the $10$-cycle of $M_\varnothing$ (speculated to be signaled by the second Chern number on the $4$-cell), and the $11$-cycle of $M_\medcirc$ can link with the $3$-cycle of $M_\varnothing$ (expected to be signaled by the three-dimensional winding number on the $3$-cycle).

\section{Symmetry class \texorpdfstring{$\textrm{D}$}{D}}
\label{sec:D_class}

The symmetry class $\textrm{D}$ is characterized by the presence of particle-hole symmetry that obeys $\mathcal{C}^2 = +1$.
Since every band at negative energy has its particle-hole symmetric partner at positive energy, the Hamiltonian has an even dimension.
We represent it as $\mathcal{C}=\mathcal{K}$; then, due to $\mathcal{K}H\mathcal{K} = - H$, the Hamiltonian is enforced to be purely imaginary. 
We write $H = \imi \widetilde{H}$, where, due to the Hermiticity of the Hamiltonian, $\widetilde{H}$ is a $2N \times 2N$ real skew-symmetric matrix.

We begin by adapting the von Neumann-Wigner counting argument.
By counting the number of linearly independent parameters, we find that the dimension of the space of real skew-symmetric $2N\times 2N$ matrices is 
\begin{subequations}
\begin{equation}
d^\textrm{D}_{(N,0)} = \tfrac{1}{2}\times 2N\times (2N-1) = N(2N-1). 
\end{equation}
This can also be interpreted as the dimension of the Lie algebra $\mathfrak{so}(2N)$; therefore, we can equivalently express
\begin{equation}
\label{eqn:class-D-general-dimension}
d^\textrm{D}_{(N,0)} = \dim[\mathsf{SO}(2N)]. 
\end{equation}
To find the dimension $d^\textrm{D}_{(N,n)}$ of such Hamiltonians that further exhibit $2n$ zero-energy eigenvalues, we utilize the rank $2r = 2N - 2n$ of the Hamiltonian and argue as follows.
First, recall that using an orthogonal $\in \mathsf{O}(2N)$ rotation it is possible to bring a real skew-symmetric matrix $\widetilde{H}$ to a canonical block-diagonal form 
\begin{equation}
\label{eqn:class-D-canonical-diag}
\widetilde{H} = \textrm{diag}\,\left\{ \left(\begin{array}{cc} 0 & \lambda_1 \\ -\lambda_1 & 0 \end{array}\right), \cdots, \left(\begin{array}{cc} 0 & \lambda_r \\ -\lambda_r & 0 \end{array}\right),\mathbb{0},\cdots\right\}
\end{equation}
\end{subequations}
where $ \lambda_1 >\lambda_2 > \ldots > \lambda_r > 0$ are real numbers encoding the non-zero eigenvalues of the Hamiltonian (assumed to be distinct).
We see that
\begin{subequations}
\begin{eqnarray}
d^\textrm{D}_{(N,n)}\! 
&=& \dim[\mathsf{O}(2N)] + r -r\dim[\mathsf{SO}(2)] \nonumber \\
&\phantom{=}& \quad - \dim[\mathsf{O}(2N-2r)] \nonumber \\
&=& N(2N-1) - (N-r)(2N-2r-1).
\label{eqn:D-Ham-dim}
\end{eqnarray}
In the first line above, the first term reproduces $d^\textrm{D}_{(N,0)}$ from Eq.~(\ref{eqn:class-D-general-dimension}), the second term counts the distinct non-zero eigenvalues, the third term corresponds to the dimension of the stabilizer of the non-zero blocks, and the last term represents the stabilizer of the $N-r$ zero blocks.
From here, the codimension for forming $(2n)$-fold zero-energy degeneracy equals
\begin{equation}
\delta^{\textrm{D}}_{(n)} = d^\textrm{D}_{(N,0)} - d^\textrm{D}_{(N,n)} = n(2n-1).    
\end{equation}
\end{subequations}
In particular, $\delta_{(1)}^\textrm{D} = 1$, meaning that an effective two-band model is expanded into a single Pauli matrix~\cite{Bzdusek:2017}.
This result implies the existence of nodal surfaces in three-dimensional crystals in symmetry class $\textrm{D}$, which are topologically stabilized by a Pfaffian invariant $\pi_0(S^0) = \ztwo$.
Physically, such nodal surfaces are realized as Bogoliubov-Fermi surfaces in spin-orbit coupled superconductors that exhibit singlet-triplet mixing due to sublattice-switching inversion symmetry~\cite{Agterberg:2017}.

We further focus on the topological characterization of the multifold degeneracy of the next available order. 
The result $\delta^\textrm{D}_{(2)} = 6$ implies the expansion of the minimal four-band Hamiltonian around the fourfold degeneracy using six Dirac matrices.
Such a Hamiltonian can be expressed as~\cite{Bzdusek:2017}
\begin{equation}
    H^\textrm{D} = \boldsymbol{k} \cdot \boldsymbol{V} + \boldsymbol{q} \cdot \boldsymbol{W}, 
\end{equation} 
where $\boldsymbol{k} = (k_1,k_2,k_3)$ and $\boldsymbol{q} = (q_1,q_2,q_3)$ parameterize momentum space $\mathbb{R}^6$ and 
\begin{subequations}
\begin{align}
    \boldsymbol{V} &= (\sigma_x \otimes \tau_y, \sigma_y \otimes \mathbb{1} , \sigma_z \otimes \tau_y) \\ 
    \boldsymbol{W} &= (\sigma_y \otimes \tau_x, \mathbb{1} \otimes \tau_y, \sigma_y \otimes \tau_z)
\end{align}
\end{subequations}
are pairwise anticommuting matrices satisfying $ \{V_i,V_j \} = \{ W_i, W_j \} = 2 \delta_{ij}$. 
For arbitrary values of $( \boldsymbol{k},\boldsymbol{q})$, the spectrum can be expressed as 
\begin{equation}
\varepsilon = \pm ||\boldsymbol{k}|| \pm ||\boldsymbol{q}||.
\end{equation}
We order the eigenvalues of the Hamiltonian such that $\varepsilon_1 \leq \dots\leq \varepsilon_4$ and define the degeneracy loci as in Eq.~(\ref{eqn:chiral-loci}).
We surround the fourfold degeneracy located in the origin ($\boldsymbol{k} = \boldsymbol{q} = \boldsymbol{0}$) with a five-dimensional sphere $S^5$ of radius $\kappa = 1$ and investigate the intersections $M_\medcirc = S^5 \cap \mathcal{L}_\medcirc$ and
$M_\varnothing = S^5 \cap \mathcal{L}_\varnothing$.

The nodal manifold $M_\medcirc$ corresponds to momenta $(\boldsymbol{k},\boldsymbol{q})$ satisfying the set of equations
\begin{subequations}
\begin{align}
    & k_1 ^2 + k_2^2 + k_3^2 + q_1^2 + q_2^2 + q_3^2= 1 \\
    & k_1^2+k_2^2 + k_3^2 - q_1^2-q_2^2 - q_3^2=0.
\end{align}
\end{subequations}
These equations define the product of two-dimensional spheres, $M\simeq S^2 \times S^2$, that have radius $1/\sqrt{2}$ and lie in the $(k_1,k_2,k_3)$ and in the $(q_1,q_2,q_3)$ subspace, respectively. 
Clearly, this space exhibits nontrivial cycles in dimensions $D\in\{0,2,4\}$. 
This space can also be obtained from the general formula derived in Appendix~\ref{app:class-space-w-zero-states} via the chain of identifications $\mathsf{O(4)}/\mathsf{O}(2)\times\mathsf{U}(1)\simeq \mathsf{SO(4)}/\mathsf{SO}(2)\times\mathsf{SO}(2)=\mathsf{Gr}^+(2,4)\simeq S^2\times S^2$, where $\mathsf{Gr}^+(2,4)$ is the Grassmannian of oriented two-planes in $\mathbb{R}^4$ that is known to be diffeomorphic to $S^2\times S^2$~\cite{Baird:1987}.

To reveal the linking of $M_\medcirc$ with $M_\varnothing$, we compute the first Chern number of the negative energy band on both of the spheres in $M_\medcirc$. 
To do this, notice that the negative energy eigenstate is a simultaneous eigenstate of the operators $\hat{A} = \hat{\boldsymbol{k}} \cdot \boldsymbol{V}$ and $\hat{B} = \hat{\boldsymbol{q}} \cdot \boldsymbol{W}$ with eigenvalues $A = -1$ and $B=-1$. 
Here $\hat{\boldsymbol{k}}$ and $\hat{\boldsymbol{q}}$ are normalized vectors $\hat{\boldsymbol{x}} = \boldsymbol{x} / || \boldsymbol{x}||$. 
Then, we can define the projector to the negative-energy line bundle as $\Pi = \frac{1}{4} (1-\hat{A})(1-\hat{B})$. 
For the first sphere, we fix $\boldsymbol{q} =(1,0,0)/ \sqrt{2}$ and vary the components $\boldsymbol{k} = (\sin\theta \cos \phi, \sin \theta \sin\phi, \cos\theta) / \sqrt{2}$ with $\theta \in [0, \pi] $ and $\phi \in [0, 2\pi]$. 
For the second sphere, we proceed analogously by exchanging the roles of $\bs{q}$ and $\bs{k}$. 
This parametrization allows us to express the projector $\Pi$ using $\theta$ and $\phi$. 
Then, the Berry curvature is straightforwardly computed and integrated using the formula $F = + \imi \mathrm{Tr}\left[ \Pi~ d\Pi \wedge d \Pi\right]$. 
Explicit computation yields  $F=+\frac{1}{2} \sin \theta ~ d\theta \wedge d \phi$ giving $C=+1$ for the negative energy bundle (depending on the orientation)~\cite{supplementary_code_data}. 
Due to the particle-hole symmetry, the positive energy band carries nontrivial Chern number of the opposite sign, while the rank-$2$ bundle spanned by the degenerate zero-energy bands carry vanishing first Chern number.

Notably, the nontrivial result for the first Chern number of the bands at non-zero energy allows us to deduce the nontrivial \emph{second} Chern number carried by the degenerate zero-energy bands.
The argument mimics the discussion of the class $\textrm{A}$ in Sec.~\ref{sec:class-A}. By the Künneth theorem~\cite{Hatcher_book}, the cohomology ring of the product of two spheres is the tensor product of their cohomology rings; therefore,
\begin{equation}
H^*(S^2\times S^2,\intg) = \intg[a,b]/(a^2,b^2).
\end{equation}
The elements of this ring are the polynomials of the form $c=\beta_0+\beta^1_1a+\beta^2_1b+\beta_2ab$ with integer coefficients, the product $ab = ba \equiv a \smile b$, and relations $a^2=b^2=0$. 
The elements $a$ and $b$ are the generators of the respective second cohomology group of each of the spheres $H^2(S^2,\intg)$, while the product $ab$ is the generator of the fourth cohomology group of their product $H^4(S^2\times S^2,\intg)$. 
The commutativity of $a$ and $b$ follows from the fact that they are both $2$-forms.

Let us denote the rank-$1$ bundle on $M_\medcirc$ spanned by the negative (positive) energy band as $E^{(-)}$ ($E^{(+)}$), while we use $E^{(0)}$ for the rank-$2$ bundle spanned by the degenerate zero-energy bands. 
The calculated first Chern numbers with the fact that $E^{(-)}$ is of rank-$1$ fix its total Chern class as
\begin{subequations}
\begin{equation}
    c(E^{(-)})=1+a+b.
\end{equation}
Since particle-hole symmetry flips the sign of first Chern numbers, we further have that
\begin{equation}
    c(E^{(+)})=1-a-b.
\end{equation}
Using the Whitney sum formula, we find
\begin{equation}
    c(E^{(-)}\oplus E^{(+)})=c(E^{(-)})\smile c(E^{(+)})=1-2ab.
\end{equation}
However, since the total bundle $E^{(-)}\oplus E^{(+)}\oplus E^{(0)}$ is trivial, inversion of the Whitney formula implies
\begin{equation}
    c(E^{(0)}) = (1 - 2ab)^{-1} = 1 + 2ab.    
\end{equation}
\end{subequations}
Therefore, the zero-energy bands carry vanishing first Chern number on the $2$-cycles of $M_\medcirc$, but a nontrivial (and even) \emph{second} Chern number on its $4$-cycle.
Utilizing standard algorithms for computing the first~\cite{Fukui:2005} and second~\cite{Mochol:2019} Chern numbers, we have verified these predictions numerically~\cite{supplementary_code_data}.

We next turn our attention to the nodal manifold $M_\varnothing$, which consists of the union of two non-intersecting spheres, satisfying either $||\boldsymbol{k}|| = 0$ or $|| \boldsymbol{q}|| =0$. 
That is,
\begin{subequations}
\begin{gather}
\quad  k_1 ^2 + k_2^2 + k_{3}^2 + q_1^2 + q_2^2 + q_3^2= 1 \quad  \\ 
 k_1^2 + k_2^2 + k_3^2 =0 \quad \mathrm{or} \quad q_1^2 + q_2^2 + q_3^2 =0.
\end{gather}
\end{subequations}
Therefore, $M_\varnothing \simeq S^2 \sqcup S^2$.
Taking $k_1^2 + k_2^2 + k_3^2=0$ results in $q_1^2+q_2^2 + q_3^2=1$, defining one of the spheres, while fixing $q_1^2 + q_2^2 + q_3^2= 0$ defines the other sphere. 
Note that the pair of $2$-spheres can also be obtained as the classifying space of class-$\textrm{D}$ $4$-band Hamiltonian with a single energy gap at half-filling, $M_\varnothing\simeq \mathsf{O(4)}/\mathsf{U(2)}$. 
Clearly, $M_\varnothing$ has nontrivial cycles in dimensions $D\in\{0,2\}$.

Let us also point out that the two $2$-spheres are \emph{linked} on the enclosing $S^5$.
This is conveniently revealed by inscribing the 3D ball 
\begin{equation}
\label{eqn:class-D-inscribed-ball}
B^3_{\bs{q}} = \left\{ (\bs{k},\bs{q}) \, \big| \, \bs{k} = (0,0,k_z), \lVert \bs{q} \rVert^2 = 1 - k_z^2, k_z > 0\right\}    
\end{equation}
to the sphere inside $\bs{q}$-subspace, and noticing that this ball intersects the sphere inside $\bs{k}$-subspace transversally at a single point $\bs{k}=(0,0,1)$ [cf.~Eq.~(\ref{eq:linkin_number}) and Fig.~\ref{fig:linking_scheme}]. 
Therefore, the geometric constellation of $M_\varnothing$ in the class $\textrm{D}$ mimics the case illustrated for the class $\textrm{BDI}$ in Fig.~\ref{fig:class-BDI-linking}, except that all circles are replaced by $2$-spheres.

To reveal the linking of $M_\varnothing$ with $M_\medcirc$, we first compute the sign of the Pfaffian of the skew-symmetric matrix $\tilde{H} = -\imi H$. 
For the $S^2$ defined by $|\boldsymbol{k}| = 0$, we have $\mathrm{sign~Pf} ~\tilde{H} = +1$, while for the other sphere, we obtain $\mathrm{sign~Pf} ~\tilde{H} = -1$. The sign of the Pfaffian can change only when the determinant of $\tilde{H}$ vanishes, which implies that the two components of $M_\varnothing$ are on opposite sides of (i.e., are linked with) $M_\medcirc$ in the enclosing $S^5$.
Furthermore, through an explicit calculation using the projector approach with $\Pi=\frac{1}{2}(1-\hat{A})$ or $\Pi=\frac{1}{2}(1-\hat{B})$ (depending on which sphere we consider), it is also possible to verify that the degenerate states carry a nontrivial Chern number on both components of $M_\varnothing$ (see~\cite{supplementary_code_data}), which provides further evidence of the linking with $M_\circ$.

Before concluding, we relate our analysis to the result of Ref.~\citenum{Kim:2021} that class-$\textrm{D}$ nodal surfaces in 3D that carry a nontrivial Chern number $\pi_2[\mathsf{O}(4)/\mathsf{U}(2)]=\intg$ on the enclosing $2$-sphere necessarily contain a Weyl point in adjacent energy gaps. 
In the context of the present treatment of the multifold degeneracy, consider inscribing a ball to either $2$-sphere of $M_\varnothing$.
According to the discussion around Eq.~(\ref{eqn:class-D-inscribed-ball}) this ball intersects the second component of $M_\varnothing$; this intersection corresponds to a Weyl point formed within the positive (and within the negative) energy bands.
Furthermore, the nontrivial first Chern number derived in the previous paragraph implies, according to the general discussion in Sec.~\ref{sec:general-linking-to-invariants}, that this ball necessarily intersects with $M_\medcirc$. This corresponds to a zero-energy node that, due to codimension reasons, is generically a nodal surface. 
The Chern number derived above can be computed as the base of the integration shrinks from the $2$-cycle in $M_\varnothing$ to any other $2$-sphere that encloses the nodal ring. 
Therefore, our topological characterization of fourfold degeneracy in class $\textrm{D}$ exactly reproduces the finding of Ref.~\citenum{Kim:2021}.

\section{Symmetry class \texorpdfstring{$\textrm{C}$}{C}}
\label{sec:class-C}

The symmetry class $\textrm{C}$ is characterized by particle-hole symmetry with $\mathcal{C}^2 = -1$ that enforces the Hamiltonian to have even dimension.
We represent $\mathcal{C} = \mathbb{1}_N \otimes (-\imi \sigma_y) \mathcal{K}$ with the block-diagonal structure in Eq.~(\ref{eqn:spinful-TRS-blocks}). 
The particle-hole symmetry anticommutes with the $2N \times 2N$ Hamiltonian, $\mathcal{C} H \mathcal{C}^{-1} = - H$; therefore, if $\ket{u^\alpha}$ is an eigenstate obeying $H \ket{u^\alpha} = \varepsilon^\alpha \ket{u^\alpha}$ (i.e., has energy $\varepsilon^\alpha$), its chiral symmetric partner is an eigenstate with $H \mathcal{C} \ket{u^\alpha} = - \varepsilon^\alpha \mathcal{C} \ket{u^\alpha}$ (i..e, has energy $-\varepsilon^\alpha$).
One can collect these eigenstates into a matrix $\Psi$ that diagonalizes the Hamiltonian. Similarly to the case of class $\textrm{AII}$, the matrix of eigenstates is compact symplectic, $\Psi \in \mathsf{Sp}(N)$.

We next adapt the von Neumann-Wigner counting.
Let us consider a class-$\textrm{C}$  Hamiltonians with $2n$-fold degeneracy at zero energy, which is equivalent to setting the Hamiltonian rank to $2r = 2N - 2n$.
Then, the eigenspectrum of a rank-$2r$ Hamiltonian in class $\textrm{C}$ consists of $r$ pairs of non-zero eigenvalues $(\varepsilon^\alpha,-\varepsilon^\alpha)$, $\varepsilon^\alpha \neq 0$ (assumed to be distinct and ordered from largest to smallest), and $n$ pairs of zero eigenvalues.
From such a characterization, we deduce that
\begin{subequations}
\begin{eqnarray}
d^\textrm{C}_{(N,n)} 
&=& \dim[\mathsf{Sp}(N)] + r - r\dim[\mathsf{U}(1)] \nonumber \\
&\phantom{=}& \quad - \dim[\mathsf{Sp}(N-r)] \nonumber \\
&=& N(2N+1) - (N-r)(2N-2r+1).\quad
\label{eqn:C-Ham-dim}
\end{eqnarray}
Here, going through the right-hand side of the first line, the first term encodes the freedom in specifying the eigenstate matrix $\Psi$, and the second term counts the freedom in specifying the non-zero eigenvalues $\varepsilon^{\alpha}$. 
On the other hand, the third term subtracts the gauge freedom of eigenvectors associated with the eigenvalue pairs $(\varepsilon^\alpha,-\varepsilon^\alpha)$, while the last term $\mathsf{Sp}(N-r)$ accounts for the gauge freedom of the zero-energy sector.
From the above, the codimension of forming a $2n$-fold zero-energy degeneracy is extracted as 
\begin{equation}
    \delta^\textrm{C}_{(n)} = d^\textrm{C}_{(N,0)} - d^\textrm{C}_{(N,n)} = n(2n+1).  
\end{equation}
\end{subequations}
In particular, $\delta_{(1)}^\textrm{C} = 3$, meaning that an effective two-band model is expanded into three Pauli matrices~\cite{Bzdusek:2017}.
This result implies the existence of nodal points in three-dimensional crystals in symmetry class $\textrm{C}$ that are topologically stabilized by the Chern number $\pi_2(S^2) = \intg$.

In the following, we investigate the multifold degeneracy of the next available order; namely, a fourfold degeneracy realized in a four-band Hamiltonian. 
The result $\delta^\textrm{C}_{(2)} = 10$ implies that the Hamiltonian can be expanded into ten Dirac matrices; specifically, into
\begin{subequations}
\begin{equation}
\{\sigma_x,\sigma_y,\sigma_z\}\otimes\{\mathbb{1}_\tau,\tau_x,\tau_z\}\,\cup \,\{\mathbb{1}_\sigma\otimes \tau_y\}.
\label{eqn:10-Diracs-in-class-C}
\end{equation}
Note that $10 = \binom{5}{2}$, which motivates us to seek a representation of the ten Dirac matrices as the basis of $2$-forms over a five-dimensional vector space. 
This is indeed possible; namely, we identify
\begin{eqnarray}
e^1 = \mathbb{1}_\sigma \otimes \tau_x 
\!\!\!\!\!\!\!\! &\qquad & \!\!\!\!\!\!\!\!
e^2 = \mathbb{1}_\sigma \otimes \tau_z
\\
e^3 = \sigma_x \otimes \tau_y 
\qquad
&e^4 = \sigma_y \otimes \tau_y&
\qquad
e^5 = \sigma_z \otimes \tau_y \nonumber 
\end{eqnarray}
as the five pair-wise anticommuting Dirac matrices, each squaring to $\mathbb{1}_4$.
Then, defining the wedge product as the commutator 
\begin{equation}
e^a\wedge e^b = -\tfrac{i}{2}[e^a,e^b] = -\tfrac{i}{2}(e^a\cdot e^b - e^b \cdot e^a),    
\end{equation} 
we obtain
\begin{eqnarray}
e^1 \wedge e^2 = -\mathbb{1}_\sigma \otimes \tau_y 
&\qquad &
e^1 \wedge e^3 = +\sigma_x \otimes \tau_z \nonumber \\
e^1 \wedge e^4 = +\sigma_y \otimes \tau_z 
&\qquad &
e^1 \wedge e^5 = +\sigma_z \otimes \tau_z \nonumber \\
e^2 \wedge e^3 = -\sigma_x \otimes \tau_x 
&\qquad &
e^2 \wedge e^4 = -\sigma_y \otimes \tau_x \quad \\
e^2 \wedge e^5 = -\sigma_z \otimes \tau_x 
&\qquad &
e^3 \wedge e^4 = +\sigma_z \otimes \mathbb{1}_\tau \nonumber \\
e^3 \wedge e^5 = -\sigma_y \otimes \mathbb{1}_\tau 
&\qquad &
e^4 \wedge e^5 = +\sigma_x \otimes \mathbb{1}_\tau\nonumber ,
\end{eqnarray}
\end{subequations}
which exactly match the ten Dirac matrices listed in Eq.~(\ref{eqn:10-Diracs-in-class-C}).
Therefore, we identify the general Hamiltonian with the $2$-form 
\begin{subequations}
\begin{equation}
\label{eqn:class-C-Ham}
H^\textrm{C} = \sum_{a<b} H_{ab} \,e^a \wedge e^b,
\end{equation} 
where the coefficients can be arranged into an antisymmetric matrix
\begin{equation}
\label{eqn:Hamiltonian-coefficients}
H = \left(\begin{array}{ccccc}
0 & H_{12} & H_{13} & H_{14} & H_{15} \\ 
- H_{12} & 0 & H_{23} & H_{24} & H_{25} \\
- H_{13} & - H_{23} & 0 & H_{34} & H_{35} \\
- H_{14} & - H_{24} & - H_{34} & 0 & H_{45} \\
- H_{15} & - H_{25} & - H_{35} & - H_{45} & 0 .
\end{array}\right)  
\end{equation}
\end{subequations}
Note that, to avoid double-counting, we restrict the sum in Eq.~(\ref{eqn:class-C-Ham}) to $a<b$.

To find the spectrum of the Hamiltonian (\ref{eqn:class-C-Ham}), consider
\begin{subequations}
\begin{equation}
\big(H^\textrm{C}\big)^2 = \sum_{a < b} \sum_{c<d} H_{ab} H_{cd} (e^a \wedge e^b)\cdot(e^c \wedge e^d).  
\end{equation}
To simplify the right-hand side, we utilize that
\begin{eqnarray}
&\phantom{=}& (e^a \wedge e^b)\cdot(e^c \wedge e^d) \\ &=&  (\delta^{ac}\delta^{bd}-\delta^{ad}\delta^{bc})\mathbb{1}_4 - \sum_e \epsilon_{abcde} e^e \nonumber  
\end{eqnarray}
where we utilized Kronecker symbols and a five-dimensional Levi-Civita symbol.
Observe that the case $a=d$ and $b=c$ (i.e., second combination of Kronecker symbols inside the parentheses) is incompatible with $a<b$ and $c<d$ and therefore never arises.
The other (i.e., the first) combination of Kronecker symbols gives simply $\left(\sum_{a<b} H_{ab}^2\right) \mathbb{1}_4$.
Finally, for the term involving $\epsilon_{abcde}$, we need to collect all the ways of obtaining distinct indices $(a,b,c,d)$ with $a<b$ and $c<d$. 
For example, when $e=5$, there are three solutions: $(1,2,3,4)$, $(1,3,2,4)$, and $(1,4,2,3)$ (with Levi-Civita symbols $\epsilon_{12345}=+1$, $\epsilon_{13245}=-1$, and $\epsilon_{14235}=+1$).
Collecting all the contributions for all choices of $e^e$, we obtain\\
\begin{widetext}
\begin{eqnarray}
&(H^\textrm{C})^2 = \left(\sum_{a<b}H_{ab}^2\right)\mathbb{1}_4 + (H_{23}H_{45}-H_{24}H_{35}+H_{25}H_{34}) e^1 + (-H_{13}H_{45}+H_{14}H_{35}-H_{15}H_{34}) e^2& \label{eqn:ok!}\\
&+ (H_{12}H_{45}-H_{14}H_{25}+H_{15}H_{24})e^3 + (-H_{12}H_{35}+H_{13}H_{25}-H_{15}H_{23})e^4 + (H_{12}H_{34}-H_{13}H_{24}+H_{14}H_{23})e^5.\nonumber &
\end{eqnarray}
\end{widetext}
\end{subequations}
Let us call the coefficient appearing in front of the Dirac matrix $e^a$ in Eq.~(\ref{eqn:ok!}) as $G_a$. 
One can interpret $G_a$ as the Pfaffian of the $4\times 4$ minor obtained by deleting the row and column $a$ in Eq.~(\ref{eqn:Hamiltonian-coefficients}). 
In fact, the entire expression $G_a e^a$ can be recognized as the Hodge dual of the $4$-form $H \wedge H$.

From these expressions, we find the spectrum of the Hamiltonian in Eq.~(\ref{eqn:Hamiltonian-coefficients}) as
\begin{subequations}
\begin{equation}
\varepsilon = \pm\sqrt{\sum_{a<b}H_{ab}^2 \pm \sqrt{\sum_a G_a^2}}    
\end{equation}
Using the interpretation of the coefficients $H_{ab}$ and $G_a$ as the components of the $2$-form $H$ and of the $4$-form $H\wedge H$, the previous expression can be alternatively expressed as 
\begin{equation}
\varepsilon=\pm\sqrt{\lVert H\rVert^2 \pm \lVert H\wedge H\rVert}. 
\end{equation}
\end{subequations}
On the enclosing sphere $S^9$, we require $||H||^2=1$.
To understand the nodal manifold $M_\medcirc$ and $M_\varnothing$, we need to understand when $\lVert H \wedge H \rVert  = \lVert H \rVert ^2$ resp.~when $\lVert H \wedge H\rVert = 0$  on $S^9$.

To describe the nodal manifold $M_\medcirc$, we transform the basis $\{e^a\}_{a=1}^5$ by an orthogonal transformation
$Q \in \mathsf{O}(5)$ to a basis $\{f^a\}_{a=1}^5$ so that the $2$-form $H$ in the new basis takes the form 
\begin{subequations}
\begin{equation}
\label{eqn:2-form-in-5D-decomposition}
H=\gamma \,f^1\wedge f^2+\delta \,f^3\wedge f^4    
\end{equation} 
with $\gamma,\delta \geq 0$. 
Such a transformation exists for any $2$-form over a five-dimensional space, as it corresponds to bringing the antisymmetric matrix~(\ref{eqn:Hamiltonian-coefficients}) to the canonical form (\ref{eqn:class-D-canonical-diag}) with $\lambda_1 = \gamma$ and $\gamma_2 = \delta$; the sole difference is that the odd matrix dimension enforces a single left-alone $0\times 0$ block.
In the rotated basis, 
\begin{equation}
H \wedge H = 2\gamma\delta \,f^1 \wedge f^2 \wedge f^3 \wedge f^4,
\end{equation}
therefore, the condition $\lVert H \wedge H \rVert  = \lVert H \rVert ^2$ translates to
\begin{equation}
    4\gamma^2\delta^2=(\gamma^2+\delta^2)^2 \,\iff \,\gamma^2=\delta^2.
\end{equation}
The condition $||H||^2=1$, in turn, fixes $\delta^2=\gamma^2=\frac{1}{2}$. 
In other words, we obtain that in the block-diagonal decomposition 
\begin{equation}
\label{eqn:block-decomposition}
    H=Q\Sigma Q^{T},
\end{equation}
\end{subequations}
the block-diagonal matrix $\Sigma$ is fixed and consists of two equal $2\times 2$ blocks, together with the enforced vanishing $1\times 1$ block. 
The stabilizer in $\mathsf{O}(5)$ that leaves this block decomposition invariant consists of two components: (\emph{i})~the $\mathsf{U}(2)\subset\mathsf{O}(4)$ freedom in the first four components due to the equal values of the eigenvalues\footnote{The same gauge freedom is utilized when deriving the classifying space $\mathsf{O}(2N)/\mathsf{U}(N)$ of Hamiltonians in the symmetry class $\textrm{D}$, cf.~Appendix~B of Ref.~\citenum{Bzdusek_thesis}. This $\mathsf{U}(2)$ freedom is the generalization to the degenerate case of the $\mathsf{SO}(2)$ freedom encountered for the non-degenerate eigenvalues in Eq.~(\ref{eqn:D-Ham-dim}) in our discussion of the class $\textrm{D}$.} $\gamma=\delta$, and~(\emph{ii}) the $\mathsf{O}(1)\simeq\mathbb{Z}_2$ freedom in flipping the orientation of the last vector in $Q$.
This reasoning allows us to identify the nodal manifold as $M_\medcirc\simeq\mathsf{O}(5)/[\mathsf{U}(2)\times\mathsf{O}(1)]\simeq\mathsf{SO}(5)/\mathsf{U}(2)$. 
It is known that the obtained symmetric space can be identified with the complex projective space\footnote{The idea for establishing this equivalence is it pass to the universal double cover $\mathsf{Spin}(5)\simeq\mathsf{Sp}(2)$ of $\mathsf{SO}(5)$. The full inverse image of $\mathsf{U}(2)\subset \mathsf{SO}(5)$ under this covering is $\mathsf{U}(1)\times \mathsf{Sp}(1)$. Since the kernel $\mathbb Z_2$ of the double covering is contained in the lifted stabilizer $\mathsf{U}(1)\times\mathsf{Sp}(1)$, we can identify the nodal manifold as $M_\medcirc\simeq\mathsf{Sp}(2)/[\mathsf{U}(1)\times\mathsf{Sp}(1)]$. Furthermore, we notice that $\mathsf{Sp}(2)$ acts transitively on the lines in $\mathbb{C}^4$, with $\mathsf{U}(1)\times\mathsf{Sp}(1)$ being a stabilizer of the action. We also note that the same nodal manifold is directly reproduced by the general formula in Appendix~\ref{app:class-space-w-zero-states}.} $\mathbb{C}P^3$~\cite{Butruille:2004,Schwahn:2022}. The cell decomposition of $\mathbb{C}P^3$ is standard~\cite{Hatcher_book}, and we find that the nodal manifold $M_\medcirc$ has cells in dimensions $D\in\{0,2,4,6\}$. 
Due to the lack of convenient parametrization, we don't proceed with the computation of the band invariants, but we conjecture that the $2$-cell should support the chirally resolved Chern number.

On the other hand, the condition $\lVert H \wedge H\rVert = 0$ that determines $M_\varnothing$ is equivalent to  $H \wedge H=0$. 
This is true if and only if the $2$-form $H$ is decomposable, i.e., if it can be represented as a wedge product of $1$-forms: $H=\alpha\wedge\beta$~\cite{Baird:2013}. 
Then, the normalization condition $\lVert H \rVert=1$ becomes $\lVert \alpha\wedge\beta \rVert=1$. 
By comparing against Eq.~(\ref{eqn:2-form-in-5D-decomposition}), this is equivalent to $H$ being decomposable as
$H=f^1\wedge f^2$, where $f^1$ and $f^2$ are orthonormal. 
In terms of the decomposition~(\ref{eqn:block-decomposition}), we find that the block-diagonal matrix $\Sigma$ consists of one fixed $2\times2$ of the form $\imi \sigma_y$ and a $3\times 3$ zero block. 
The $2\times 2$ block is preserved by $\mathsf{U}(1)\simeq \mathsf{SO}(2)$ transformation, while the $3\times 3$ zero block is preserved under rotations with the $\mathsf{O}(3)$ group. 
Hence, we can identify the nodal manifold as $M_\varnothing\simeq \mathsf{O}(5)/\mathsf{SO}(2)\times \mathsf{O}(3)\simeq \mathsf{SO}(5)/\mathsf{SO}(2)\times \mathsf{SO}(3)$, i.e., as the oriented Grassmann manifold $\mathsf{Gr}^+(2,5)$.  

An alternative route to identify $M_\varnothing$ is to notice that the eigenvalues on this nodal manifold are fixed and equal to $\pm1$. 
Thus, we can use the standard result for the classifying spaces of class-$C$ Hamiltonians to obtain\footnote{The identification of the symmetric spaces $\mathsf{Gr}^{+}(2,5)\simeq\mathsf{Sp}(2)/\mathsf{U}(2)$ is well-known and can be found, for example, in Ref.~\citenum{Wolf:1969}.} $M_\varnothing\simeq \mathsf{Sp}(2)/\mathsf{U}(2)$. 
This space can be identified with the complex quadric threefold $Q^3$~\cite{Kups:2005} and has nontrivial homology groups with $\intg$ coefficients in dimensions $D\in\{0,2,4,6\}$~\cite{Borel:1958,Zibrowius:2011}. 
The dimensions of the available cells for both nodal manifolds $M_\medcirc$ and $M_\varnothing$ indicate the potential linking. 
In analogy with $M_\medcirc$, we conjecture that the relevant band invariant is the Chern number on $2$-cell of $M_\varnothing$; however, owing to the more elaborate structure of this nodal manifold and the lack of an explicit parameterization, we do not proceed with an explicit verification of this claim. 
Thus, we leave the predicted correspondence between the fourfold zero energy degeneracy and the linking of the nodal manifolds $M_\medcirc$ and $M_\varnothing$ on the enclosing sphere $S^9$ unverified for the symmetry class~$\textrm{C}$.

\section{Symmetry class \texorpdfstring{$\textrm{CI}$}{CI}}
\label{sec:CI_class}

The symmetry class $\textrm{CI}$ is characterized by spinless time-reversal symmetry ($\mathcal{T}^2 = +1$) and by a particle-hole symmetry ($\mathcal{C}^2 = -1$). 
The latter enforces the Hamiltonian to have an even dimension, and their composition gives the chiral symmetry ($\mathcal{S}=\mathcal{T}\mathcal{C}$).
We represent 
\begin{subequations}
\label{eqn:CI-AZ-sym-representation}
\begin{eqnarray}
\mathcal{T} &=& \sigma_x \otimes \mathbb{1}_N \,\mathcal{K} \\    
\mathcal{C} &=& -\imi \sigma_y  \otimes \mathbb{1}_N  \,\mathcal{K} \\
\mathcal{S} &=& \sigma_z \otimes \mathbb{1}_N,
\end{eqnarray}
\end{subequations}
where the first matrix encodes all additional degrees of freedom and specifies the Hamiltonian size as $2N\times 2N$. 
The chiral symmetry implies a block-off-diagonal form of the Hamiltonian, analogous to Eq.~(\ref{eqn:block-off-diag-Ham}) for the class $\textrm{AIII}$, and time-reversal symmetry further enforces that $h = h^\top$ is a symmetric (but complex) matrix.

We begin by adapting the von Neumann-Wigner codimension counting. 
First, using the rank-factorization theorem, we find that the space of symmetric complex $N \times N$ rank-$r$ matrices has (real) dimension 
\begin{subequations}
\begin{equation}
\label{eqn:CI-Ham-dim}
d^\textrm{CI}_{(N,N-r)} = 2Nr-r(r-1)
\end{equation}
The reason is that any complex symmetric matrix can be rank-factorized as $h = A A^\top$ with $A \in \cmplx^{N\times r}$, which corresponds to $2Nr$ real parameters, with non-uniqueness specified by the replacement $A \mapsto \tilde{A} = A U$ where $U\in \cmplx^{r \times r}$ obeys $U U^\top = \mathbb{1}_r$.
The conditions on $U$ specify it as an element of the \emph{complex orthogonal group}, $\mathsf{O}(r,\cmplx)$; its dimension is given by
\begin{equation}
\dim[\mathsf{O}(r,\cmplx)] = 2r^2 - 2r - r(r-1) = r(r-1)
\end{equation}
where the first term specifies the number of (real) matrix components of $U$, the second corresponds to the normalization of the individual columns of $U$, and the third term encodes the orthogonality of each pair of distinct columns of $U$.
Equation~(\ref{eqn:CI-Ham-dim}) follows by combining the parameters in $A$ with the freedom in $U$.
Substituting through the rank-nullity relation, $r = N-n$, we obtain
\begin{equation}
d^\textrm{CI}_{(N,n)} = (N-n)(N+n+1)
\end{equation}
Therefore, the codimension of forming a $(2n)$-fold zero-energy degeneracy in the class $\textrm{CI}$ is
\begin{equation}
\delta_{(n)}^\textrm{CI} = d^\textrm{CI}_{(N,0)}-d^\textrm{CI}_{(N,n)} = n^2 + n.
\end{equation}
\end{subequations}
In particular, $\delta_{(1)}^\textrm{CI} = 2$, meaning that an effective two-band model is expanded into two matrices.
This result implies the existence of nodal lines in three-dimensional crystals in symmetry class $\textrm{CI}$, which are topologically stabilized by the winding number $\pi_1(S^1) = \intg$~\cite{Bzdusek:2017}.

We further investigate the topological characterization of the multifold degeneracy with $n=2$, whose minimal model is given by a  $4\times 4$ Hamiltonian.
We find that the general such Hamiltonian can be conveniently decomposed using six Dirac matrices as
\begin{equation}
\label{eq:Hamiltonian_CI}
H^\textrm{CI} = \bs{a} \cdot \bs{V} + \bs{b} \cdot \bs{W}  
\end{equation}
where $\bs{a}=(a_1,a_2,a_3)$ and $\bs{b}=(b_1,b_2,b_3)$ parameterize the momentum space, and
\begin{subequations}
\begin{eqnarray}
\bs{V}&=&(
 \sigma_x \otimes \tau_z,
-\sigma_x \otimes \tau_x,
 \sigma_y \otimes \mathbb{1}_\tau
) \\
\bs{W}&=&(
-\sigma_y \otimes \tau_z, 
 \sigma_y \otimes \tau_x,
 \sigma_x \otimes \mathbb{1}_\tau
) 
\end{eqnarray}
are collections of Dirac matrices such that 
\begin{eqnarray}
\{V_a,V_b\}=\{W_a,W_b\} &=& 2\delta_{ab} \\
\textrm{and}\quad [V_a,W_b]&=&-2\imi \delta_{ab} \mathcal{S}
\end{eqnarray}
\end{subequations}
It can be shown that the spectrum of such a Hamiltonian is
\begin{equation}
\varepsilon = \pm \sqrt{\bs{a}^2 + \bs{b}^2 \pm 2 \lVert \bs{a}\times \bs{b} \rVert}    
\end{equation}
where $\lVert\bs{x}\rVert$ is the norm of $\bs{x}$ and we write $\bs{x}^2 = \lVert \bs{x}\rVert^2$ for the norm squared.

We consider the Hamiltonian~(\ref{eq:Hamiltonian_CI}) on the sphere 
\begin{equation}
\label{eqn:CI-S5}
S^5 = \{ (\bs{a},\bs{b})\,|\,\bs{a}^2 + \bs{b}^2 = 1\} \subset \reals^6.
\end{equation}
On this sphere, $\tr[H^2] = \sum_{j=1}^4 \varepsilon_j^2 = 4$ is constant.
In analogy with the other symmetry classes, we define the nodal manifolds $M_\medcirc \subset S^5$ and $M_\varnothing \subset S^5$ as the intersections of the degeneracy loci with the enclosing sphere following the general prescription in Eqs.~(\ref{eqn:chiral-loci}). 
Our next goal is to characterize both nodal manifolds and to equip them with explicit coordinates.

The case $M_\medcirc$ corresponds to requiring $\bs{a}^2 + \bs{b}^2 = 2\lVert \bs{a}\times \bs{b} \rVert$. 
This is possible only if simultaneously $\rVert \bs{a} \lVert = \rVert \bs{b} \lVert$ and $\bs{a}\perp \bs{b}$. 
The first condition, in combination with Eq.~(\ref{eqn:CI-S5}), fixes the magnitude of both vectors to ${1}/{\sqrt{2}}$. 
Furthermore, the orthogonality condition implies that $\bs{a}$ can be chosen as $\bs{n}/\sqrt{2}$ with any unit vector $\bs{n}\in S^2$, together with $\sqrt{2}\bs{b}$ a normalized vector in the tangent space $\mathrm{T}_{\bs{n}}S^2$. 
Therefore, we recognize $M_{\medcirc}$ as the \emph{unit tangent bundle} $\mathrm{UT}S^2$. 
Interestingly, this is the same as the $\mathsf{SO}(3)$ space of orthonormal $3$-frames, since we can use $\bs{a}$ and $\bs{b}$ to construct the frame $\mathfrak{e}=(\sqrt{2}\bs{a},\sqrt{2}\bs{b},2\bs{a}\times\bs{b})$. 
Using standard identifications, we therefore recognize $M_\medcirc$ as either of $\mathsf{SO}(3) \simeq \reals P^3 \simeq S^3/\ztwo$. 
The same result can be derived from the general formula in Appendix~\ref{app:class-space-w-zero-states} since $M_\medcirc\simeq\mathsf{U(2)}/\mathsf{U}(1)\times \mathsf{O}(2)\simeq S^3/\ztwo$.
We can parameterize $M_\medcirc$ explicitly as 
\begin{subequations}
\begin{align}
a_1 &= \sin \theta \cos \phi / \sqrt{2}, \\ 
a_2 &= \sin \theta \sin \phi /\sqrt{2}, \\ 
a_3 &= \cos \theta /\sqrt{2}, \\ 
b_1 &= (- \cos \chi \cos \theta \cos \phi + \sin \chi \sin \phi) /\sqrt{2}, \\ 
b_2 &= (- \cos \chi \cos \theta \sin\phi - \sin \chi \cos \phi) /\sqrt{2}, \\ 
b_3 &= \cos \chi \sin \theta /\sqrt{2}.
\end{align}
\end{subequations}
where $\phi\in[0,2\pi]$ and $\theta = [0,\pi]$ are using $2$-spherical coordinates characterizing $\bs{a}$, while $\chi \in [0,2\pi]$ specifies the orientation of $\bs{b}$ in the tangent plane $T_{\bs{a}} S^2$.

To reveal the linking of $M_\medcirc$ and $M_\varnothing$, we compute the first Stiefel-Whitney invariant on the $1$-cell of $M_\medcirc$. To do this, we fix $\theta = \pi/2 = - \chi$, yielding 
\begin{subequations}
\begin{align}
    \bs{a} &= (\cos\phi, \sin \phi,0)/\sqrt{2}, \\
    \bs{b} &= (-\sin \phi, \cos \phi, 0)/\sqrt{2}.
\end{align}
On the nontrivial $1$-cycle, the off-diagonal block of the Hamiltonian in Eq.~(\ref{eqn:block-off-diag-Ham}) reduces to 
\begin{equation}
    h = \sqrt{2} e^{-\imi \phi} u u^T, \quad u = \frac{1}{\sqrt{2}} \begin{pmatrix}
        1 \\ -\imi
    \end{pmatrix}.
\end{equation}
Using this structure of the off-diagonal block, the positive and negative energy eigenstates can be conveniently written as 
\begin{equation}
    \psi_\pm(\phi) = \frac{1}{\sqrt{2}} \begin{pmatrix}
        \pm e^{-\imi \phi} u \\
        u^*
    \end{pmatrix}.
\end{equation}
\end{subequations}
A straightforward calculation yields that along the nontrivial $1$-cycle of $M_\medcirc$ the Berry phase of these states is nontrivial. 
This result signals that this $1$-cycle is linked with the $3$-cycle~of~$M_\varnothing$.

Remarkably, the nontriviality of the Berry phase of the non-degenerate bands determines the Stiefel-Whitney invariants of the degenerate $0$-energy band on $M_\medcirc$. 
We follow the strategy outlined in Sec.~\ref{sec:class-AI} for AI class. On $M_\medcirc\simeq \reals P^3$, the bundles are spanned by the bands of the Hamiltonian~(\ref{eq:Hamiltonian_CI}). 
Let us denote the bundle spanned by the negative and positive energy non-degenerate bands as $E^{(-)}$ and $E^{(+)}$, and for the rank-2 bundle spanned by the zero energy bands, we use $E^{(0)}$. 
The cohomology ring of $\reals P^3$ is~\cite{Hatcher_book}
\begin{subequations}
\begin{equation}
H^*(\reals P^3,\ztwo) = \ztwo[a]/(a^4).
\end{equation}
The symbol on the right-hand side indicates sums of the form $w = \beta_0 + \beta_1 a + \beta_2 a^2+\beta_3 a^3$ with $\beta_i \in \ztwo = \{0,1\}$, where $a$ is the generator of $H^1(\reals P^3,\ztwo)$, $a^2$ is the generator of $H^2(\reals P^3,\ztwo)$, $a^3$ is the generator of $H^3(\reals P^3,\ztwo)$, and $a^4 = 0$ vanishes. 
The nontriviality of the Berry phases on rank-$1$ bundles $E^{(-)}$ and $E^{(+)}$, as well as the triviality of the total bundle $E^{\mathrm{tot}}=E^{(-)}\oplus E^{(0)}\oplus E^{(+)}$ give us the following relations on the total Stiefel-Whitney classes
\begin{gather}
    w(E^{(-)})=w(E^{(+)})=1+a\\ 
    w(E^{(0)})\smile w(E^{(-)}\oplus E^{(+)}) 
    =1.
\end{gather}
From the first equation, we have
\begin{equation}
w(E^{(-)}\oplus E^{(+)})  = 1 + a^2.
\end{equation}
Then, by inverting the second relation, we obtain the total Stiefel-Whitney class of $E^{(0)}$ as
\begin{equation}
    w(E^{(0)})=(1+a^2)^{-1}=1+a^2.
\end{equation}
\end{subequations}
Thus, we have found that the second Stiefel-Whitney number of the bundle $E^{(0)}$ is nontrivial, indicating the linking of $2$-cells of $M_{\medcirc}$ and $M_{\varnothing}$.

The case $M_\varnothing$ corresponds to imposing $ \bs{a}\times \bs{b}  = \bs{0}$, implying the parallelism $\bs{a} \parallel \bs{b}$. 
We therefore write $\bs{a} = a \bs{n}$ and $\bs{b} = b \bs{n}$ with $\bs{n} \in S^2$ a unit vector. 
It follows from Eq.~(\ref{eqn:CI-S5}) that we can define $a= \cos\chi$ and $b = \sin\chi$ with $\chi \in [0,2\pi]\simeq S^1$. 
Using further the spherical coordinates $\varphi\in[0,2\pi]$ and $\theta\in[0,\pi]$ to characterize the unit vector, we can parameterize $M_\varnothing$~as 
\begin{subequations}
\begin{eqnarray}
\label{eq:class-CI-non-zero-parametrization}
a_1 = \cos \chi \sin \theta \cos \phi &\quad& 
b_1 = \sin \chi \sin \theta \cos \phi \quad \\
a_2 = \cos \chi \sin \theta \sin \phi &\quad& 
b_2 = \sin \chi \sin \theta \sin \phi \quad \\
a_3 = \cos \chi \cos \theta  &\quad & 
b_3 = \sin \chi \cos \theta . \quad 
\end{eqnarray}
\end{subequations}
However, there is a redundancy in this description; namely, changing $(\chi,\bs{n})\mapsto (\chi+\pi,-\bs{n})$ reproduces the same point $(\bs{a},\bs{b})$. 
Therefore, we conclude that
\begin{equation}
M_\varnothing \simeq (S^1 \times S^2)/\ztwo    
\end{equation}
where $\ztwo$ acts as a simultaneous antipodal map on $S^1$ and $S^2$. 
Low-dimensional topology of this manifold was considered in the context of non-Hermitian Hamiltonians in Ref.~\citenum{Wojcik:2020}. 
One can equivalently express $M_\varnothing \simeq \mathsf{U}(2)/\mathsf{O}(2)$, i.e., as the space of spectrally flattened class-$\textrm{CI}$ $4$-band Hamiltonians with an energy gap at half-filling. 
Crucially, the manifold exhibits cells in all dimensions $D\in \{0,1,2,3\}$.

We next consider band invariants on $M_\varnothing$.
To first diagnose the linking of the $1$-cell in $M_{\varnothing}$ with the $3$-cell in $M_{\medcirc}$, we investigate the winding number on the $1$-cell. 
To obtain the winding number along the $1$-cell, we fix $\theta = \phi = 0$ and vary $\chi \rightarrow \chi + 2 \pi$. The Hamiltonian reduces to $\boldsymbol{a} = (0,0,\cos \chi)$ and $\boldsymbol{b} = (0, 0, \sin \chi)$. 
Hence, the off-diagonal block of the Hamiltonian reduces to 
\begin{subequations}
\begin{equation}
    h(\chi) = \begin{pmatrix}
        -\imi e^{\imi\chi} & 0 \\ 0 & -\imi e^{\imi \chi}
    \end{pmatrix}.
\end{equation}
Accounting for the fact that the chosen loop covers the nontrivial cycle twice we have 
\begin{equation}
    2 \times \nu = \frac{1}{2 \pi i} \int_0^{2 \pi} d\chi ~ \partial_\chi \log \det h(\chi) = 2. 
\end{equation}
\end{subequations}

Furthermore, we reveal that the second Stiefel-Whitney invariant is nontrivial on the $2$-cell of $M_\varnothing$ defined by fixing any value of $\chi$ and allowing $\theta \in [0,\pi]$ and $\phi \in [0,2\pi]$ to change. 
We compute the second Stiefel-Whitney invariant by inspecting the Wilson loop flow. 
For each Wilson loop, we fix $\theta $ and find the Wilson loop matrix by using the rotating-frame trick along $\phi$. 
The invariant is deduced by counting how many times the phase of the Wilson loop eigenvalues crosses the value $\pi$. 
 
To proceed with this calculation, observe that on $M_\varnothing$ any normalized vector $v \in \mathbb{C}^2$ defines a negative energy eigenstate of the Hamiltonian via the construction 
\begin{subequations}
\begin{equation}
    \psi = \frac{1}{\sqrt{2}} \begin{pmatrix}
        -h v \\ v
    \end{pmatrix}, \quad H \psi = - \psi
\end{equation}
owing to the offdiagonal structure of the Hamiltonian and $h^\dagger h = \mathbb{1}$, with $h$ being the offdiagonal block 
\begin{equation}
    h \!=\! e^{\imi \chi } \!
    \begin{pmatrix}
    \sin \theta \cos\phi - \imi \cos\theta \! & -\sin\phi \sin\theta \\ 
    - \sin \phi \sin \theta & \! -\sin \theta \cos\phi - \imi \cos\theta
    \end{pmatrix}
\end{equation}
Therefore, we can define an orthonormal frame for the negative energy subspace as 
\begin{equation}
    \Psi_-=\frac{1}{\sqrt{2}} \begin{pmatrix}
        -h \\ \mathbb{1}
    \end{pmatrix}.
\end{equation}
Using the parametrization of $M_\varnothing$ in Eq.~(\ref{eq:class-CI-non-zero-parametrization}), the non-Abelian Berry connection of the negative energy subspace (defined as $\mathcal{A_\phi} = \imi \Psi_-^\dagger \partial_\phi \Psi_-$) is 
\begin{equation}
\label{eqn:CI-2cell-connection}
    \mathcal{A}_\phi = \frac{1}{2} \left[ s^2 \tau_y 
    +s c \left( \cos(\phi) \tau_x + \sin(\phi) \tau_z \right) \right]
\end{equation}
\end{subequations}
where we used $s = \sin(\theta)$ and $c = \cos(\theta)$ for brevity.

Note that we cannot directly integrate the connection~(\ref{eqn:CI-2cell-connection}) to get the Wilson loop matrix [defined in Eq.~(\ref{eq:Wilson-matrix-definition})] due to the fact that $\left[\mathcal{A}_\phi(\phi_1), \mathcal{A}_\phi(\phi_2) \right] \neq 0$ for general 
$\phi_{1,2}$. 
Instead, we solve the parallel-transport equation 
\begin{subequations}
\begin{equation}
\label{eq:parallel-transport-ODE}
    \partial_\phi U(\phi) = -\imi  U(\phi) \mathcal{A}_\phi
\end{equation}
where $U(0) = \mathbb{1}$ and $U(2\pi) = W(\theta)$. 
To do this, we recognize that the $\phi$ dependence of the connection is of a special form
\begin{equation}
    \mathcal{A}_\phi(\phi) = R^{\dagger}(\phi) \mathcal{A}_\phi(0) R(\phi)
\end{equation}
where 
\begin{equation}
    \mathcal{A}_\phi(0) = \frac{1}{2} \left[ s^2 \tau_y + sc \tau_x \right]  \;\;\; \textrm{and} \;\;\; R(\phi) = e^{-\imi \phi \tau_y /2}
\end{equation}
The rotating-frame trick allows us to write the Ansatz 
\begin{equation}
    U(\phi) = e^{-\imi \phi G} R(\phi)
\end{equation}
where $G$ is a constant unknown matrix. We plug this Ansatz back into Eq.~(\ref{eq:parallel-transport-ODE}) and compare the two sides to deduce that 
\begin{equation}
    G = \mathcal{A}_\phi(0) - \frac{\tau_y}{2} = \frac{c}{2} \left[ s \tau_x - c \tau_y \right].
\end{equation}
With this, we identified the general solution 
\begin{equation}
    U(\phi) = e^{\frac{-\imi \phi c}{2} \left[s \tau_x - c \tau_y \right]} e^{-\imi \phi \tau_y/2} .
\end{equation}
At $\phi= 2\pi$, we have $e^{-\imi \pi \tau_y} = -\mathbb{1}_2$; therefore, the Wilson loop matrix is 
\begin{equation}
    W(\theta) = - \textrm{exp}\left(- \imi \pi c \left[s \tau_x - c \tau_y \right] \right).
\end{equation}
\end{subequations}
The eigenvalues of $W(\theta)$ are $\lambda_\pm(\theta) = - e^{\mp i \pi \cos(\theta)}$ with phases $\nu_\pm(\theta) = \pm \pi(1-\cos(\theta))$. For $\theta \in [0,\pi]$, $\nu_{+}(\theta)$ increases from $0$ to $2\pi$, crossing $\pi$ once at $\theta = \pi/2$. 
The Wilson loop flow therefore has an odd parity, implying that the second Stiefel-Whitney invariant on the $2$-cell of $M_\varnothing$ is nontrivial.

Before concluding, we point out a relation of our discussion above to the finding of Ref.~\citenum{Kim:2021} that zero-energy class-$\textrm{CI}$ nodal rings in 3D that carry a nontrivial second Stiefel-Whitney invariant on the enclosing $2$-sphere $S^2 \subset \reals^3$ are necessarily linked with nodal lines in an adjacent energy gap. The situation is, in fact, exactly analogous to the case of the symmetry class $\textrm{AI}$~\cite{Ahn:2018}, with the sole difference being that the additional chiral symmetry pins the considered nodal rings to $\varepsilon = 0$. 
Correspondingly, the relation to our present analysis proceeds in analogy with the discussion at the end of Sec.~\ref{sec:class-AI}.

\section{Symmetry class \texorpdfstring{$\textrm{DIII}$}{DIII}}
\label{sec:class-DIII}

The symmetry class $\textrm{DIII}$ is characterized by spinful time-reversal symmetry ($\mathcal{T}^2 = -1$) and by a particle-hole symmetry ($\mathcal{C}^2 = +1$). 
The two symmetries together enforce the Hamiltonian to be of dimension $4N \times 4N$ (over complex numbers), and their composition gives the chiral symmetry ($\mathcal{S}=\mathcal{T}\mathcal{C}$).
We represent 
\begin{subequations}
\label{eqn:DIII-AZ-sym-representation}
\begin{eqnarray}
\mathcal{T} &=& \imi \sigma_y \otimes \mathbb{1}_{2N} \,\mathcal{K} \\    
\mathcal{C} &=& \sigma_x  \otimes \mathbb{1}_{2N}  \,\mathcal{K} \\
\mathcal{S} &=& \sigma_z \otimes \mathbb{1}_{2N}.
\end{eqnarray}
\end{subequations}
Despite the spinful time-reversal symmetry, and in contrast to the discussion of classes $\textrm{AII}$ and $\textrm{CII}$, we do not utilize quaternion numbers to discuss the class $\textrm{DIII}$.

We begin with adapting the codimension counting argument to the class $\textrm{DIII}$.
Owing to chiral symmetry, the Hamiltonian matrix $H$ takes the block-off-diagonal form of Eq.~(\ref{eqn:block-off-diag-Ham}) with blocks $h$ and $h^\dagger$ of dimension $2N\times 2N$.
Time-reversal further implies that the off-diagonal block is skew-symmetric, $h = - h^\top$. 
Generally, a skew-symmetric $2N\times 2N$ matrix $h$ can be expressed as
\begin{subequations}
\begin{equation}
\label{eqn:DIII-offblock-decomposition}
h = \Psi\cdot\Sigma\cdot \Psi^\top
\end{equation}
where $\Psi \in \mathsf{U}(2N)$ and $\Sigma$ is a diagonal matrix of $2\times 2$ blocks as listed in Eq.~(\ref{eqn:class-D-canonical-diag})~\cite{Youla:1961}. 
For simplicity, we assume that the non-zero eigenvalues in $\Sigma$ are all distinct.
Then, to understand the non-uniqueness of the decomposition~(\ref{eqn:DIII-offblock-decomposition}), we first study the stabilizer [as a subgroup of $\mathsf{U(2)}$] of a single off-diagonal block. 
Since the block is proportional to the symplectic form $\imi \sigma_y$, this subgroup is exactly $\mathsf{U}(2)\cap \mathsf{Sp}(2,\cmplx) = \mathsf{Sp}(1)$. 
Furthermore, in the presence of $n$ zero blocks in $\Sigma$ (which physically corresponds to a $4n$-fold zero-energy degeneracy of $H$), the gauge freedom over the zero block is enhanced to \emph{all} $\mathsf{U}(2n)$ matrices. 
Therefore, the dimension of rank $r=(2N-2n)$ skew-symmetric $\cmplx^{2N \times 2N}$ matrices $h$ equals
\begin{eqnarray}
d^\textrm{DIII}_{(N,n)} &=& \dim[\mathsf{U}(2N)]+(N-n) \nonumber \\
&\phantom{=}& \quad - (N-n)\dim[\mathsf{Sp}(1)]-\dim[\mathsf{U}(2n)] \nonumber \\
&=& 4N^2 - 2(N-n) - 4n^2.
\end{eqnarray}
In the first line of the above, the first term corresponds to the freedom in $\Psi$, the second term specifies the non-zero eigenvalues, the third term corresponds to the gauge freedom in the non-zero blocks, and the last term represents the freedom in the zero block.
From here, the sought node codimension is 
\begin{equation}
\delta^\textrm{DIII}_{(n)} =  d^{\textrm{DIII}}_{(N,0)} - d^{\textrm{DIII}}_{(N,n)} = 2n(2n-1).   
\end{equation}
\end{subequations}
In particular, $\delta_{(1)}^\textrm{DIII} = 2$, meaning that an effective four-band model is expanded into two Dirac matrices.
This result implies the existence of fourfold degenerate nodal lines in three-dimensional crystals in symmetry class $\textrm{DIII}$ that are topologically stabilized by the winding number $\pi_1(S^1) = \intg$~\cite{Bzdusek:2017}.

In the following, we focus on the multifold degeneracy with $n=2$. This corresponds to an eightfold degeneracy, and the minimal Hamiltonian corresponds to an $8 \times 8$ matrix.
To analyze the minimal model, we observe that it can be decomposed using twelve Dirac matrices as
\begin{subequations}
\begin{equation}
\label{eqn:DIII-Hamiltonian}
H^\textrm{DIII} = \bs{a} \cdot \bs{V} + \bs{b} \cdot \bs{W}  
\end{equation}
where 
\begin{eqnarray}
\bs{a} &=& (a_1,a_2,a_3, a_4, a_5, a_6) \\
\textrm{and}\;\;  
\bs{b}&=&(b_1,b_2,b_3,b_4,b_5,b_6)
\end{eqnarray}
\end{subequations}
are collections of real-valued parameters and
\begin{subequations}
\begin{eqnarray}
\bs{V}{=}\begin{pmatrix}
\sigma_x\otimes \mathbb{1}_\tau\otimes\mu_y\\
\sigma_y\otimes \tau_x\otimes\mu_y\\
\sigma_x\otimes \tau_y\otimes\mu_x \\
\sigma_x\otimes \tau_y\otimes\mu_z\\
\sigma_y\otimes \tau_y\otimes\mathbb{1}_\mu\\
\sigma_y\otimes \tau_z\otimes\mu_y
\end{pmatrix},\quad
\bs{W}{=}\begin{pmatrix}
\sigma_y\otimes \mathbb{1}_\tau\otimes\mu_y\\
-\sigma_x\otimes \tau_x\otimes\mu_y\\
\sigma_y\otimes \tau_y\otimes\mu_x\\
\sigma_y\otimes \tau_y\otimes\mu_z\\
-\sigma_x\otimes \tau_y\otimes\mathbb{1}_\mu\\
-\sigma_x\otimes \tau_z\otimes\mu_y
\end{pmatrix}
\end{eqnarray}
are collections of Dirac matrices such that 
\begin{eqnarray}
\{V_a,V_b\} &=&\{W_a,W_b\} = 2\delta_{ab} \quad \\{}
[ V_a , W_b ] &=& 2\imi \delta_{ab}\mathcal{S}\quad \\
\textrm{and for $a\neq b$:}\quad V_a W_b &=&-W_a V_b, \quad 
\end{eqnarray}
\end{subequations}
where the commutation relations follow from the properties $W_a=-\imi\mathcal{S}V_a$ and $\{\mathcal{S},V_a\}=0$, similarly to class $\mathrm{AIII}$ (see Sec.~\ref{sec:class-AIII}).

Using these properties, one can show that the spectrum of the Hamiltonian~(\ref{eqn:DIII-Hamiltonian}) is
\begin{equation}
\varepsilon = \pm \sqrt{\bs{a}^2 + \bs{b}^2 \pm 2 \lVert \bs{a}\wedge \bs{b} \rVert}    
\end{equation}
where the norm of the $2$-vector $\bs{a}\wedge \bs{b}$ corresponds to the norm of a vector consisting of all the $(6 \times 5)/2 = 15$ independent coefficients components of the $2$-vector \footnote{The norm of a $2$-vector can be equivalently written as the determinant of the Gram matrix, which gives $\lVert \bs{a}\wedge \bs{b} \rVert^2=\lVert\bs{a}\rVert^2 \lVert\bs{b}\rVert^2-|\bs{a}\cdot\bs{b}|^2$~\cite{Greub:1978}.}. We have also numerically checked the choice of $\bs{V}$ and $\bs{W}$ together with the analytical formula for the spectrum~\cite{supplementary_code_data}.

We consider the Hamiltonian on the sphere 
\begin{equation}
\label{eqn:DIII-S}
S^{11} = \{ (\bs{a},\bs{b})\,|\,\bs{a}^2 + \bs{b}^2 = 1\} \subset \reals^{12}.
\end{equation}
The sum of the squared eigenenergies is constant on this sphere: $\sum_{j=1}^8 \varepsilon_j^2 = 8$.
In analogy with the formerly discussed symmetry classes, we define the degeneracy loci by adapting Eq.~(\ref{eqn:chiral-loci}), and we construct the nodal manifolds $M_\medcirc \subset S^5$ and $M_\varnothing \subset S^5$ as the intersections of the degeneracy loci with the enclosing sphere.
Our goal is to inspect the linking of $M_\medcirc$ with $M_\varnothing$ by considering topological band invariants over their nontrivial cycles.

The case of $M_\medcirc$ requires $\bs{a}^2 + \bs{b}^2 = 2\lVert \bs{a} \wedge \bs{b}\rVert$, which translates to the simultaneous conditions $\lVert \bs{a} \rVert = \lVert \bs{b} \rVert$ and $\bs{a}\perp \bs{b}$.
The first condition, in combination with Eq.~(\ref{eqn:DIII-S}), fixes the magnitude of both six-component vectors to $\bs{n}/\sqrt{2}$ with $\bs{n}\in S^5$, together with $\bs{b}$ a normalized vector in the tangent space $\mathrm{T}_{\bs{n}}S^5$. 
Therefore, we recognize $M_\medcirc$ as the unit tangent bundle $\mathrm{UT}S^5$. 
We can identify this manifold by constructing a $2$-frame from the vectors $\bs{a}$ and $\bs{b}$. The unit tangent bundle is obtained by varying the vectors $\bs{a}$ and $\bs{b}$; hence, it is the space of orthonormal 2-frames in $\reals^6$, which corresponds to the Stiefel manifold: 
\begin{equation}
M_\medcirc \simeq \mathsf{O}(6)/\mathsf{O}(4).    
\end{equation}
The same result follows from the general formula in Appendix~\ref{app:class-space-w-zero-states}, together with the chain of standard diffeomorphisms $\mathsf{U}(4)/\mathsf{U}(2)\times \mathsf{Sp}(1)\simeq \mathsf{SU}(4)/\mathsf{SU}(2)\times \mathsf{SU}(2)\simeq \mathsf{Spin}(6)/\mathsf{Spin}(4)\simeq \mathsf{O}(6)/\mathsf{O}(4)$~\cite{Totaro:2002}.
The bundle structure $M \simeq \mathrm{UT}S^5$
of the Stiefel manifold implies that nontrivial cycles exist in dimensions $D\in\{0,4,5,9\}$. 
We anticipate that the $4$-cell should carry the second Chern number, as this provides a natural spinful counterpart of the second Stiefel-Whitney invariant identified for the $2$-cell of $M_\medcirc$ for the spinless class $\textrm{CI}$.
However, due to the lack of a convenient parametrization, we did not proceed with an explicit verification of this statement.

The case $M_\varnothing$ corresponds to requiring $ \lVert \bs{a} \wedge \bs{b}\rVert =0$, which translates to the parallelism $\bs{a}\parallel \bs{b}$. 
We therefore write $\bs{a} = a \bs{n}$ and $\bs{b} = b \bs{n}$ with $\bs{n} \in S^5$ a unit vector. 
It follows from Eq.~(\ref{eqn:DIII-S}) that we can define $a= \cos\chi$ and $b = \sin\chi$ with $\chi \in [0,2\pi]\simeq S^1$. 
However, there is a redundancy in this description; namely, changing $(\chi,\bs{n})\mapsto (\chi+\pi,-\bs{n})$ reproduces the same point $(\bs{a},\bs{b})$. 
Therefore, we conclude that
\begin{equation}
M_\varnothing \simeq (S^1 \times S^5)/\ztwo    
\end{equation}
where $\ztwo$ acts as a simultaneous antipodal map on $S^1$ and $S^5$. 
The nodal manifold $M_\varnothing$ can be equivalently expressed as the classifying space of class-$\textrm{DIII}$ eight-band Hamiltonian with a gap at half-filling, $M_\varnothing \simeq \mathsf{U}(4)/\mathsf{Sp}(2)$. 
The nodal manifold $M_{\varnothing}$ clearly has cells in dimensions $D\in\{0,1,5,6\}$ that can potentially be linked with cells of $M_{\medcirc}$.

We diagnose the linking of $1$-cell in $M_\varnothing$ and $9$-cell in $M_{\medcirc}$ by calculating the winding number on the $1$-cell.
The calculation of the winding number closely mimics the one presented for the class $\textrm{CI}$ in Sec.~\ref{sec:CI_class}, with one technical difference that the unit vector $\boldsymbol{n}$ presently has six (rather than three) components. 
Thus, we consider the Hamiltonian $H = a \, \boldsymbol{n} \cdot \boldsymbol{V}+ b \, \boldsymbol{n} \cdot \boldsymbol{W}$ with $a = \cos \chi$, $b = \sin \chi$, and $\boldsymbol{n} = (0,0,0,0,0,1)$, and we vary $\chi \rightarrow \chi + 2\pi$. 
With this choice, the Hamiltonian~(\ref{eqn:DIII-Hamiltonian}) reduces to $H = \cos \chi V_6 + \sin \chi W_6$. The off-diagonal block of the Hamiltonian becomes 
\begin{equation}
    h(\chi) = -\imi e^{-\imi\chi} \cdot \sigma_z \otimes \sigma_y
\end{equation}
and the winding number is
\begin{equation}
    2\times \nu = \frac{1}{2 \pi \imi} \int_{0}^{2\pi} d\chi \, \partial_\chi \log \det q(\chi) = -4, 
\end{equation}
where we took into account the fact that the chosen loop traverses the nontrivial $1$-cycle twice.
Contrasting this result with that of Sec.~\ref{sec:CI_class}, we observe a doubling in the winding number, which is due to the spinful nature of $\mathcal{T}$ for class $\textrm{DIII}$ compared to the spinless time-reversal symmetry of class $\textrm{CI}$.

\section{Summary and outlooks}
\label{sec:conclude}

In this work, we have developed a topological characterization of generic $n$-fold band degeneracies whose stability derives solely from Altland-Zirnbauer (AZ) symmetries acting locally in momentum space.
As a first step, we derived the codimension of forming such multifold band degeneracies in a general parameter space [Table~\ref{tab:codims-WD+balanced}].
By adapting the von Neumann-Wigner counting~\cite{vonNeumann:1929} from the three Wigner-Dyson classes to the ten AZ classes, we find that the codimension of forming an $n$-fold degeneracy generally grows quadratically with the order $n$.
Generic multifold nodes are thus naturally situated in enlarged parameter spaces that supplement momenta with tuning parameters, pumping cycles, or synthetic dimensions.
From this perspective, the previously studied crystalline-symmetry-protected multifold fermions, such as those considered in Refs.~\cite{Young:2012,Yang:2015b,Armitage:2018,Wieder:2016,Bradlyn:2016,Zhu:2016,Lenggenhager:2022a,Yu:2022}, correspond to symmetry-constrained slices of this higher-dimensional universal setting.

Second, and more fundamentally, we investigated topological invariants that characterize the robustness of the $n$-fold band nodes.
This task, which has been previously carried out only for minimal band degeneracies (i.e., the first non-zero entry in each row of Table~\ref{tab:codims-WD+balanced})~\cite{Bzdusek:2017}, faces a nontrivial obstruction: any sphere used to enclose such $n$-fold degeneracy lacks a uniform spectral gap. 
Correspondingly, beyond the minimal case, the topological stability of $n$-fold nodes cannot be captured by homotopy groups of the known classifying spaces of gapped Hamiltonians~\cite{Kitaev:2009,Ryu:2010,Lundell:1992}.
To overcome this challenge, we developed a mathematical characterization that elevates the obstruction into a marker of the multifold band degeneracy.
Specifically, complementary gap conditions are recovered on two nodal manifolds (labeled $M_1$ and $M_2$ in the Wigner-Dyson classes, resp.~$M_\medcirc$ and $M_\varnothing$ in the chiral and BdG classes) defined as the intersection of $(n\,{-}\,1)$-fold degeneracy loci with the enclosing sphere [Fig.~\ref{fig:main_scheme}].
We have established a two-way correspondence, where (1)~the presence of an $n$-fold degeneracy is encoded in robust linking of the two nodal manifolds; in turn, (2)~linking is encoded in conventional topological invariants (such as Chern numbers, Stiefel-Whitney invariants, and winding numbers) evaluated on nontrivial cycles of the~nodal~manifolds. 

The codimension counting ensures that, away from fine-tuned situations, the linking of the enclosing nodal manifolds is robust, and their band invariants characterize the protection of the multifold nodal point. 
Nevertheless, some care is required. 
In non-generic cases, the linking detected by the band invariants may be unstable and disappear under an arbitrarily small perturbation of the Hamiltonian, as discussed in Appendix~\ref{app:unstable-linking}. 
Even when the linking is robust, its particular configuration may still be accidental and change through deformations involving cobordisms, as analyzed in Appendix~\ref{app:cobordism-transform}. 
The latter possibility also points toward a broader formulation of the correspondence in terms of cobordism invariants, to which we return at the end of this section.

Following the general mathematical formulation in Sec.~\ref{sec:general-section} that discusses the linking of higher-dimensional manifolds, as well as its relation to topological band invariants computed from Bloch bundles on cycles of the nodal manifolds, we specifically focus on the first nontrivial instance of multifold band degeneracies (i.e., the second non-zero entry in each row of Table~\ref{tab:codims-WD+balanced}).
In Secs.~\ref{sec:class-AI}--\ref{sec:class-DIII}, we therefore construct the minimal Hamiltonian of such multifold degeneracy for each of the ten AZ symmetry classes. 
These models are `minimal' in the sense that they have the smallest number of bands that allows for the formation of such $n$-fold degeneracy, and they are generally decomposable into a suitably adapted set of Gell-Mann matrices (Secs.~\ref{sec:class-AI}--\ref{sec:class-AII}) or Dirac matrices (Secs.~\ref{sec:class-BDI}--\ref{sec:class-DIII}).
For each symmetry class, we present analytical arguments that allow us to identify the nodal manifolds and formulate their homological cell-decomposition [third and fourth columns in Table~\ref{tab:nodal_manifolds}].
Using explicit parameterization (achieved for all classes except $\textrm{CII}$ and $\textrm{C}$), and by combining analytical reasoning with elementary numerics, we have established nontrivial topological invariants on nontrivial cycles of the nodal manifolds (a small number of entries in Table~\ref{tab:nodal_manifolds}, marked explicitly with the symbol $\ddagger$, remain conjectural).
Owing to the robustness of topological band invariants under symmetry-preserving perturbations, the derived invariants on cycles of the nodal manifolds [last column in Table~\ref{tab:nodal_manifolds}] characterize the $n$-fold degeneracy for arbitrary Hamiltonians with the same number of energy bands.
Our methodology thus relates the topological stability of multifold band degeneracies to the linking of nodal manifolds supporting lower-order degeneracies while weaving together cohomological and homotopy-theoretic approaches to topological band theory.

Our results open several directions for future investigation.
Some are concrete extensions of the present framework: to many-band models, to higher-fold nodes, and to nodal lines and surfaces.
Others point toward structural connections with non-Hermitian topology, cobordisms, and higher gauge~structures.

We begin with the topological characterization of $n$-fold band nodes \emph{beyond minimal models}, i.e., in models with a larger number of energy bands.
To that end, note that the $n$-fold nodal point in the minimal model is uniquely defined as the intersection of the only two available degeneracy loci of order $(n\,{-}\,1)$. 
This construction admits a natural generalization to multiband systems. 
For simplicity, let us explicitly consider the case of threefold nodal points of codimension $\delta_{(3)}$ in Wigner-Dyson symmetry classes. 
In analogy with the minimal models, we enclose the nodal point by a sphere $S^{\ell}$ with $\ell=\delta_{(3)}-1$ inside momentum space. 
In contrast to minimal models, multiband systems can exhibit \emph{multiple} (i.e., more than two) loci of twofold degeneracy $\mathcal{L}_i$, each encoding the degeneracy of bands $\{i,i+1\}$. 
Threefold nodal points are thus formed as intersections of any consecutive pair of loci, $\mathcal{L}_i\cap \mathcal{L}_{i+1}$, in which case the triplet of bands $\{i,i+1,i+2\}$ becomes degenerate.
Therefore, to study threefold degeneracy in a model with four (or more) energy bands, we construct three (or more) nodal manifolds $M_i=S^{\ell}\cap \mathcal{L}_i$ and evaluate their topological band invariants.

If two nodal manifolds $M_i$ and $M_{i+1}$ extend to a threefold nodal point inside $S^\ell$, then this multifold band node contributes to band invariants computed on cycles of $M_i$ and $M_{i+1}$ following the information summarized in Table~\ref{tab:nodal_manifolds}.
However, band invariants on $M_i$ are also influenced by triple points formed by the intersection $\mathcal{L}_{i-1}\cap \mathcal{L}_i$ inside $S^\ell$, while the band invariants on $M_{i+1}$ exhibit a contribution from triple points similarly formed by the intersection $\mathcal{L}_{i+1}\cap \mathcal{L}_{i+2}$.
Note that the situation is analogous to the familiar case of Weyl points (twofold degeneracies in the class $\textrm{A}$) in models with more than two bands: a Weyl point formed by bands $\{i,i+1\}$ generates a Chern number $C_i = -C_{i+1}$ (i.e., of opposite sign) on both bands on the enclosing $2$-sphere. 
Correspondingly, the Chern number $C_i$ exhibits a contribution from Weyl points formed by either the pair of bands $\{i-1,i\}$ or the pair $\{i,i+1\}$. 
For this reason, the Chern number $C_i$ does not automatically signal the presence of a Weyl point in any of the two adjacent energy gaps.
However, if we use $n_{i,i+1}$ to denote the total chiral charge of all Weyl points formed by the bands $\{i,i+1\}$ inside the $2$-sphere, then $C_i = n_{i,i+1}-n_{i-1,i}$. 
As this is a linear relation, the knowledge of all Chern numbers allows us to extract the total charge of Weyl points inside the individual energy gaps as $n_{i,i+1} = \sum_{j=1}^i C_j$.
In the same spirit, we anticipate that one can extract the total charge of threefold degeneracies formed by bands $\{i,i+1,i+2\}$ inside $S^\ell$ from the knowledge of band invariants on all nodal manifolds $M_i$. 
What makes these two situations somewhat different is that, in the case of Weyl points, we consider the \emph{same} enclosing $S^2$ for each band $i$, whereas in the case of triple degeneracies, we compute topological invariants for the bands $\{i,i+1\}$ on \emph{distinct subsets} $M_i$ of $S^\ell$.
We leave the detailed discussion of these topological relations to future studies.

\begin{figure}[!t]
    \centering
    \includegraphics[width=\linewidth]{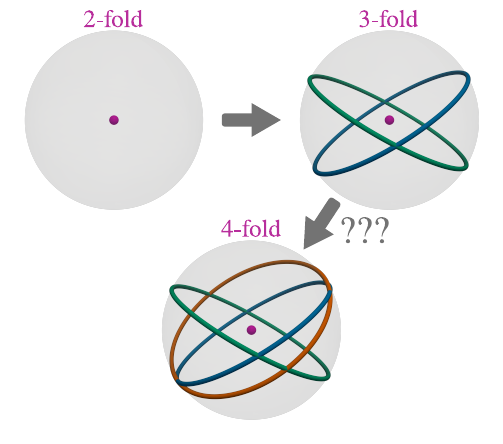}
    \caption{
    Generalization of the enclosing-sphere construction to band nodes of increasing order.
    For a minimal ($2$-fold) band degeneracy, one asks where on the enclosing sphere the degeneracy splits as $(1\,{+}\,1)$.
    Since this splitting occurs everywhere around the $2$-fold node, the analog of the nodal manifolds $M_1$ and $M_2$ is the entire $S^\ell$, and the extraction of bundle invariants on nontrivial cycles reduces to the conventional approach rooted in homotopy groups of the standard classifying spaces of Hamiltonians.
    At the first nontrivial order ($3$-fold band degeneracy), which is the principal subject of the present work, the two partial splittings $(1\,{+}\,2)$ and $(2\,{+}\,1)$ are possible; their linked nodal manifolds are revealed by bundle invariants supported on their nontrivial cycles.
    Generalizing the strategy, we anticipate that the next ($4$-fold) band degeneracy is characterized by bundle invariants on three nodal manifolds, corresponding to the splittings $(1\,{+}\,3)$, $(2\,{+}\,2)$,~and~$(3\,{+}\,1)$.
    }
    \vspace{-4 mm}
    \label{fig:conclusion}
\end{figure}

The extension to \emph{higher-fold nodal points} (i.e., the third and subsequent non-zero entries in each row of Table~\ref{tab:codims-WD+balanced}) is less direct, but we expect a similar principle to apply.
In this case, we anticipate that the construction involves several nodal manifolds, one for each pattern in which the multifold degeneracy can partially split.
For example, a fourfold nodal point may require three nodal manifolds, corresponding to the splittings $(1\,{+}\,3)$, $(2\,{+}\,2)$, and $(3\,{+}\,1)$ of the four bands (see Fig.~\ref{fig:conclusion}).
The resulting network of linked nodal manifolds should then protect the associated multifold nodal point.
Realizing this program concretely, and determining whether the associated band invariants can also detect genuinely higher linking phenomena, such as those captured by Massey products, remain interesting questions for future work.
We also remark that this perspective naturally subsumes the characterization of minimal band degeneracies.
For a minimal (twofold) node, the analogous question asks where on the enclosing sphere the degeneracy splits as $(1\,{+}\,1)$.
Since this splitting occurs everywhere on the sphere, the corresponding nodal manifold is the entire $S^{\ell}$, and the construction reduces to the conventional homotopy-theoretic classification~\cite{Bzdusek:2017}.

It should further be possible and interesting to relate our construction to recent studies of how multifold degeneracies, whether fine-tuned or symmetry-protected, resolve into generic configurations of Weyl points upon perturbations~\cite{Frank:2026}.
Mathematically, such considerations are related to restricting the parameter space to dimensions lower than the determined codimension $\delta^\textrm{CL}_{(n)}$, and asking how such a parameter space generically intersects the attached nodal loci.

Our work leaves unanswered the question of whether the classifying spaces capturing $n$-fold degeneracies beyond the minimal case admit, in any of the ten AZ symmetry classes, an expression as homogeneous spaces of the form $M=\mathsf{G}/\mathsf{H}$, where $\mathsf{G}$ is a group of transformations that generates all relevant Hamiltonians, and $\mathsf{H}<\mathsf{G}$ is the stabilizer of a reference Hamiltonian.
With such a classifying space we mean, e.g., the space of Hamiltonians in which either bands $\{i,i\,{+}\,1\}$ or bands $\{i\,{+}\,1,i\,{+}\,2\}$ may be degenerate, but \emph{not both simultaneously}.
Such quotient expressions are established for band structures exhibiting a single-energy gap~\cite{Kitaev:2009,Ryu:2010}, as well as for delicate and multigap topological insulators~\cite{Wu:2019,Bouhon:2020b,Lim:2025,Lapierre:2021}. 
Alternatively, one may seek these generalized classifying spaces not as homogeneous spaces but, instead, by virtue of an explicit cell decomposition.
Knowledge of either formulation would make it possible to investigate not only multifold \emph{point} degeneracies, but also the topology of more intricate nodal objects in higher-dimensional momentum spaces, such as multifold \emph{nodal lines} and \emph{surfaces}.
In a separate work~\cite{Iliasov:2026}, we discuss how the approach involving an explicit cell decomposition can be applied to at least some of the ten AZ symmetry classes. 
Such an explicit construction allows us to extract all homology groups of the classifying space and, by virtue of the Hurewicz theorem, at least the lowest nontrivial homotopy group of this classifying space.
Unfortunately, this understanding of the topological obstruction does not provide an explicit connection to the Bloch bundle invariants as investigated in the present work; establishing this connection would unify the two complementary~perspectives.

We also briefly contrast the case of real Hermitian Hamiltonians (symmetry class $\textrm{AI}$) with their non-Hermitian (nH) analogs.
The natural counterparts of $n$-fold band degeneracies in the context of non-Hermitian Hamiltonians are $n$-fold exceptional points (EP$n$)~\cite{Berry:2004,Demange:2012} where $n$ eigenstates coalesce to form a single Jordan block.
Using resultant analysis, Refs.~\citenum{Delplace:2021,Yoshida:2025} found that the codimension of forming EP$n$ grows linearly with $n$,
\begin{equation}
\delta_{(n)}^\textrm{nH} = (n-1)\delta_{(2)}^\textrm{nH},    
\end{equation}
and characterized EP$n$s in minimal (i.e., $n$-band) models with integer winding numbers of their resultant vectors in the non-Hermitian classes $\textrm{A}$ and $\textrm{AI}$.
In the lowest nontrivial case, the relation $\delta_{(3)}^\textrm{nH} = 2\times\delta_{(2)}^\textrm{nH}$ suggests that exceptional points formed by two overlapping pairs of bands can be moved through the parameter space independently of each other. 
This linear dependence on $n$ should be contrasted against the quadratic formula for the Hermitian class $\textrm{AI}$, shown in Eq.~(\ref{eqn:AI-codim}), which implies $\delta_{(3)}^\textrm{AI} > 2\times\delta_{(2)}^\textrm{AI}$.
It was discussed in Ref.~\citenum{Tiwari:2020} that this inequality, in the context of 3D systems, prevents lines of twofold degeneracy in adjacent energy gaps from passing across each other, providing an alternative insight into the non-Abelian band topology in the symmetry class AI~\cite{Ahn:2019,Wu:2019,Bouhon:2020}.
Owing to the generally quadratic dependence of $\delta_{(n)}^\textrm{CL}$ on $n$ in all Hermitian symmetry classes [Table~\ref{tab:codims-WD+balanced}], we anticipate that similar constraints preventing the crossing of higher-fold nodal manifolds arise in appropriate higher dimensions in all the symmetry classes considered in this work.

Further echoing this line of non-Hermitian research, we note that the classifying space of EP$n$s in minimal (i.e., $n$-band) models is a sphere whose dimension grows linearly with $n$~\cite{Yoshida:2025}.
Owing to the rich pattern of higher homotopy groups of spheres, interpreted in Ref.~\citenum{Yoshida:2026} as generalized Hopf invariants, minimal non-Hermitian models in dimensions exceeding the codimension of the EP$n$ were shown to support an enriched family of so-called \emph{Hopf exceptional points}.
Since the classifying space of minimal models of an $n$-fold band degeneracy in the presently studied Hermitian setting is likewise a sphere $S^{\ell}$, with $\ell = \delta^{\textrm{CL}}_{(n)}-1$, an analogous enrichment is expected whenever the  parameter space exceeds the codimension $\delta^{\textrm{CL}}_{(n)}$, with extended nodal structures characterized by elements of the higher homotopy groups $\pi_{d}(S^{\ell})$ for $d>\ell$.
In the lowest-dimensional instance of such a construction, the homotopy group $\pi_4(S^3) = \ztwo$~\cite{Freudenthal:1938} (related to the Witten anomaly~\cite{Witten:1983,Grinevich:1988}) enables a Hopf-like $\ztwo$ characterization of a \emph{ring} of fourfold zero-energy
degeneracies of class-$\textrm{BDI}$ Hamiltonians in five-dimensional parameter space, enclosed by a four-dimensional sphere.
We leave the systematic study of such generalized Hopf invariants of Hermitian multifold band degeneracies to future work.

We finally list several more speculative directions that merit deeper investigation.
First, we observe a conspicuous feature shared by all the studied minimal models; namely, that their nodal manifolds appear to be \emph{maximally} linked.
With this, we mean the following two properties.
(1)~The nontrivial cycles of the two nodal manifolds always come in complementary pairs: for each nontrivial $p$-cycle with $p>0$ of one nodal manifold, there exists a nontrivial $q$-cycle of the other nodal manifold, such that the two obey $p+q+1=\ell$ [Eq.~(\ref{eq:linkin_constr})]. 
(In classes $\textrm{BDI}$ and $\textrm{D}$, one needs to explicitly invoke the $0$-cycle, corresponding to the Pfaffian sign invariant.) 
Each nontrivial cycle thus has the potential to be linked with another cycle inside the enclosing sphere.
(2)~For every complementary pair of cycles that we were able to investigate explicitly, the linking is indeed nontrivial, as revealed by a nontrivial topological band invariant [last column of Table~\ref{tab:nodal_manifolds}].
While a nontrivial invariant on a single cycle of either nodal manifold is, in principle, sufficient to signal the linking (and, thus, to imply the presence of the multifold node inside $S^\ell$), the observed pattern leads us to conjecture that the linking is nontrivial for \emph{all} complementary pairs of nontrivial cycles [including those not explicitly listed in Table~\ref{tab:nodal_manifolds}].
Proving or disproving this statement is beyond the scope of our present study.

Second, it may be illuminating to formulate the topology of the nodal manifolds from a cobordism viewpoint~\cite{Thom:1954,Stong:1968,Milnor:1974}.
Note that when a multifold nodal point is unstable, the associated nodal manifolds are unlinked and can be eliminated by a smooth deformation of the Hamiltonian, resulting in $M_1 = M_2 = \varnothing$.
Since the empty set is the trivial element of any cobordism group, this suggests that the instability of the multifold node forces each nodal manifold to be null-cobordant; conversely, a nodal manifold representing a nontrivial cobordism class cannot be removed, indicating a topologically protected multifold node.
Furthermore, since null-cobordism is obstructed by nonvanishing characteristic numbers~\cite{Milnor:1974}, cobordism invariants of the nodal manifolds provide a candidate diagnostic of the topological protection that does not require the explicit computation of band invariants on cycles.
The nodal manifolds identified in our work [Table~\ref{tab:nodal_manifolds}] are consistent with this expectation: $\reals P^2$, obtained for the symmetry class $\textrm{AI}$, generates the unoriented cobordism group $\Omega^{\mathsf{O}}_2 = \ztwo$ (its nonorientability, encoded in the first Stiefel-Whitney class, directly mirrors the nontrivial Berry phase computed in
Sec.~\ref{sec:class-AI}); $\cmplx P^2$, obtained for the class $\textrm{A}$, generates the oriented cobordism group $\Omega^{\mathsf{SO}}_4 = \intg $; and $\quats P^2$, obtained for the class $\textrm{AII}$, is one of the two generators of the cobordism group $\Omega^{\mathsf{Spin}}_8 =\intg^2$.
A systematic assignment of cobordism invariants to the nodal manifolds in all ten symmetry classes may therefore provide a complementary and mathematically rich characterization of multifold band degeneracies.

Lastly, a promising direction is to explore prospective relations between multifold nodal points and bundle gerbes.
The latter have recently been applied to reformulate topological band invariants in three-dimensional systems in the classes $\textrm{AI}$ and $\textrm{AIII}$~\cite{Jankowski:2025,Palumbo:2019,Bzdusek:2026}.
The motivation is analogous to the familiar description of band topology in terms of characteristic classes. 
In that setting, characteristic classes can be represented by differential forms (possibly with distributional coefficients) equipped with geometric data such as a connection and its curvature. 
However, this construction presupposes a globally well-defined vector bundle over the entire parameter space, which generally fails on the enclosing sphere in the presence of multifold points. 
Bundle gerbes may provide a natural alternative, since their curvature data need only be defined locally on patches, with a globally consistent structure enforced by cocycle (coherence) conditions on overlaps. 
This perspective aligns with our description of multifold points, where band invariants are naturally defined only on the nodal manifolds and their cycles rather than on the entire enclosing sphere.
It is therefore plausible that minimal multifold points admit an associated gerbe invariant, and that more general settings may require higher-gerbe analogues~\cite{Moore:2025}.

\begin{acknowledgments}
We would like to thank 
Jing-Yuan Chen, 
Shu Hamanaka, 
Hyeongmuk Lim, 
Joseph Maciejko, 
Mykhailo Pavliuk, and 
Bohm-Jung Yang 
for valuable discussions. 
A.I., Z.G., and T.B.~were supported by the Starting Grant No.~211310 by the Swiss National Science Foundation.
A.I.~and Z.G.~acknowledge funding from the University of Zurich under the UZH Candoc and Postdoc Grants, grants No.~FK-24-104 and No.~FK-26-087.
T.Y.~is grateful for the support from the ETH Pauli
Center for Theoretical Studies and from the Yamada Science Foundation.
T.Y.~is supported by JSPS KAKENHI Grant Nos.~JP23KK0247,
JP25K07152, and JP25H02136, as well as JSPS Bilateral Program No.~JSBP120249925.
A.T.~is funded by Villum Fonden Grant no. VIL60714.
\end{acknowledgments}


\appendix

\section{Explicit counting of codimensions}\label{app:explicit_counting}

We compute the codimensions by explicitly counting the perturbations.
An $n$-tiple point is described by a Hermitian $n\times n$-matrix whose matrix elements are $a_{ij}$ $i,j=1,\ldots,n$.
Choosing the energy of $n$-tiple point to be zero, the matrix becomes traceless.
The necessary and sufficient condition of such an $n$-tiple point is that all the matrix elements vanish, $a_{ij}=0$ for $i,j=1,\ldots,n$ [see Appendix~\ref{sec: iff aij=0}].
Therefore, by counting the number of symmetry-allowed perturbations of $n\times n$ matrices, we obtain codimensions for each symmetry class.

Symmetry-allowed perturbations for each symmetry class are summarized in Tables~\ref{tab: 3x3 ptb matrix}, \ref{tab: 4x4 ptb matrix} and \ref{tab: 6x6 ptb matrix}.
In these tables, 
we consider the following symmetries
\begin{eqnarray}
\label{eq: ExC PT}
\mathcal{T} H(\bm{k}) \mathcal{T}^{-1} &=  H(\bm{k}), \\
\mathcal{C} H(\bm{k}) \mathcal{C}^{-1} &=  -H(\bm{k}), \\
\mathcal{S} H(\bm{k}) \mathcal{S}^{-1} &=  -H(\bm{k}),
\end{eqnarray}
where $\mathcal{T}$ and $\mathcal{P}$ are antiunitary operators, and $\mathcal{S}$ is a unitary operator satisfying $\mathcal{S}^2=1$.
Operator $\mathcal{T}$ ($\mathcal{T}^2=\pm 1$) represents the time-reversal operation, while $\mathcal{C}$ ($\mathcal{C}^2=\pm 1$) represents the charge-conjugation operation. 
We suppose that these operations keeps $\bm{k}$ invariant because $\bm{k}$ describes parameters rather than physical momentum.

Here, as a demonstration, we take a triple point of class AI.
We consider a traceless $ 3\times3$-Hamiltonian $H(\bm{k})$ with momentum $\bm{k}$. In general, such a Hamiltonian is expanded by Gell-Mann matrices $\lambda_i$ ($i=1,2,\ldots, 8$) with real coefficients.
Under time-reversal symmetry [Eq.~\eqref{eq: ExC PT}], we see that $\lambda_i$ ($i=1,3,4,6,8$) are compatible with the symmetry constraint.
In other words, for class AI, the $3\times 3$-Hamiltonian describing all perturbations is written as
\begin{eqnarray}
H &=& \sum_{i} a_i \Lambda_i
\end{eqnarray}
with $a_i$ ($i=1,\ldots,5$) being real numbers and 
$\bm{\Lambda}=(\lambda_1,\lambda_3,\lambda_4,\lambda_6,\lambda_8)$.
Thus, for class AI, the codimension of the triple point is $\delta_{(3)}=5$ [see Table~\ref{tab: 3x3 ptb matrix}], which coincides with the result obtained based on von Neumann-Wigner arguments [see Table~\ref{tab:codims-WD+balanced}].
On the sphere enclosing the triple point in the five-dimensional momentum space, vector $\bm{R}=\bm{a}/\sqrt{\bm{a}\cdot\bm{a}}$ with $\bm{a}=(a_1,\ldots,a_5)$ defines the map to $S^4$. Because the map from $S^4$ to $S^4$ can be nontrivial $\pi_4(S^4)=\mathbb{Z}$, the triple point can be topological. Its topology is characterized by the four-dimensional winding number $W_4$. 
For other values of $n$,
the construction proceeds in the same way. We obtain a map from $S^{\delta_{n}-1}$ to $S^{\delta_{n}-1}$, the topology of which is characterized by the winding number $W_{\delta_{(n)}-1}$.

The toy model of $W_4=1$ is given by
\begin{eqnarray}
\label{eq: 3x3 Hami AI}
H &=& \sum_{i} k_i \Lambda_i,
\end{eqnarray}
with momentum $\bm{k}=(k_1,\ldots,k_5)$.

In a similar way, we compute the codimensions of each AZ class for $n=3,4,6$. The results are summarized in Tables~\ref{tab: 3x3 ptb matrix}, \ref{tab: 4x4 ptb matrix} and \ref{tab: 6x6 ptb matrix} which coincide with the codimensions in Table~~\ref{tab:codims-WD+balanced}. 
Winding numbers and toy models are also obtained in a similar way to the above argument.

\begin{table*}
    \centering
    \begin{tabular}{cccccl}
          \hline\hline
       ~~      & ~$\mathcal{T}$~                          & ~$\mathcal{C}$~             & ~$\mathcal{S}$~               & ~Perturbations of $3\times 3$ matrices~                                                 & ~$\delta$~  \\      \hline 
        A      & ~$0$~	                                  & ~$0$~                       & ~$0$~                    &  $\{\lambda_1,\lambda_2,\lambda_3,\lambda_4,\lambda_5,\lambda_6,\lambda_7,\lambda_8 \}$ &  $8$        \\ 
        AIII   & ~$0$~	                                  & ~$0$~                       & ~$\lambda_{\mathcal{S}}$~     &  $\{\lambda_4,\lambda_5,\lambda_6,\lambda_7 \}$                                         &  $4$        \\ \hline  
        AI     & ~$\mathcal{K}$~	                      & ~$0$~                       & ~$0$~                    &  $\{\lambda_1,\lambda_3,\lambda_4,\lambda_6,\lambda_8 \}$                               &  $5$        \\ 
        BDI    & ~$\lambda_{\mathcal{S}}\mathcal{K}$~	      & ~$0$~                       & ~$\lambda_{\mathcal{S}}$~     &  $\{\lambda_5,\lambda_7 \}$                                                             &  $2$        \\ 
        D      & ~$0$~	                                  & ~$\mathcal{K}$~             & ~0~                      &  $\{\lambda_2,\lambda_5,\lambda_7 \}$                                                   &  $3$        \\ 
        DIII   & ~--~	                                  & ~--~                        & ~--~                     &  --                                                                                     &             \\ 
        AII    & ~--~	                                  & ~--~                        & ~--~                     &  --                                                                                     &             \\ 
        CII    & ~--~                                     & ~--~                        & ~--~                     &  --                                                                                     &             \\ 
        C      & ~--~   	                              & ~--~                        & ~--~                     &  --                                                                                     &             \\ 
        CI     & ~--~	                                  & ~--~                        & ~--~                     &  --                                                                                     &             
\\ \hline\hline
    \end{tabular}
     \caption{ 
List of symmetry-allowed perturbations of $3\times 3$-matrices. Gell-Mann matrices are denoted by $\lambda_i$ ($i=1,2,\ldots,8$), and $\lambda_{\mathcal{S}}$ is defined as $\lambda_{\mathcal{S}}=\mathrm{diag}(1,1,-1)$. ``0" denotes the absence of the corresponding symmetry. For symmetry classes DIII, AII, and CII, $3\times 3$-matrices are incompatible with $\mathcal{T}^2=-1$. For symmetry classes C, and CI, $3\times 3$-matrices are incompatible with $\mathcal{C}^2=-1$.             .
      }
    \label{tab: 3x3 ptb matrix}
\end{table*}

\begin{table*}
\centering
\begin{tabular}{cccccl}
\hline\hline
      & ~$\mathcal{T}$~           & ~$\mathcal{C}$~         & ~$\mathcal{S}$~          & ~Perturbations of $4\times 4$ matrices~                                                                                                                                                                                       & ~$\delta$~         \\  \hline 
 A    & ~$0$~	                 & ~$0$~                   & ~$0$~               & $\{\sigma_x,\sigma_y,\sigma_z\}\otimes\tau_0 \cup \{\sigma_0,\sigma_x,\sigma_y,\sigma_z\}\otimes\tau_x \cup \{\sigma_0,\sigma_x,\sigma_y,\sigma_z\}\otimes\tau_y \cup \{\sigma_0,\sigma_x,\sigma_y,\sigma_z\}\otimes\tau_z$   & ~$15$~  \\ 
 AIII & ~$0$~	                 & ~$0$~                   & ~$\tau_z$~          & $\{\sigma_0,\sigma_x,\sigma_y,\sigma_z\}\otimes\tau_x \cup \{\sigma_0,\sigma_x,\sigma_y,\sigma_z\}\otimes\tau_y$                                                                                                              & ~$8^{\,\textrm{a}}$~  \\  \hline
 AI   & ~$\mathcal{K}$~	         & ~$0$~                   & ~$0$~               & $\{\sigma_x,\sigma_z\}\otimes\tau_0 \cup \{\sigma_0,\sigma_x,\sigma_z\}\otimes\tau_x \cup \{\sigma_y\tau_y\} \cup \{\sigma_0,\sigma_x,\sigma_z\}\otimes\tau_z$                                                                & ~$9$~  \\ 
 BDI  & ~$\mathcal{K}$~	         & ~$0$~                   & ~$\tau_z$~          & $\{\sigma_0,\sigma_x,\sigma_z\}\otimes\tau_x \cup \{ \sigma_y\tau_y \}$                                                                                                                                                       & ~$4^{\,\textrm{a}}$~  \\ 
 D    & ~$0$~	                 & ~$\mathcal{K}$~         & ~0~                 & $\{ \sigma_y\tau_0\} \cup \{\sigma_y\tau_x\} \cup \{\sigma_0,\sigma_x,\sigma_z\}\otimes\tau_y \cup \{ \sigma_y\tau_z \}$                                                                                                      & ~$6^{\,\textrm{a}}$~  \\ 
 DIII & ~$\imi\sigma_y\mathcal{K}$~  & ~$\mathcal{K}$~         & ~$\sigma_y$~        & $\{\sigma_x,\sigma_z\}\otimes\tau_y$                                                                                                                                                                                          & ~$2^{\,\textrm{a,b}}$~  \\ 
 AII  & ~$\imi\sigma_y\mathcal{K}$~  & ~$0$~                   & ~$0$~               & $\{\sigma_0\tau_x\} \cup \{\sigma_x,\sigma_y,\sigma_z\}\otimes\tau_y \cup \{ \sigma_0\tau_z\}$                                                                                                                                & ~$5^{\,\textrm{b}}$~  \\ 
 CII  & ~$\imi\sigma_y\mathcal{K}$~  & ~$\imi\tau_y\mathcal{K}$~  & ~$\sigma_y\tau_y$~  & $\{\sigma_0\}\otimes\tau_x \cup \{\sigma_x,\sigma_z\}\otimes\tau_y \cup \{\sigma_0\}\otimes\tau_z$                                                                                                                            & ~$4^{\,\textrm{a,b}}$~  \\ 
 C    & ~$0$~                     & ~$\imi\tau_y\mathcal{K}$~  & ~$0$~               & $\{\sigma_y\tau_0 \} \cup \{\sigma_0,\sigma_x,\sigma_z\}\otimes\tau_x \cup \{\sigma_0,\sigma_x,\sigma_z\}\otimes\tau_y \cup \{\sigma_0,\sigma_x,\sigma_z\}\otimes\tau_z$                                                      & ~$10^{\,\textrm{a}}$~  \\ 
 CI   & ~$\mathcal{K}$~	         & ~$\imi\tau_y\mathcal{K}$~  & ~$\tau_y$~          & $\{\sigma_0,\sigma_x,\sigma_z\}\otimes\tau_x \cup \{\sigma_0,\sigma_x,\sigma_z\}\otimes\tau_z$                                                                                                                                & ~$6^{\,\textrm{a}}$~
\\ \hline\hline
    \end{tabular}
     \caption{
List of symmetry-allowed perturbations of $4\times 4$-matrices. Pauli matrices are denoted by $\sigma_i$ and $\tau_i$ ($i=x,y,z$). $2\times 2$-identity matrices are denoted by $\sigma_0$ and $\tau_0$.
``0" denotes the absence of the corresponding symmetry. 
In the column of $\delta$, entries without superscripts denote $\delta_{(4)}$.
Entries with superscript ``${\,\textrm{a}}$" (``${\,\textrm{b}}$") denote $\delta_{(2)}$ due to particle-hole (Kramers) doubling.
Entries with superscript ``${\,\textrm{a,b}}$" denote $\delta_{(1)}$ due to both particle-hole and Kramers doubling.
      }
    \label{tab: 4x4 ptb matrix}
\end{table*}

\begin{table*}
    \centering
    \begin{tabular}{cccccl}
          \hline\hline
       ~~              & ~$\mathcal{T}$~                         & ~$\mathcal{C}$~                           & ~$\mathcal{S}$~               & ~Perturbations of $6\times 6$ matrices~                                                                                                                                                                                                                               & ~$\delta$~               \\  \hline 
 \multirow{2}{*}{A}    & ~\multirow{2}{*}{$0$}~	                 & ~\multirow{2}{*}{$0$}~                    & ~\multirow{2}{*}{$0$}~   &  $\{\lambda_1,\lambda_2,\lambda_3,\lambda_4,\lambda_5,\lambda_6,\lambda_7,\lambda_8\}\otimes \sigma_0 \cup \{\lambda_0,\lambda_1,\lambda_2,\lambda_3,\lambda_4,\lambda_5,\lambda_6,\lambda_7,\lambda_8\}\otimes \sigma_x$                                             & ~\multirow{2}{*}{$35$}~  \\ 
                       &                                         &                                           &                          &  $~\cup \,\{\lambda_0,\lambda_1,\lambda_2,\lambda_3,\lambda_4,\lambda_5,\lambda_6,\lambda_7,\lambda_8\}\otimes \sigma_y \cup \{\lambda_0,\lambda_1,\lambda_2,\lambda_3,\lambda_4,\lambda_5,\lambda_6,\lambda_7,\lambda_8\}\otimes \sigma_z $                            &                          \\
        AIII           & ~$0$~	                                 & ~$0$~                                     & ~$\sigma_{z}$~           &  $\{\lambda_0,\lambda_1,\lambda_2,\lambda_3,\lambda_4,\lambda_5,\lambda_6,\lambda_7,\lambda_8\}\otimes \sigma_x \cup \{\lambda_0,\lambda_1,\lambda_2,\lambda_3,\lambda_4,\lambda_5,\lambda_6,\lambda_7,\lambda_8\}\otimes \sigma_y$                                   &  ~$18^{\,\textrm{a}}$~   \\ \hline 
 \multirow{2}{*}{AI}   & ~\multirow{2}{*}{$\mathcal{K}$}~        & ~\multirow{2}{*}{$0$}~                    & ~\multirow{2}{*}{$0$}~   &  $\{\lambda_1,\lambda_3,\lambda_4,\lambda_6,\lambda_8\}\otimes \sigma_0 \cup \{\lambda_0,\lambda_1,\lambda_3,\lambda_4,\lambda_6,\lambda_8\}\otimes \sigma_x$                                                                                                         &  ~\multirow{2}{*}{$20$}~ \\  
                       &         	                         &                                           &                          &  $\cup \,\{\lambda_2,\lambda_5,\lambda_7\}\otimes \sigma_y  \cup \{\lambda_0,\lambda_1,\lambda_3,\lambda_4,\lambda_6,\lambda_8\}\otimes \sigma_z$                                                                                                                       &                          \\ 
        BDI            & ~$\mathcal{K}$~	                 & ~$\sigma_z\mathcal{K}$~                   & ~$\sigma_z$~             &  $\{\lambda_0,\lambda_1,\lambda_3,\lambda_4,\lambda_6,\lambda_8\}\otimes \sigma_x \cup \{\lambda_2,\lambda_5,\lambda_7\}\otimes \sigma_y$                                                                                                                             &  ~$9^{\,\textrm{a}}$~                   \\
        D              & ~$0$~	                                 & ~$\mathcal{K}$~                           & ~0~                      &  $\{\lambda_2,\lambda_5,\lambda_7\}\otimes \sigma_0 \cup \{\lambda_2,\lambda_5,\lambda_7\}\otimes \sigma_x \cup \{\lambda_0,\lambda_1,\lambda_3,\lambda_4,\lambda_6,\lambda_8\}\otimes \sigma_y \cup \{\lambda_2,\lambda_5,\lambda_7\}\otimes \sigma_z$               &  ~$15^{\,\textrm{a}}$~                  \\ 
        DIII           & ~$\imi\sigma_y\mathcal{K}$~	         & ~$\mathcal{K}$~                           & ~$\sigma_y$~             &  $\{\lambda_2,\lambda_5,\lambda_7\}\otimes \sigma_x \cup \{\lambda_2,\lambda_5,\lambda_7\}\otimes \sigma_z$                                                                                                                                                           &  ~$6^{\,\textrm{a,b,c}}$~                   \\ 
        AII            & ~$\imi\sigma_y\mathcal{K}$~	         & ~$0$~                                     & ~0~                      &  $\{\lambda_1,\lambda_3,\lambda_5,\lambda_6,\lambda_8\}\otimes \sigma_0 \cup \{\lambda_2,\lambda_4,\lambda_7\}\otimes \sigma_x \cup \{\lambda_2,\lambda_4,\lambda_7\}\otimes \sigma_y \cup \{\lambda_2,\lambda_4,\lambda_7\}\otimes \sigma_z$                         &  ~$14^{\,\textrm{b}}$~                  \\ 
        CII            & ~$\imi\lambda_\mathcal{S} \sigma_y\mathcal{K}$~ & ~$\imi\sigma_y\mathcal{K}$~                  & ~$\lambda_{\mathcal{S}}$~     &  $\{\lambda_5,\lambda_7\}\otimes \sigma_0 \cup \{\lambda_4,\lambda_6,\lambda_8\}\otimes \sigma_x \cup \{\lambda_4,\lambda_6,\lambda_8\}\otimes \sigma_y \cup \{\lambda_4,\lambda_6,\lambda_8\}\otimes \sigma_z$                                                       &  ~$11^{\,\textrm{a,b,c}}$~                  \\ 
\multirow{2}{*}{C}     & ~\multirow{2}{*}{$0$}~                  & ~\multirow{2}{*}{$\imi\sigma_y\mathcal{K}$}~ & ~\multirow{2}{*}{$0$}~   &  $\{\lambda_2,\lambda_5,\lambda_7\}\otimes \sigma_0 \cup \{\lambda_0,\lambda_1,\lambda_3,\lambda_4,\lambda_6,\lambda_8\}\otimes \sigma_x $                                                                                                                            &  ~\multirow{2}{*}{$21^{\,\textrm{a}}$}~ \\ 
                       &                                         &                                           &                          &  $\cup \, \{\lambda_0,\lambda_1,\lambda_3,\lambda_4,\lambda_6,\lambda_8\}\otimes \sigma_y \cup \{\lambda_0,\lambda_1,\lambda_3,\lambda_4,\lambda_6,\lambda_8\}\otimes \sigma_z$                                                                                          &                          \\ 
        CI             & ~$\mathcal{K}$~	                 & ~$\imi\sigma_y\mathcal{K}$~                  & ~$\sigma_y$~             &  $\{\lambda_0,\lambda_1,\lambda_3,\lambda_4,\lambda_6,\lambda_8\}\otimes \sigma_x \cup \{\lambda_0,\lambda_1,\lambda_3,\lambda_4,\lambda_6,\lambda_8\}\otimes \sigma_z$                                                                                               &  ~$12^{\,\textrm{a}}$~                 
\\ \hline\hline
    \end{tabular}
     \caption{
List of symmetry-allowed perturbations of $6\times 6$-matrices. Pauli matrices are denoted by $\sigma_i$ ($i=x,y,z$). The $2\times 2$-identity matrix is denoted by $\sigma_0$.
Gell-Mann matrices are denoted by $\lambda_i$ ($i=1,2,\ldots,8$), and $\lambda_{\mathcal{S}}$ is defined as $\lambda_{\mathcal{S}}=\mathrm{diag}(1,1,-1)$.
``0" denotes the absence of the corresponding symmetry. 
In the column of $\delta$, entries without superscripts denote $\delta_{(6)}$. Entries with superscript ``${\,\textrm{a}}$" (``${\,\textrm{b}}$") denote $\delta_{(3)}$ due to particle-hole (Kramers) doubling. 
Entries with superscript ``${\,\textrm{a,b}}$" due to both particle-hole and Kramers doubling.
Entries with superscript ``${\,\textrm{c}}$" denote the presence of flat bands.
}
    \label{tab: 6x6 ptb matrix}
\end{table*}

\subsection{Necessary and sufficient condition of $n$-tiple points}
\label{sec: iff aij=0}

We show that a $n\times n$-traceless Hamiltonian hosts a $n$-tiple point if and only if all of the matrix elements $a_{ij}$ vanish; $a_{ij}=0$ for all $i,j=1,\ldots,n$.

The sufficient condition is easy to see; if all of the matrix elements are zero, we have $n$-fold degeneracy.  
The necessary condition is observed as follows. We consider the characteristic polynomial 
\begin{eqnarray}
P(\varepsilon) &=& \mathrm{det}[H-\varepsilon\mathbb{1}], \nonumber \\
&=& (-1)^n\left[ \varepsilon^n +\alpha_{n-1}\varepsilon^{n-1}+\alpha_{n-2}\varepsilon^{n-2}+\cdots+\alpha_0 \right], \nonumber \\
\end{eqnarray}
The necessary condition of an $n$-tiple root is all of the coefficients $\alpha$'s vanish.
In particular, the coefficient of $\varepsilon^{n-2}$ is written as
\begin{eqnarray}
\label{eq: |aij|=0}
\alpha_{n-2} &=& -\frac{1}{2} \sum_{i,j=1,\ldots,n} |a_{ij}|^2.
\end{eqnarray}
indicating that all of the matrix elements need to be zero to have $\alpha_{n-2}=0$. 
Therefore, for a traceless Hermitian $n\times n$-matrix, the necessary and sufficient condition of the $n$-fold band touching is $a_{ij}=0$ for all $i,j=1,\ldots,n$.
Equation~\eqref{eq: |aij|=0} is derived in Appendix~\ref{sec: derivation of |aij|=0}.

\subsubsection{
Derivation of \texorpdfstring{Eq.~\eqref{eq: |aij|=0}
}{Eq.?}}
\label{sec: derivation of |aij|=0}
We compute the coefficient of $(n-2)$-th power of the characteristic polynomial
\begin{eqnarray}
P(\varepsilon) &=& \mathrm{det}(H-\varepsilon\mathbb{1}).
\end{eqnarray}

The contributions to the $(n-2)$-th power are categorized into two types: one is from all of the diagonal elements, and the other is from $n-2$ of the diagonal elements.

(i) The contribution from all of the diagonal elements is given by
\begin{eqnarray}
\alpha^{(\mathrm{i})}_{n-2} &=& \frac{1}{2} 
\sum_{i\neq j} a_{ii}a_{jj} \nonumber \\
&=& -\frac{1}{2}  
\sum_{i } a_{ii}\Big(a_{ii}-\sum_j a_{jj}\Big) \nonumber \\
&=& -\frac{1}{2} 
\sum_{i } a^2_{ii}.
\end{eqnarray}
From the second to the third line we have used the traceless-ness of the matrix.

(ii) The contribution from $n-2$ of the diagonal elements is given by
\begin{eqnarray}
\alpha^{(\mathrm{ii})}_{n-2} &=& 
\sum_{i<j}\mathrm{sgn}
\left(
\begin{array}{ccccccc}
1 &\cdots & i& \cdots & j &\cdots & n \\
1 &\cdots & j& \cdots & i &\cdots & n 
\end{array}
\right)
a_{ij}a_{ji} 
\nonumber \\
&=& -\frac{1}{2} 
\sum_{i\neq j} |a_{ij}|^{2}. 
\end{eqnarray}
Here $\mathrm{sgn}(\cdots)$ denotes the sign corresponding to permutations. If the bottom row is obtained by an even (odd) permutation of the top row, $\mathrm{sgn}(\cdots)$ takes $1$ ($-1$). Combining the above two contributions, we obtain Eq.~\eqref{eq: |aij|=0}.

\section{Classifying spaces with zero-energy states}
\label{app:class-space-w-zero-states}

In this section, we analyze the space of Hamiltonians with a particular gap condition for the chiral and the BdG symmetry classes. 
First, we consider in detail the symmetry class $\textrm{AIII}$, and later we adapt the methodology for other classes, briefly noting the important differences.

To restate the well-known facts, recall that the (single-gap) classifying space of Hamiltonians in symmetry class $\textrm{AIII}$ and with an energy gap at $\varepsilon=0$ is 
\begin{equation}
\label{eqn:M-AIII-n}
\mathfrak{X}^\textrm{AIII}_N = \mathsf{U}(N)\times \mathsf{U}(N)/\mathsf{U}(N) \simeq \mathsf{U}(N).
\end{equation}
Here, $N$ is the number of occupied as well as the number of unoccupied bands (i.e., we consider a model with $2N$ bands with a gap at half-filling).
The unitary group in the quotient acts diagonally on both $\mathsf{U}(N)$ factors.
In the presence of additional gaps that divide the $N$ positive-energy bands into $k$ groups of $(n_1,n_2,\ldots,n_k$) bands, with $\sum_{j=1}^k n_j=N$, the stabilizer is reduced, extending the classifying space to~\cite{Lim:2025}
\begin{equation}
\label{eqn:M-AIII-nj}
\mathfrak{X}^\textrm{AIII}_{(n_1,n_2,\ldots,n_k)} = \mathsf{U}(N)\times\mathsf{U}(N) / \textstyle{\prod_{j=1}^k} \mathsf{U}(n_j)
\end{equation}
with each quotient $\mathsf{U}(n_j)$ acting diagonally on the two copies of $\mathsf{U}(N)$.
Homotopy groups of these spaces can be determined by studying certain long exact sequences~\cite{Lim:2025}.

In the following discussion, we aim to derive the classifying space of Hamiltonians in symmetry class $\textrm{AIII}$ that, likewise, exhibit multiple energy gaps among the positive-energy bands, but we further assume the existence of several states at $\varepsilon=0$ (i.e., without a gap at half-filling). 
The manifold of such Hamiltonians turns out to be important in our discussion of multifold degeneracies in symmetry class $\textrm{AIII}$.

For concreteness, we assume a total of $2N$ bands with $2n_0$ states at $\varepsilon=0$ and with additional $k$ energetically separated groups of $n_j$ bands at negative energies such that
\begin{equation}\label{eq:band_partition}
n_1 + \ldots + n_k = N - n_0.
\end{equation}
Due to chiral symmetry, one finds a corresponding energetically separated group of bands at positive energies. 
By continuous deformations that preserve the energy gaps, the bands within each multiplet can be made to coincide, resulting in $k$ unique positive eigenvalues $\varepsilon_1 < \ldots < \varepsilon_k$. 
By further continuous deformations, the energy of these multiplets can be standardized; for example, we may assume $\varepsilon_j = j$.

Owing to the chiral symmetry $\mathcal{S} = \sigma_z$, the Hamiltonian acquires a block-off-diagonal form
\begin{equation}
\label{eqn:SVD-offblock}
H = \left(\begin{array}{cc}
\mathbb{0}  &   h           \\
h^\dagger   &   \mathbb{0}  
\end{array}\right)    
\end{equation}
where $\mathbb{0}$ is an $N\times N$ zero matrix and $h$ is an $N\times N$ matrix with complex entries. 
Importantly, the block-off-diagonal form allows us to relate the eigenvalues of $H$ to singular values of $h$.
To see this, let us start with the singular value decomposition
\begin{equation}
\label{eqn:off-block-SVD}
h = U \Sigma V^\dagger = \sum_{j=1}^n \sigma_j u_j v_j^\dagger    
\end{equation}
where $U,V\in\mathsf{U}(N)$, and $\Sigma$ is a diagonal matrix of singular values $\sigma_j \geq 0$. 
In the second step, we further introduced $u_j$ for columns of $U$ and $v_j$ for columns of $V$.
Since the columns of unitaries are orthogonal to each other, we rewrite Eq.~(\ref{eqn:SVD-offblock})~as
\begin{equation}
h v_i = \sum_{j=1}^n \sigma_j u_j \underbrace{v_j^\dagger v_i}_{\delta_{ij}} = \sigma_i u_i
\end{equation}
where $\delta_{ij}$ is the Kronecker symbol. 
Starting with the Hermitian conjugate of Eq.~(\ref{eqn:SVD-offblock}), we similarly show that $h^\dagger u_i = \sigma_i v_j$.
With these results, we find 
\begin{equation}
\left(\begin{array}{cc}
\mathbb{0}  &   h           \\
h^\dagger   &   \mathbb{0}  
\end{array}\right)\left(\begin{array}{c}
u_i \\ \pm v_i 
\end{array}\right) = \pm \sigma_i \left(\begin{array}{c}
u_i \\ \pm v_i 
\end{array}\right),
\end{equation}
i.e., a singular value $\sigma_i$ of $h$ corresponds to the pair of eigenvalues $\pm \sigma_i$ of $H$.
Therefore, we can directly translate the gap condition on eigenvalues of $H$ into an analogous gap condition on singular values of $h$. 
Specifically, $h$ contains $n_0$ zero singular values, followed by groups of $n_j$ non-zero singular values separated by spectral gaps. 
Through continuous deformations that preserve the gaps, the $n_j$ singular values within each multiplet can be made to coincide, and further continuous deformations can standardize the singular values to the integer values $\{0,1,\ldots,k\}$.

Given a singular value decomposition in Eq.~(\ref{eqn:off-block-SVD}), the Hamiltonian is uniquely determined.
However, several distinct singular value decompositions can potentially result in the \emph{same} Hamiltonian with $n_j$-fold multiplets with the prescribed energies.
Therefore, to derive the sought space of Hamiltonians with the specified band multiplets, we need to understand the degree of non-uniqueness of the decomposition in Eq.~(\ref{eqn:off-block-SVD}).
First, to remove much of the redundancy, we assume that the singular values are arranged in ascending order, $\sigma_1 \leq \sigma_2 \leq \ldots \sigma_n$, i.e., Eq.~(\ref{eqn:off-block-SVD}) takes the form
\begin{equation}
\label{eqn:block-SVD}
h = U \, \textrm{diag}\left(\mathbb{0}_{n_0},\sigma_1 \mathbb{1}_{n_1},\ldots,\sigma_k \mathbb{1}_{n_k}\right) V^\dagger     
\end{equation}
where we wrote a block-diagonal matrix in the center of the expression, and the subscripts $n_j$ indicate the dimensions of the zero matrix and of the identity matrices.
For the standardized choice of the singular value, the entire freedom in Eq.~(\ref{eqn:block-SVD}) reduces to the two unitary matrices $U,V\in\mathsf{U}(N)$.

However, we also read from the expression in Eq.~(\ref{eqn:block-SVD}) that the mixing of certain columns of $U$ and of certain columns of $V$ (equivalently: rows of $V^\dagger$) leaves the form of $h$ unaffected.
Specifically, within the zero sector, we can perform independent $\mathsf{U}(n_0)$ rotations of the $n_0$ columns of $U$ and of the $n_0$ columns of $V$. 
On the other hand, in each non-zero sector, the form of $h$ stays invariant if the columns of both $U$ and $V$ are rotated by the same $\mathsf{U}(n_j)$ matrix. 
Therefore, we find the sought space of chiral-symmetric Hamiltonians as
\begin{equation}
\label{eqn:M-AIII-n0-nj}
\mathfrak{X}^\textrm{AIII}_{(n_0;n_1,\ldots,n_k)} = \frac{\mathsf{U}(N) \times \mathsf{U}(N)}{\left[\mathsf{U}(n_0)\times \mathsf{U}(n_0)\right] \times \textstyle{\prod_{j=1}^k} \mathsf{U}(n_j)}. 
\end{equation}
Here, the two quotients $\mathsf{U}(n_0)$ inside the brackets act respectively on the two copies of $\mathsf{U}(N)$, while the quotients $\mathsf{U}(n_j)$ with $j\in\{1,\ldots,k\}$ act diagonally on both copies of $\mathsf{U}(N)$. 
One can see that the classifying spaces in Eqs.~(\ref{eqn:M-AIII-n}) and~(\ref{eqn:M-AIII-nj}) correspond to special cases of Eq.~(\ref{eqn:M-AIII-n0-nj}).

Through similar considerations over the field of real resp.~of symplectic matrices, we anticipate the analogous result for the chiral orthogonal class,
\begin{equation}
\label{eqn:M-BDI-n0-nj}
\mathfrak{X}^\textrm{BDI}_{(n_0;n_1,\ldots,n_k)} =  \frac{\mathsf{O}(N) \times \mathsf{O}(N)}{\left[\mathsf{O}(n_0)\times \mathsf{O}(n_0)\right] \times \textstyle{\prod_{j=1}^k} \mathsf{O}(n_j)}, 
\end{equation}
and for the chiral symplectic class, 
\begin{equation}
\label{eqn:M-CII-n0-nj}
\mathfrak{X}^\textrm{CII}_{(n_0;n_1,\ldots,n_k)} =  \frac{\mathsf{Sp}(N) \times \mathsf{Sp}(N)}{\left[\mathsf{Sp}(n_0)\times \mathsf{Sp}(n_0)\right] \times \textstyle{\prod_{j=1}^k} \mathsf{Sp}(n_j)}.  
\end{equation}
In the last equation, one should bear in mind that $N$-fold unitary matrices over quaternion numbers translate to $2N$ bands, i.e., the values $n_j$ need to be doubled to obtain the physical multiplicities of the band groups.

We further turn our attention to the BdG classes.
We can again adopt the band partition introduced in Eq.~\eqref{eq:band_partition}, owing to the spectral pairing enforced by chiral or particle-hole symmetry. 
The derivation mostly follows a similar strategy as for class $\mathrm{AIII}$, with the key difference that each symmetry class requires an appropriate symmetry-adapted matrix decomposition and corresponding stabilizer groups.

We begin with the class $\mathrm{D}$. 
As discussed in Sec.~\ref{sec:D_class}, a class-$\mathrm{D}$ Hamiltonian can be written as $H=\imi A$, where $A$ is a real antisymmetric matrix. 
The matrix $A$ admits the decomposition
\begin{equation}
A=\Psi\Sigma \Psi^{\mathrm{T}},
\end{equation}
where $\Psi\in\mathsf{O}(2N)$ is orthogonal and $\Sigma$ is block diagonal, with $2\times 2$ antisymmetric blocks whose upper off-diagonal entries are given by the energies $\varepsilon_j$. 
The zero-energy sector is preserved by $\mathsf{O}(2n_0)$, whereas the stabilizer associated with a pair of bands at energies $\pm\varepsilon_j$, each with multiplicity $n_j$, is $\mathsf{U}(n_j)$. 
We therefore obtain the classifying space
\begin{equation}
\label{eqn:M-D-n0-nj}
\mathfrak{X}^{\textrm{D}}_{(n_0;n_1,\ldots,n_k)}
=\frac{\mathsf{O}(2N)}
{\mathsf{O}(2n_0)\times\textstyle\prod_{j=1}^{k}\mathsf{U}(n_j)}.
\end{equation}

We next consider the class $\mathrm{C}$. 
The particle-hole symmetry operator may be represented as $\mathcal{C}=\Omega\mathcal{K}$, where $\Omega$ is the symplectic form. 
Accordingly, the eigenvector basis is parametrized by a symplectic matrix $\Psi\in\mathsf{Sp}(N)$. 
As discussed in Sec.~\ref{sec:class-C}, particle-hole symmetry arranges the eigenvalues into pairs of opposite energies. 
The stabilizer of the zero-energy sector of complex dimension $2n_0$ is $\mathsf{Sp}(n_0)$, while the stabilizer associated with each pair of non-zero-energy bands of multiplicity $n_j$ is $\mathsf{U}(n_j)$. 
Taking these degrees of freedom into account, we find
\begin{equation}
\label{eqn:M-C-n0-nj}
\mathfrak{X}^{\textrm{C}}_{(n_0;n_1,\ldots,n_k)}
=\frac{\mathsf{Sp}(N)}
{\mathsf{Sp}(n_0)\times\textstyle\prod_{j=1}^{k}\mathsf{U}(n_j)}.
\end{equation}

We now turn to the class $\mathrm{CI}$, for which the relevant matrix decomposition is the Autonne-Takagi factorization of complex symmetric matrices~\cite{HornJohnson:2012}. 
Applying this factorization to the off-diagonal block $h$ introduced in Sec.~\ref{sec:CI_class}, we write
\begin{equation}
h=\Psi\Sigma \Psi^{\mathrm{T}},
\end{equation}
where $\Psi\in\mathsf{U}(N)$, and $\Sigma$ is diagonal with non-negative entries $\varepsilon_j$. 
The zero-energy block is preserved by an arbitrary unitary transformation in $\mathsf{U}(n_0)$. 
In contrast, preserving a non-zero-energy block at energy $\varepsilon_j$ and with multiplicity $n_j$ requires the corresponding basis transformation $\Phi$ to satisfy $\Phi\Phi^{\mathrm{T}}=\mathbb{1}_{n_j}$,
and hence to belong to the orthogonal group $\mathsf{O}(n_j)$. 
The classifying space for class $\mathrm{CI}$ therefore takes the form
\begin{equation}
\label{eqn:M-CI-n0-nj}
\mathfrak{X}^{\textrm{CI}}_{(n_0;n_1,\ldots,n_k)}
=\frac{\mathsf{U}(N)}
{\mathsf{U}(n_0)\times\textstyle\prod_{j=1}^{k}\mathsf{O}(n_j)}.
\end{equation}

Finally, we consider class $\mathrm{DIII}$. 
The appropriate matrix decomposition is the skew-Takagi, or Youla, decomposition~\cite{Youla:1961}, which was also utilized when deriving the codimension counting in Sec.~\ref{sec:class-DIII}. 
As discussed there, the off-diagonal block $h$ is a complex skew-symmetric matrix, and it admits the decomposition
\begin{equation}
h=\Psi\Sigma\Psi^{\mathrm{T}},
\end{equation}
where $\Psi\in\mathsf{U}(N)$ and $\Sigma$ are block-diagonal, analogous to class $\mathrm{D}$. 
In contrast to the preceding classes, every non-zero singular value of a complex skew-symmetric matrix has even multiplicity. 
We therefore let $n_j$ count the number of Kramers pairs at either $+\varepsilon_j$ or $-\varepsilon_j$. 
The physical multiplicity of each band at a fixed sign of the energy is thus $2n_j$. 
We also note that a zero-energy degeneracy can be produced only by bringing together two particle-hole-related Kramers pairs, i.e., the zero-energy states come in multiples of four. 
Hence, disregarding the Kramers degeneracy and splitting the zero-energy states equally among the occupied and unoccupied sectors, we obtain the same relation for the band partition as in other classes: $n_0 + n_1 + \ldots + n_k = N$ [Eq.~(\ref{eq:band_partition})].

Then, the stabilizer of the zero-energy sector is $\mathsf{U}(2n_0)$. Each non-zero block of $\Sigma$ is instead preserved by the compact symplectic group $\mathsf{Sp}(n_j)$, acting on the $2n_j$-dimensional complex space formed by a chiral pair $\pm\varepsilon_j$. 
This group is $\mathsf{Sp}(n_j)$ rather than $\mathsf{U}(2n_j)$, because the corresponding transformations should preserve the symplectic form given by blocks corresponding to the $n_j$-multiplet in $\Sigma$. 
Combining these stabilizers, we obtain the classifying space
\begin{equation}
\label{eqn:M-DIII-n0-nj}
\mathfrak{X}^{\textrm{DIII}}_{(n_0;n_1,\ldots,n_k)}
=\frac{\mathsf{U}(2N)}
{\mathsf{U}(2n_0)\times\textstyle\prod_{j=1}^{k}\mathsf{Sp}(n_j)}.
\end{equation}

\section{Linking numbers and differential forms}\label{app:linking}

In this appendix, we explain in detail the definition of linking number via differential forms. 
We start by introducing a few important notions, such as Poincaré duality and Thom form, and continue by applying them to the computations of linking numbers. 
Throughout the section, we assume that the (co)homological groups have coefficients in real numbers $\reals$. 
This allows us to invoke the de Rham isomorphism, which states that the cohomology groups can be identified with the de Rham cohomology of the differential forms~\cite{Warnerbook:1983}. 
In our discussion of the linking number, we mostly follow the book by Bott and Tu~\cite{Bott:1982}.

We start by reminding the notion of a Poincaré dual of a compact manifold. Let us introduce an ambient manifold $L$ (not necessarily compact) of dimension $\ell$ and a compact submanifold $M\subset L$ of dimension $p$. The Poincaré dual $\mathrm{PD}(M)$ of the manifold $M$ is a closed $(\ell-p)$-form with compact support that satisfies the identity \footnote{It is important to note that the manifold $M$ comes with an orientation, i.e., its fundamental class can be taken with both signs. The switching of orientation changes the sign of the integral on the right side of the equation. More formally, this can be expressed by introducing the inclusion map $i_{M}:M\to L$, and one should write the integral of the pullback form $i^*_M\omega$, not just $\omega$. To make notation shorter, we write just $\omega$ assuming the pullback induced by the inclusion map $i_M$.}
\begin{equation}\label{eq:poincare_dual}
    \int_L w\wedge \mathrm{PD}(M)=\int_M w,
\end{equation}
for any closed $p$-form $w$, not necessarily with compact support. The necessity for introducing the differential forms with compact support comes from the fact that the ambient manifold $L$ may not be compact. As we will see later, this is important for defining the linking numbers.

The identity~\eqref{eq:poincare_dual} is naturally a (co)homological identity. Namely, if $M$ represents an element of the homology group $H^p(L,\reals)$, its Poincaré dual $\mathrm{PD}(M)$ represents an element of the compactly supported de Rham cohomology group $H^{\ell-p}_{c,\mathrm{dR}}(L)$. Let us recall that the elements of the compactly supported de Rham cohomology group are represented by the closed differential forms with compact support. In other words, one can construct a complex of differential forms with compact support
\begin{equation}
0 \longrightarrow \Omega_c^0(L)\xrightarrow{d}\Omega_c^1(L)\xrightarrow{d}\cdots\xrightarrow{d}\Omega_c^n(L)\xrightarrow{d}\cdots,
\end{equation}
where $\Omega_c^n(L)$ denotes the space of compactly supported smooth $n$-forms on $L$. The cohomology groups of this complex are the compactly supported de Rham cohomology groups $H^{*}_{c,\mathrm{dR}}(L)$.

Intuitively, a Poincaré dual encodes the orientation of the manifold that comes from the normal component. The concrete construction can be carried out via the so-called Thom form, which is a closed form representing an element in $H_{c, \mathrm{dR}}^{\ell-p}(L)$, and is defined as follows. First, we make a tubular neighborhood $N_M$ around the given submanifold $M\subset L$. The tubular neighborhood can be identified with a normal bundle of $M$, which we denote by $\nu_M$. Then, we define a Thom form as a closed differential form with support in $N_M$, such that an integral over each normal fiber $F$ of $\nu_M$ is equal to $1$. The orientation of the ambient manifold $L$ and of submanifold $M$
determine an orientation of the normal bundle $\nu_M$ by the convention
\begin{equation}\label{eq:thom_orientation}
  \operatorname{or}(T L|_M)
  = \operatorname{or}(T M)\wedge \operatorname{or}(\nu_M),
\end{equation}
where the orientation is encoded by the wedge product of coordinate differentials. By construction, the Thom form is a closed $(\ell-p)$-form with compact support; hence, it indeed represents an element of $H_{c, \mathrm{dR}}^{\ell-p}(L)$.

Let us show that the Thom form satisfies the integral relation for Poincaré duality given by Eq.~\eqref{eq:poincare_dual}. First, we should notice that the restriction of the closed form $\omega$ on the tubular neighborhood can be written as the sum of the differential form with constant normal component $\tilde\omega_{N_M}$\footnote{One can write it formally, by introducing projection from the tubular neighborhood $N_M$ to $M$, $\pi:N_M\to M$; from the point of view of normal bundle this is the projection from the total space to base space. Then the form with constant values on fibers can be written as the pullback of the projection of the form $i^*_M\omega$ defined on manifold $M$, i.e. double pullback of the form $\omega$, as $\tilde \omega_{N_M}=\pi^*i^*_M\omega$.} and the exact form
\begin{equation}\label{eq:homotopy-exact-correction}
    \omega_{N_M}=\tilde\omega_{N_M}+d\alpha.
\end{equation}
This relation holds because $M$ is a deformation retract of $N_M$, and homology groups are homotopy invariant. However, we can show it more explicitly using the fact that both forms $\omega_{N_M}$ and $\tilde\omega_{N_M}$ have the same integrals for all submanifolds $\Gamma\subset N_M$. Since both forms are closed, their integrals are the same for homologous submanifolds. However, one can always project $\Gamma$ to $M$, and since  $\omega_{N_M}$ and $\tilde\omega_{N_M}$ take the same values on $M$, the integral over projected $\Gamma$ is the same for both manifolds. In other words, both forms lie in the same cohomology group; hence their difference is an exact form.

Now, let us check the Poincaré relation \eqref{eq:poincare_dual}. Since $\eta_M$ is supported in $U_M$, Eq.~\eqref{eq:homotopy-exact-correction} gives
\begin{align}
  \int_L \omega\wedge\eta_M
  &=\int_{U_M}\omega\wedge\eta_M \\
  &=\int_{U_M}\tilde\omega_{N_M}\wedge\eta_M
    +\int_{U_M}d\alpha\wedge\eta_M .
  \label{eq:PD-proof-step}
\end{align}
Let us show that the second term vanishes. First, we notice that
\begin{equation}
  d(\alpha\wedge\eta_M)=d\alpha\wedge\eta_M,
\end{equation}
since $d\eta_M=0$. Then, due to the fact that $\alpha\wedge\eta_M$ has compact support in $U_M$, Stokes' theorem implies
\begin{equation}
  \int_{U_M}d \alpha\wedge\eta_M=0,
\end{equation}
since one can take a large enough boundary outside of the support of $\eta_M$.
For the first term in Eq.~\eqref{eq:PD-proof-step}, the defining property of fiber integration gives
\begin{equation}
  \int_{U_M}\tilde\omega_{N_M}\wedge\eta_M
  =\int_M \omega\wedge \int_{F}\eta_M
  =\int_M \omega,
\end{equation}
which shows that the Thom form is a Poincaré dual of the submanifold $M$. Let us note that when integrated over $M$, the form $\omega$ should be understood as the pullback $\omega$ onto $M$, via $M \hookrightarrow L$, which carries information about the orientation of the manifold $M$.

Locally, the Thom form can be represented in a quite simple manner. In a trivialized piece of the normal bundle (tubular neighborhood), choose coordinates
\begin{equation}
  (x^1,\ldots,x^p;y^1,\ldots,y^r),
  \qquad
  M=\{y^1=\cdots=y^r=0\},
  \label{eq:local-coordinates}
\end{equation}
where $(x^1,\ldots,x^p)$ are oriented coordinates on $M$ and $(y^1,\ldots,y^r)$ are oriented normal coordinates.  The orientation of $L$ is then represented by the form
\begin{equation}
  d  x^1\wedge\cdots\wedge d  x^p
  \wedge
  d  y^1\wedge\cdots\wedge d  y^r .
\end{equation}
Choose a compactly supported bump function $\rho(y)$ such that
\begin{equation}
  \int_{\mathbb{R}^r}\rho(y)\, d ^r y=1 .
\end{equation}
Then, the local model for a Thom form is
\begin{equation}
  \eta_M=\rho(y)\, d  y^1\wedge\cdots\wedge d  y^r .
  \label{eq:local-thom-form}
\end{equation}
It is closed, has compact support in the normal directions, and has a unit fiber integral. This normalization is independent of the particular tubular neighborhood: integrating $\eta_M$ over any sufficiently small oriented transverse disk gives $1$. Different choices of tubular neighborhood give representatives of the same cohomological class, differing by a compactly supported exact form. On a nontrivial normal bundle, a global Thom form can be obtained by patching such local expressions using an atlas on the manifold $M$.

After studying the general properties of the Thom form, let us consider a few simple examples.
The easiest example is when the manifold $M$ is a point $*$. In this case, the Thom form  is $\eta_{*}=\psi dV$, where $dV$ is the volume form and the function $\psi$ is compactly supported around the point $*$, such that its integral equals to $1$
\begin{equation}\label{eq:thom_point_example}
    \int_L \psi d  V=1.
\end{equation}
The class of closed $0$-forms on a manifold only consists of constant functions, so the relation in Eq.~\eqref{eq:poincare_dual} follows straightforwardly from the normalization of the integral over a fiber in Eq.~\eqref{eq:thom_point_example}. Interestingly, in this case, the Poincaré dual of a point acts similarly to a $\delta$-function
\begin{equation}
    \int_{L} w\wedge \eta_{*}=w(*).
\end{equation}

Another example is a circle $S$ around the point $\{0\}$ in the plane with excluding point $\mathbb{R}^2\backslash\{0\}$. In this case, one can write the Thom form explicitly as
\begin{equation}\label{eq:thomform_circle}
\eta_{S}=-\rho(r) dr
\end{equation}
where $\rho$ is the function compactly supported in the neighborhood of $S$ such that $\int^{\infty}_0 \rho(r)dr=1$. The {sign in Eq.~\eqref{eq:thomform_circle} (i.e. orientation of the normal bundle)} agrees with the counterclockwise orientation of $S$ given by angle form $d\phi$ and the standard orientation in $\mathbb{R}^2$ given by the form $dr\wedge d\phi$.

It is worth noting that in the limit, when compactly supported differential forms become strictly localized on the manifold, they correspond to de Rham currents, i.e., become differential forms with coefficients in distributions. Therefore, in the calculations involving Thom forms, one can choose to work either with compactly supported differential forms or with de Rham currents.

Before applying the Thom forms to the linking of manifolds, we briefly recall the definition of the linking number in terms of an intersection number; see also Sec.~\ref{sec:general_linking}. Let $A$ and $\mathcal{B}$ be compact oriented submanifolds of an oriented $\ell$-dimensional manifold $L$, possibly with boundary, whose dimensions satisfy$\mathrm{dim}(A){+}\mathrm{dim}(\mathcal{B}){=}\ell$.
After an arbitrarily small perturbation, if necessary, the submanifolds may be assumed to intersect transversely. Their intersection then consists of a finite set of isolated points ${p_i}$. Transversality implies that, at every intersection point, the tangent space of $S^\ell$ is isomorphic to the direct sum of tangent spaces of $A$ and $\mathcal{B}$: $T_{p_i}L=T_{p_i}A\oplus T_{p_i}\mathcal{B}$.
The orientations of $A$ and $\mathcal{B}$ therefore induce an orientation of $T_{p_i}L$, which may be written in terms of differential forms as
\begin{equation}
  \operatorname{\widetilde {or}}(T_{p_i} L)
  = \operatorname{or}(T_{p_i}A)\wedge \operatorname{or}(T_{p_i}\mathcal{B}).
\end{equation}
We define a function $\mathrm{sign}(p_i)=\pm 1$, which is equal to $+1$ if the induced orientation $\operatorname{\widetilde {or}}(T_{p_i} L)$ is the same as the original orientation on $L$, i.e. $\operatorname{\widetilde {or}}(T_{p_i} L)=\operatorname{or}(T_{p_i} L)$, and equal to $-1$ if these orientations are opposite. 
The intersection number of $A$ and $\mathcal{B}$ is the sum of all signs of intersection points\footnote{We should also note that while this definition is given for the intersection of manifolds, it also works for general chains (singular or simplicial) with slight modification. More concretely, we assume that the intersection points belong to the interior of the simplices, the orientation can be defined for these simplices and should be weighted with their multiplicities.}
\begin{equation}
\mathrm{Int}(A,\mathcal{B})=\sum_{i}\mathrm{sign}(p_i).
\end{equation}

Now, we are ready to define the linking number. Let $A$ and $B$ be disjoint oriented closed null-homologous manifolds in the ambient manifold $L$. Additionally, their dimensions should satisfy $p+q+1=\ell$, where we denote $\dim A=p$, $\dim B=q$ and $\dim L=\ell$.  Then the linking number of $A$ and $B$ in $L$ is defined as
\begin{equation}\label{eq:linking_as_intersection_app}
\mathrm{Lk}(A,B)=\mathrm{Int}(A,\mathcal{B}),
\end{equation}
where $\mathcal{B}$ is a manifold (generally, a chain), such that $B=\partial \mathcal{B}$ is the boundary of $\mathcal{B}$. Up to a possible sign change, the linking number is symmetric under the exchange of $A$ and $B$.

Having defined the necessary notions, we are ready to write down the linking number of two manifolds $A$ and $B$ in the ambient manifold $L$ in terms of the differential forms as

\begin{equation}\label{eq:Lk_diffform_app}
    \mathrm{Lk}(A,B)=\int_{L}\omega_A \wedge \eta_{B}.
\end{equation}
In this expression, $\eta_A$ and $\eta_B$ are Thom forms associated with the manifolds $A$ and $B$ as above, and the forms $\omega_A$ and $\omega_B$ satisfy identities $d\omega_A=\eta_A$ and $d\omega_B=\eta_B$. Equivalently, one can generally speak about Poincaré duals of $A$ and $B$ in the corresponding disjoint neighborhoods $N_A$ and $N_B$ \footnote{It is important in this definition that the forms $\eta_A$ and $\eta_B$ are Poincaré duals in neighborhoods $N_A$ and $N_B$. While these forms are also Poincaré duals in the whole ambient manifold $L$, the definition of linking number may not work if one considers a general Poincaré dual form in $L$ in the definition.}. One can define the forms $\omega$, because Thom forms are exact in $L$. This follows from the fact that $A$ and $B$ are null-homologous in $L$, hence the corresponding dual forms correspond to trivial elements in cohomology groups.

Intuitively, it is easier to understand the linking formula in a slightly different form \footnote{It should be noted that the chain $\mathcal{B}$ may not be represented by a smooth manifold. In this case, one obtains the intersection number on the level of chains. All the discussion works in the general case as well, with only minor adjustments.}
\begin{equation}
    \mathrm{Lk}(A,B)=\int_{B}\omega_A=\int_{\mathcal{B}} \eta_A,
\end{equation}
where we used several tricks: first, we restrict the area of integration to a tubular neighborhood $N_B$ of $B$, where the support of $\eta_B$ is nontrivial; then we use Eq.~\eqref{eq:poincare_dual} and the Stokes formula. The last term represents the intersection number of $A$ and $\mathcal{B}$, which is defined via the relation $\partial \mathcal{B}=B$. Each contribution to the integral comes from a normal fiber of $A$ at the intersection point. Due to the fiber normalization of the Thom form $\eta_A$, an integral of $\eta_A$ around each intersection point is equal to $\pm1$, where the appropriate sign of the intersection is induced by the mutual orientation of $A$ and $\mathcal{B}$.

One can see it from the local model of the Thom form. Let us consider the neighborhood $\mathcal{B}_{p_i}$ of one intersection point $p_i$ and introduce the local coordinate system $(\bf{x},\bf{y})$, such that $(x^1,\ldots,x^p)$ are oriented coordinates on $A$ and $(y^1,\ldots,y^r)$ are oriented normal coordinates that agree with orientation of $L$ by Eq.~\eqref{eq:thom_orientation}, and that in the neighborhood of $p_i$, the manifold $\mathcal{B}$ is given by equations $\mathcal{B}=\{x^1=\cdots=x^p=0\}$. Depending on the mutual orientation of the normal bundle of $A$ and $\mathcal{B}$, the integral around the intersection point $p_i$ should be multiplied by the $\epsilon_{p_i}=\pm1$. The multiplier $\epsilon_{p_i}$ is equal to $1$ if the aforementioned orientations agree, and equal to $-1$ otherwise. Thus, in the neighborhood of $p_i$, we have
\begin{equation}
  \int_{\mathcal{B}_{p_i}}\eta_A
  =\epsilon_{p_i}\int_{\mathbb{R}^{\ell-p}}\rho(y)\,d ^{\ell-p}y
  =\epsilon_{p_i} .
\end{equation}
Summing over all intersection points gives the intersection number of $A$ and $\mathcal{B}$
\begin{equation}
  \int_{\mathcal{B}}\eta_A
  =\sum_{p_i\in A\cap \mathcal{B}}\epsilon_{p_i}
  =I(A,\mathcal{B}),
  \label{eq:thom-form-counts-intersections}
\end{equation}
which agrees with the definition of linking number in Eq.~\eqref{eq:linking_as_intersection_app}.

Notice that the linking number gives a nontrivial result despite the manifolds $A$ and $B$ lying in trivial homology classes in $L$. The reason is that the form $\omega_A$ does not belong to the cohomology group of the ambient manifold $L$, since it is not closed. However, it is closed in $L \backslash  \mathrm{supp}(\eta_A)$, which is the complement of the support of $\eta_A$, and belongs to the cohomology group of $L \backslash  \mathrm{supp}(\eta_A)$. Therefore, the linking number can be interpreted as the honest (co)homological pairing in $L \backslash  \mathrm{supp}(\eta_A)$
\begin{equation}
    \mathrm{Lk}(A,B)= \omega_A ([B]),
\end{equation}
where $[B]$ is the fundamental class of $B$ in $L \backslash  \mathrm{supp}(\eta_A)$.
This ensures that the linking number is a topological invariant, which stays constant under the continuous deformations of manifolds $A$ and $B$ (i.e. to change the linking number, $A$ and $B$ should intersect under the deformation).

Let us also directly check that the definition in Eq.~\eqref{eq:Lk_diffform_app} does not depend on particular choices of $\omega$'s and $\eta$'s. First fix $\eta_A$ and $\eta_B$, and let $\omega_A'$ be
another primitive of $\eta_A$.  Then
the form $\lambda=\omega_A'-\omega_A$ is closed in $L$.  Hence
\begin{equation}
  \int_L \lambda\wedge\eta_B
  = \int_B \lambda = 0,
\end{equation}
because $B$ is homologous to zero in $L$.  Thus, the definition is independent of the primitive $\omega_A$.

Next, replace the Thom form $\eta_B$ by another Thom form $\eta_B'=\eta_B+\mathrm d\beta_B$,
where $\beta_B$ is compactly supported in the tubular neighborhood of
$B$.  Since this neighborhood is disjoint from the support of $\eta_A$, we have
\begin{gather}
  \int_L \omega_A\wedge(\eta_B'-\eta_B)=
  \int_L \omega_A\wedge \mathrm d\beta_B =\\\nonumber
  (-1)^{\ell-p}\int_L \mathrm d\omega_A\wedge \beta_B =
  (-1)^{\ell-p}\int_L \eta_A\wedge \beta_B=  0,
\end{gather}
and the linking number is independent of the representative $\eta_B$.

Finally, replace $\eta_A$ by another Thom form $\eta_A'=\eta_A+\mathrm d\gamma_A$,
with $\gamma_A$ compactly supported in the tubular neighborhood of $A$.
Then $\omega_A+\gamma_A$ is a primitive of $\eta_A'$.  The corresponding change in Eq.~\eqref{eq:Lk_diffform_app} is
\begin{equation}
  \int_L \gamma_A\wedge\eta_B=0,
\end{equation}
again because the supports are disjoint. We confirmed that the linking number
$\textrm{Lk}(A,B)$ is well defined.

To conclude the discussion on linking numbers, let us take a look at a concrete low-dimensional example. We will show that in $S^3$, the linking number coincides with the Hopf invariant. Let us define the function $f:S^3\to S^2$. We first choose the disjoint neighborhoods $N_a$ and $N_b$ of two points in $a$ and $b$ in $S^2$, which are also regular values of $f$. We are interested in the linking number of $A=f^{-1}(a)$ and $B=f^{-1}(b)$. We choose the corresponding Thom forms in $S^3$ to be the pullbacks of Thom forms of $a$ and $b$ in $S^2$. Let us denote the Thom forms in $S^2$ as $\alpha_a \in \Omega^2_c(N_a)$ and $\alpha_b\in \Omega^2_c(N_b)$, whereas $\eta_A=f^*\alpha_a$ and $\eta_B=f^*\alpha_b$. The linking number of $A$ and $B$ is
\begin{equation}
    \mathrm{Lk}(A,B)=\int_{S^3}\omega_A\wedge\eta_B,
\end{equation}
where $d\omega_A=\eta_A$. The Hopf invariant, in turn, is defined as
\begin{equation}
    H(f)=\int_{S^3}\omega\wedge d\omega,
\end{equation}
where the form $\omega$ is defined by relation $d\omega=f^*\alpha$, with the form $\alpha$ being a generator of $H^2_{\textrm{dR}}(S^2)$.

Let us notice that forms $\alpha_a$ and $\alpha_b$ are both representatives for the generator of $H^2_{\textrm{dR}}(S^2)$, so we can write the Hopf invariant as
\begin{equation}
H(f)=\int_{S^3}\omega_A\wedge \eta_A.
\end{equation}
Because $\alpha_a$ and $\alpha_b$ belong to the same cohomology class in $S^2$, their difference is an exact form
\begin{equation}
\alpha_a-\alpha_b=d\beta.
\end{equation}
Hence,
\begin{gather}\label{eq:Hopf_Lk_calc}
    \omega_A\wedge(\eta_A-\eta_B)=\omega_A\wedge f^* d\beta=\\
    =-d(\omega_A\wedge f^*\beta)+d\omega_A\wedge f^*\beta.
\end{gather}
By the property of pullback, the last term in Eq.~\eqref{eq:Hopf_Lk_calc} is equal to
\begin{equation}
    \eta_A\wedge f^*\beta=f^*(\alpha_a\wedge \beta).
\end{equation}
However, the form $\alpha_a\wedge \beta$ is a $3$-form in $S^2$, and therefore vanishes. Using the Stokes theorem, we obtain that the linking number and the Hopf invariant are equal
\begin{equation}
        \mathrm{Lk}(A,B)=\int_{S^3}\omega_A\wedge\eta_B=\int_{S^3}\omega_A\wedge \eta_A=H(f).
\end{equation}

\section{Unstable linking}\label{app:unstable-linking}

In this appendix, we show that non-trivial bundle invariants can also indicate unstable linking of nodal manifolds. 
For such unstable linkings, continuously deforming the Hamiltonian allows for removing the higher-order degeneracy from the enclosing sphere. 
We illustrate this for the minimal model of class $\textrm{BDI}$ by demoting one of the momentum variables of the minimal Hamiltonian (in Eq.~(\ref{eqn:BDI-4band-Ham-Dirac})) to a parameter value and investigating the linking of nodal manifolds for fixed values of this parameter. 
Specifically, we treat $q_2 = \lambda$ as a fixed parameter of the Hamiltonian and investigate how continuous changes in this parameter cause abrupt changes in the linking of the nodal manifolds. 

\begin{figure}
    \centering
    \includegraphics[width=\linewidth]{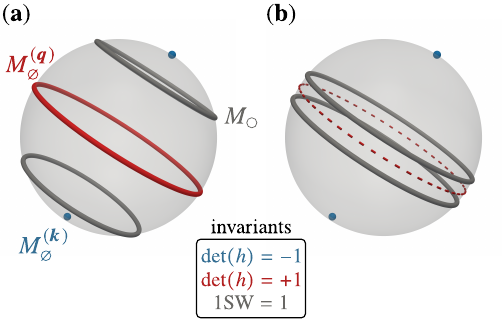}
    \caption{Unstable linking of nodal manifolds in class $\textrm{BDI}$. (\textbf{a})~Nodal manifolds on the enclosing sphere $S^2$ for $\lambda=0$. The same notation and color code is used as for Fig.~\ref{fig:class-BDI-linking}.
    (\textbf{b})~Nodal manifolds for $\lambda = 0.98$. The equatorial component of $M_\varnothing$ disappears for non-zero $\lambda$, allowing the two circles of $M_\medcirc$ to pairwise annihilate while trivializing the linking of the nodal manifolds. 
    }
    \label{fig:class-BDI-unstable-linking}
\end{figure}

First, we set $\lambda=0$ and analyze the nodal manifolds and their band invariants. 
For $\lambda=0$, the four-fold degeneracy sitting at $k_1 = k_2 = q_1 = 0$ is enclosed by an $S^2$ on which the momentum variables satisfy 
\begin{equation}
\label{eq:BDI-unstable-enclosing-sphere}
    k_1^2 + k_2^2 + q_1^2 = 1.
\end{equation}
The nodal manifolds are derived using Eq.~(\ref{eq:BDI-spectrum}) with the constraint $q_2 = 0$ (corresponding to $\lambda=0$) on the $2$-sphere specified by Eq.~(\ref{eq:BDI-unstable-enclosing-sphere}). 
On one hand, the points on $M_\medcirc$ satisfy $q_1^2=k_1^2+k_2^2$, which combined with Eq.~(\ref{eq:BDI-unstable-enclosing-sphere}) results in the conditions
\begin{equation}
    q_1^2 = \tfrac{1}{2} \quad \textrm{and} \quad k_1^2+k_2^2=\tfrac{1}{2} 
\end{equation}
which is the union of two circles on the enclosing sphere, one at a positive and one at a negative value of $q_1$. 
On the other hand, the points on $M_\varnothing$ satisfy
\begin{align}
\label{eq:BDI-unstable-nonzeroenergy-unperturbed}
    q_1^2 &= 0 \quad  \textrm{or} \quad  k_1^2 + k_2^2 = 0
\end{align}
combined with the constraint~(\ref{eq:BDI-unstable-enclosing-sphere}). 
The $q_1^2=0$ condition defines a circle along the equator of the enclosing sphere [labeled $M_\varnothing^{(\bs{q})}$], while $k_1^2+k_2^2=0$ corresponds to the north and the south pole of the sphere [labeled $M_\varnothing^{(\bs{k})}$]. 
The nodal manifolds and their linking are shown in Fig.~\ref{fig:class-BDI-unstable-linking}(\textbf{a}).

Now we establish that the nodal manifolds host nontrivial band invariants. 
On one hand, the topological invariant on $M_\varnothing$ is $\textrm{sign}[\det(h)]$, where $h$ is the real-valued off-diagonal Hamiltonian block [cf.~Eq.~(\ref{enq:Ham-BDI-off-blocks})].\footnote{As discussed in Sec.~\ref{sec:class-BDI} (and as also displayed in Table~\ref{tab:nodal_manifolds}), by instead setting $\mathcal{C} = \mathcal{K}$, the same invariant can be equivalently expressed as $\sign [\Pf (H)]$ of the full Hamiltonian. Here, we adopt the determinant formulation.}
Recalling the set of the basis matrices specified in Eq.~(\ref{eqn:BDI-basis-matrices}), we express $\det(h) = k_1^2 +k_2^2 -q_1^2$. 
We find that the determinant equals
$-1$ at the poles [i.e., on $M_\varnothing^{(\bs{k})}$], while
it is $+1$ on the equator [i.e., on $M_\varnothing^{(\bs{q})}$].
The opposite signs imply that the two components together exhibit a non-trivial value of the zeroth-homotopy $\sign [\det(h)]$ invariant.
On the other hand, on $M_\medcirc$, one instead computes the first Stiefel-Whitney class of one of the non-degenerate bands. 
For example, we can consider one of the circles in $M_\medcirc$ at $q_1 = 1/\sqrt{2}$ and examine the positive energy eigenstate of the Hamiltonian  
\begin{equation}
    \ket{u_+} = \frac{1}{\sqrt{2}} \left( \cos \alpha, -\sin\alpha, \cos\alpha, \sin\alpha \right),
\end{equation}
where $\alpha=\phi/2$ and we adopted the parameterization of the nodal manifold as $(q_1,k_1,k_2)=\frac{1}{\sqrt2} (1, \cos\phi, \sin\phi)$. 
By noting that $\ket{u_+(2\pi)} = - \ket{u_+(0)}$, we conclude that the first Stiefel-Whitney invariant is nontrivial. 
A similar calculation applied to the circle at $q_1 = -1/\sqrt{2}$ reveals the same nontrivial Stiefel-Whitney invariant for the non-degenerate eigenstates. 
We illustrate both the nodal manifolds $M_\medcirc$ and $M_\varnothing$ and their nontrivial band invariants in Fig.~\ref{fig:class-BDI-unstable-linking}(\textbf{a}).

The previous analysis has revealed that for fixed $\lambda=0$: (\emph{i})~the nodal manifolds $M_\medcirc$ and $M_\varnothing$ carry nontrivial band invariants on the $S^2$ that encloses the fourfold degeneracy, and (\emph{ii})~the two manifolds are linked as shown in Fig.~\ref{fig:class-BDI-unstable-linking}(\textbf{a}). 
Next, we demonstrate that, although the nodal manifolds carry nontrivial band invariants, the fourfold degeneracy (and hence the linking itself) is unstable and can be removed by continuously changing the parameter $\lambda$.
To this end, we now set $0 \neq \lambda \leq1$ and investigate how the nodal manifolds change upon this deformation of the Hamiltonian.

\begin{figure}[!t]
    \centering
    \includegraphics[width=\linewidth]{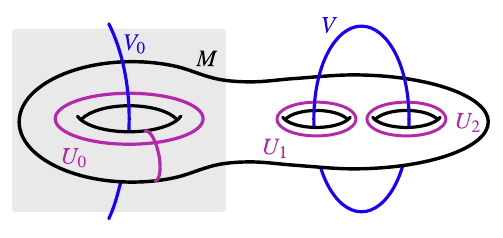}
    \caption{Stable versus accidental linking. 
    The left side (gray shaded region) illustrates a stable link between the nontrivial cycle $U_0$ (purple) of the nodal manifold $M$ and the nontrivial cycle $V_0$ (blue) of another nodal manifold (not shown for simplicity). 
    The right side illustrates an accidental link, where the cycles $U_1$ and $U_2$ of $M$ are linked with the cycle $V$.
    The removal of the accidental link using a sequence of transformations is illustrated in Fig.~\ref{fig:unstable-2}.
    }
    \label{fig:unstable-1}
\end{figure}

\begin{figure*}[!t]
    \centering
    \includegraphics[width=\linewidth]{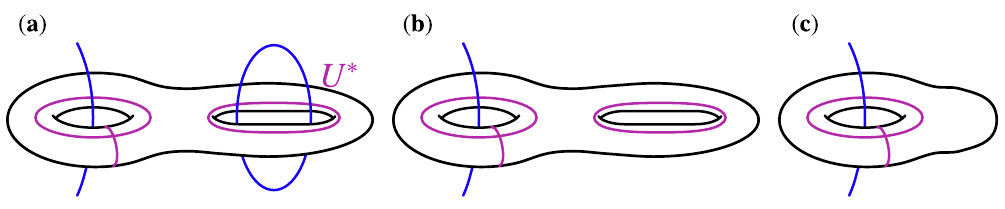}
    \caption{
    Removal of the accidental 
    linking in Fig.~\ref{fig:unstable-1}. 
    (\textbf{a})~The holes corresponding to the cycles $U_1$ and $U_2$ are merged into a single hole panned by cycle $U^*$. 
    (\textbf{b})~Now the loop $V$ can now be continuously collapsed to a single point without intersecting $M$ (i.e., without causing the appearance of a multifold point on the enclosing sphere). 
    (\textbf{c})~Finally, since $V$ no longer passes through the hole corresponding to $U^*$ it can be continuously closed.}
    \label{fig:unstable-2}
\end{figure*}

First of all, using the dispersion relation for the minimal Hamiltonian in Eq.~(\ref{eq:BDI-spectrum}), we notice that setting $\lambda \neq 0$ removes the fourfold degeneracy inside the enclosing sphere $S^2$. 
Correspondingly, the linking of the nodal manifolds should be removable. 
For the nonzero-energy nodal manifold $M_\varnothing$, using Eq.~(\ref{eq:BDI-spectrum}), we have 
\begin{align}
    q_1^2 + \lambda^2 &= 0, \quad  \textrm{or} \quad  k_1^2 + k_2^2 = 0.
\end{align}
For $\lambda=0$, the first condition previously resulted in the nodal manifold along the equator. 
This component of $M_\varnothing$ vanishes for any nonzero value of $\lambda$.
The second condition gives the north and south poles as the only components in $M_\varnothing$ irrespective of the choice of $\lambda$. 
Furthermore, for the zero-energy nodal manifold $M_\medcirc$ we have
\begin{equation}
    k_1^2+k_2^2 = q_1^2+\lambda^2.
\end{equation}
Combined with Eq.~(\ref{eq:BDI-unstable-enclosing-sphere}),
this results in the set of points 
\begin{equation}
\label{eq:BDI-unstable-perturbed-zeroenergy}
    M_\medcirc = \left\{  q_1 = \pm \sqrt{\frac{1-\lambda^2}{2}}, k_1^2 + k_2^2 = \frac{1+\lambda^2}{2} \right\}.
\end{equation}
From this, we see that as we increase $\lambda$ from $0$ to $1$, the two circles of $M_\medcirc$ move towards the equator of the enclosing sphere.
The nodal manifolds for $\lambda = 0.98$ are shown in Fig.~\ref{fig:class-BDI-unstable-linking}(\textbf{b}).

Having derived the nodal manifolds, we can describe how their linking is removed under continuous deformation parametrized by $\lambda$. 
For $\lambda = 0$, the equatorial circle of $M_\varnothing$ lies between the two circles of $M_\medcirc$, while the north and south poles lie outside the region bounded by the circles of $M_\medcirc$. 
For any nonzero $\lambda$, the equatorial subset of $M_\varnothing$ vanishes immediately since its defining condition $q_1^2 + \lambda^2 = 0$ cannot be satisfied any longer. 
Therefore, the only components of $M_\varnothing$ are the north and south poles. 
With the equatorial circle removed from the enclosing sphere, the two circles of $M_\medcirc$ are now allowed to meet along the equator (as derived in Eq.~(\ref{eq:BDI-unstable-perturbed-zeroenergy}). 
Since the two circles both carry the nontrivial value of the $\mathbb{Z}_2$ first Stiefel-Whitney invariant, their total topological charge is trivial. 
Therefore, their pairwise annihilation is topologically allowed at $\lambda=1$.
For $\lambda>1$, the only remaining nodal manifold is $M_\varnothing$, consisting of the north and south poles; as the only remaining nodal manifold, it is not linked with anything.

The analysis above demonstrates that nontrivial bundle invariants do not necessarily detect a stable multifold degeneracy inside the enclosing sphere. For that to be the case, the nodal manifolds themselves should be stable under symmetry-preserving perturbations. 
In the example above, the equatorial component of $M_\varnothing$ exists on the enclosing sphere only at the fine-tuned $\lambda=0$ point and disappears for an arbitrary small perturbation in $\lambda$. 
The absence of this component then allows the two $\mathbb{Z}_2$-charged circles of $M_\medcirc$ to approach each other and annihilate at $\lambda =1$.

\section{Removal of accidental linking through transformations of nodal manifolds} \label{app:cobordism-transform}

In this appendix, we further discuss accidental linking of nontrivial cycles of nodal manifolds. 
We exemplify that some apparently stable links can be removed by deformations of the nodal manifolds without creating multifold degeneracies on the enclosing sphere. In contrast to the `unstable' linking discussed in Appendix~\ref{app:unstable-linking}, which arises due to fine-tuning of the Hamiltonian parameters, the `accidental' linking discussed here is robust against infinitesimal deformations of the Hamiltonian; however, it can be removed by large deformations (more precisely: cobordisms) of the nodal manifolds.

To describe such deformations more precisely, let $\mathcal{M}$ denote the space of matrices satisfying the relevant $\bs{k}$-preserving
symmetries, so that the Hamiltonian is defined by a map from the parameter space to the matrix space 
\begin{equation}
H: \mathrm{BZ}\to\mathcal{M}.   
\end{equation} 
Let us further consider the restriction of the Hamiltonian map $H$ to a small sphere $S^\ell\subset \mathrm{BZ}$ enclosing the multifold nodal point in the parameter space
\begin{equation}
\mathfrak{H}:S^\ell\to\mathcal{M}.
\end{equation}
If $\mathcal{L}_i\subset\mathcal{M}$ denotes the locus where the corresponding energy gap closes, then the nodal manifold on the enclosing sphere is its preimage $M_i=\mathfrak{H}^{-1}(\mathcal{L}_i)$. 
Continuous deformations of the Hamiltonian map define a homotopy $\mathfrak{H}_t$ of the restricted map $\mathfrak{H}$. 
When $\mathfrak{H}_t$ intersects $\mathcal{L}_i$ transversely\footnote{Here, transversality means that the images of the Hamiltonian maps cross the degeneracy locus in general position rather than touching it tangentially. More precisely, the directions generated by varying the parameters, together with the directions tangent to the degeneracy locus, span the ambient matrix space at each intersection point.}, the corresponding nodal manifolds vary smoothly and are related by an isotopy within the enclosing sphere. If transversality is lost at an intermediate stage, however, the nodal manifold may undergo a change of topology. The full trajectory of $M_i$ during such a deformation can then be viewed as a cobordism between the nodal manifolds before and after the transition. Such cobordism transformations may change the particular linking configuration and thus produce accidental linking. 
However, as long as $M_1$ and $M_2$ do not intersect throughout the deformation and their linking is robust, the topological protection of the multifold nodal point persists.

For concreteness, we consider the schematic scenario shown in Fig.~\ref{fig:unstable-1}. 
The nodal manifold $M$ contains several `holes' corresponding to nontrivial cycles. On the left-hand side of the figure (gray shaded region), we illustrate an unremovable linking between the cycle $U_0 \in M$ and a nontrivial cycle $V_0$ of another nodal manifold. Since $U_0$ and $V_0$ are linked, nontrivial band invariants can be computed on either of the cycles. 
Removing the link requires $V_0$ to pass through $U_0$ which would create a multifold degeneracy on the enclosing sphere.
In contrast, the linking on the right-hand side of  Fig.~\ref{fig:unstable-1} can be removed by transformations of the nodal manifold~$M$, and therefore should be treated as accidental.

The removal of the accidental linking is illustrated through several steps shown in Fig.~\ref{fig:unstable-2}. 
First, one merges the two holes corresponding to cycles $U_1$ and $U_2$. 
The resulting scenario is shown in Fig.~\ref{fig:unstable-2}(\textbf{a}), where we indicate the merged cycle as $U^*$. 
Initially, both cycles $U_1$ and $U_2$ support nontrivial band invariants, owing to the linking of each with the cycle $V$. 
However, after the merging, the resulting cycle $U^*$ is unlinked with $V$ and therefore does not carry nontrivial band invariants. 
Then, since it is no longer linked with the nodal manifold $M$, the cycle $V$ can be continuously contracted without intersecting $M$, yielding the configuration in Fig.~\ref{fig:unstable-2}(\textbf{b}). 
Finally, the hole corresponding to $U^*$ can itself be closed, resulting in the configuration shown in Fig.~\ref{fig:unstable-2}(\textbf{c}) where only the unremovable link remains.

\bibliography{bib}

\end{document}